\documentclass[12pt]{JHEP3} 
\usepackage{amsmath,amsthm,amsfonts,amssymb}
\usepackage{graphicx}\usepackage{psfrag}
\numberwithin{table}{section}
\def\be{\begin{equation}}\def\ee{\end{equation}}
\def\bea{\begin{eqnarray}}\def\eea{\end{eqnarray}}
\def\subsubsecb#1{\noindent{{\bf #1}} \\}
\def\subsubsec#1{\noindent{{\it #1}} \\}
\def\p{\partial}\def\cL{{\cal L}}\def\cF{{\cal F}}
\def\IC{\mathbb{C}}\def\IP{\mathbb{P}}\def\IZ{\mathbb{Z}}\def\IQ{\mathbb{Q}}
\def\eps{\epsilon}\def\om{\omega}\def\DL{\mathfrak{L}}\def\cc{\mathfrak{c}}\def\qq{\mathfrak{q}}
\def\FFs{f_{S^2}(q,\eps)}\def\FFd{f_{D^2}(q,\eps)}

\def\hG{{{\Gamma}_X}}\def\hGq{{{\Gamma}_{X,q}}}
\def\hGqb{{\overline{\Gamma}_{X,q}}}
\def\td{\textrm{td}}\def\ch{\textrm{ch}}
\def\qcor#1{\langle #1 \rangle}
\def\qcord#1{\langle\!\langle #1 \rangle\!\rangle}
\def\cx#1{{\cal#1}}\def\bx#1{{\bar#1}}
\def\X{\Upsilon}\def\tD{\tilde D}\def\tN{\tilde N}\def\kh{{\hat k}}\def\nh{{\hat n}}
\def\ket#1{|#1\rangle}\def\bra#1{\langle#1|}
\def\tLX{\widetilde{LX}}

\def\X{\Upsilon}\def\tD{\tilde D}\def\tN{\tilde N}\def\kh{{\hat k}}\def\nh{{\hat n}}\def\x{u}\def\gq{{f}}
\def\tr{\textrm{\,tr\,}}\def\ff{\mathfrak{f}}\def\EC{\cx E}
\def\hAq{\hat A_{S^1}}\def\chq{\ch_{S^1}}
\def\nabt{\nabla^t}\def\nabQ{\nabla^Q}
\def\Ii{\mathbb I}\def\pp{\Phi}
\def\TT{{\cal T}}\def\TTh{{\hat{\cal T}}}
\def\ann{\circledcirc}
\def\ffe{E}
\def\Ws{W^{\textrm {shifted}}}
\def\AA{\Upsilon}
\def\DLb{{\cal{V}}}
\def\IR{\mathbb{R}}
\def\gstring{{g_s}}
\def\Ch{{\textrm{Ch}}}
\def\qm{\text{QM}}

\newdimen\tableauside\tableauside=1.0ex
\newdimen\tableaurule\tableaurule=0.4pt
\newdimen\tableaustep
\def\phantomhrule#1{\hbox{\vbox to0pt{\hrule height\tableaurule width#1\vss}}}
\def\phantomvrule#1{\vbox{\hbox to0pt{\vrule width\tableaurule height#1\hss}}}
\def\sqr{\vbox{%
  \phantomhrule\tableaustep
  \hbox{\phantomvrule\tableaustep\kern\tableaustep\phantomvrule\tableaustep}%
  \hbox{\vbox{\phantomhrule\tableauside}\kern-\tableaurule}}}
\def\squares#1{\hbox{\count0=#1\noindent\loop\sqr
  \advance\count0 by-1 \ifnum\count0>0\repeat}}
\def\tableau#1{\vcenter{\offinterlineskip
  \tableaustep=\tableauside\advance\tableaustep by-\tableaurule
  \kern\normallineskip\hbox
    {\kern\normallineskip\vbox
      {\gettableau#1 0 }%
     \kern\normallineskip\kern\tableaurule}%
  \kern\normallineskip\kern\tableaurule}}
\def\gettableau#1{\ifnum#1=0\let\next=\null\else
\squares{#1}\let\next=\gettableau\fi\next}

\title{A 3d Gauge Theory/Quantum K-Theory Correspondence}
\author{Hans Jockers$^{1}$ and Peter Mayr$^{2}$\\
\\$^1\,$Bethe Center for Theoretical Physics,\\
Physikalisches Institut, Universit\"at Bonn, 53115 Bonn, Germany
\\$^2\,$Arnold Sommerfeld Center for Theoretical Physics,
Ludwig-Maximilians-Universit\"at,
80333 Munich, Germany}
\abstract{The 2d gauged linear sigma model (GLSM) gives a UV model for quantum cohomology on a K\"ahler manifold $X$, which is reproduced in the IR limit. We propose and explore a 3d lift of this correspondence, where the UV model is the $\mathcal{N}=2$ supersymmetric 3d gauge theory and the IR limit is given by Givental's permutation equivariant quantum K-theory on $X$. This gives a one-parameter deformation of the 2d GLSM/quantum cohomology correspondence and recovers it in a small radius limit. We study some novelties of the 3d case regarding integral BPS invariants, chiral rings,  deformation spaces and mirror symmetry.}

\preprint{BONN--TH--2018--12\\LMU-ASC 47/18}

\begin{document}
\baselineskip16pt 

\section{Introduction and summary}
The quantum product of certain chiral operators in the 2d topological A-model~\cite{Witten:1988xj} defines a deformation of the classical intersection ring  
$$
\Phi_\alpha \cdot \Phi_\beta = C_{\alpha\beta}^{\ \gamma}(Q)\, \Phi_\gamma = \omega_\alpha\wedge \omega_\beta +\cx O(Q)\ ,
$$
where $\Phi_\alpha$ is an operator corresponding to the element $\omega_\alpha\in H^{2*}(X)$ and $Q$ are the exponentiated K\"ahler parameters. The structure constants $C_{\alpha\beta}^{\ \gamma}(Q)$ of the quantum cohomology ring, which is related by a topological twist to the chiral ring \cite{LVW} of the underlying $\cx N=2$ theory, encodes the Gromov--Witten invariants of a K\"ahler manifold $X$ and connects many beautiful results in mathematics and physics, such as mirror symmetry, 2d $tt^*$ equations and topological strings \cite{CK,MB}. For $X$ the quintic 3-fold, the Gromov--Witten invariants $N^{GW}_d$ at low degree $d$, computed from mirror symmetry in ref.~\cite{Candelas:1990rm}, are 
$$
N^{GW}_1= 2\,875,\quad N^{GW}_2 = \frac{4\,876\,875} 8,\quad N^{GW}_3=\frac {8\,564\,575\,000} {27},\ \dots
$$
These fractional numbers can be related to integral numbers $n_d$ that ``count'' the number of rational curves of degree $d$ in $X$ \cite{Candelas:1990rm,Aspinwall:1991ce}: 
$$
n_1= 2\,875,\quad n_2 = 609\,250,\quad n_3=317\,206\,375,\ \dots \ .
$$ 
A physics way to define the numbers $n_d$ is to consider an M-theory compactification on $X$, where membranes wrapped on curves represent BPS states in 5d. The integral degeneracies of these BPS states in the target space theory are the Gopakumar--Vafa invariants \cite{GV98}. 

The purpose of this note is to describe and explore a similar correspondence between the quantum product of operators in 3d gauge theory and quantum deformations of the tensor product $\otimes$ on vector bundles $E,F$ over $X$
$$
\Phi_\alpha * \Phi_\beta = e_\alpha\otimes e_\beta +\cx O(Q)\ ,
$$
where $\Phi_\alpha$ is an operator in the 3d theory related to an element $e_\alpha\in K(X)$. A simple physical UV model for quantum cohomology is the gauged linear sigma model (GLSM) \cite{WitPhases,MP}, a $\mathcal{N}=(2,2)$ supersymmetric 2d gauge theory, which flows in certain phases to the non-linear sigma model at low energies. We consider 3d $\cx N=2$ supersymmetric lifts of the GLSM and study the ring structure associated to them. A natural question is, whether this 3d UV gauge theory also computes a topological theory in the IR, which replaces the side of quantum cohomology in the 2d correspondence. We show that the answer is yes and the IR theory in question is the permutation equivariant quantum K-theory constructed by Givental in ref.~\cite{Giv15all}. The K-theoretic Gromov--Witten theory studies holomorphic Euler numbers of bundles over the moduli space $\cx M$ of stable maps to $X$, instead of the intersection theory computed by the cohomological theory. The product~$*$ satisfies the WDVV equation \cite{GivWDVV,Giv15all} (see also ref.~\cite{IMT}), and it is a commutative, associative Frobenius algebra as expected from the TFT point of view.

A novelty of the 3d theory is that the associated invariants have an interpretation in terms of degeneracies of BPS objects {\it on the 3d world-volume} and are thus integral from the start. More precisely there are (at least) two different integral expansions, one associated with the UV phase and another one with the IR phase.\footnote{In the mathematical framework of refs.~\cite{Giv15all,OkL} these phases are related to the theory of quasi-maps and stable maps, respectively.} The two are related by a K-theoretic mirror map that preserves integrality. The integrality of these BPS indices on the world-volume holds for any target space~$X$, implying, e.g., integral expansions for Calabi--Yau $n$-folds of any dimension $n$.
As an illustration of how the 3d theory modifies the non-integral 2d expansion, consider certain invariants in the quantum K-theory of \cite{Giv15all} computed by the 1-point function. In sect.~\ref{sec:EQK} we find  for the quintic, in the IR variables
\bea
N^{QK}_1 &=& 2\,875\cdot \left(\frac{3}{1-q}-\frac{2}{(1-q)^2}\right)  =  2\,875 +\cx O (q)\ , \nonumber\\
N^{QK}_2 &=&-\frac{4\,876\,875}{4
   (1-q)^2}+\frac{77\,625}{8 (q+1)}+\frac{2\,875}{(q+1)^2}-\frac{2\,875}{2 (q+1)^3}+\frac{14\,630\,625}{8 (1-q)} \nonumber\\
   &=& 620\,750+\cx O (q) \ .\nonumber 
\eea
The 3d integral invariants are obtained by an expansion in small $q$, which is a new parameter in the 3d theory; it  enters as a  twisting parameter for the 3d GLSM on 3d world-volumes of the form $S^1\times_q C$. The small radius limit of the 3d theory compactified on $S^1$ corresponds to $q\to 1$ and it connects the correlators of quantum-K-theory continuously to the cohomological theory. In this sense, quantum K-theory can be viewed as a $q$-deformation of quantum cohomology. The leading poles for $q=1$ in the above expressions give back the fractional Gromov--Witten invariant at degree $d$ (up to a combinatorical factor from the insertion).~The subleading terms, which make the 3d invariants integral, arise from contributions of orbifold strata in the moduli space of stable maps \cite{GivTon,GivWDVV}. There is also a permutation equivariant version of 3d (integral) invariants labeled by Young tableaux of size~$d$ \cite{Giv15all}. These invariants provide a refinement of the counting at fixed degree $d$, and we compute these invariants for the quintic and other examples. Empirically, we find for Calabi--Yau target spaces at low degrees universal refinement formulas as functions of the Gopakumar--Vafa invariants $n_d$ spelled out in app.~\ref{app:Inv}. This suggests a permutation equivariant K-theoretic multicovering formula for the Gopakumar--Vafa invariants.

Another important difference compared to 2d  is the deformation dependence of the 3d theory and the flat connections associated to them. There are two types of deformations, K\"ahler parameters $Q$ and mass parameters $t$. The central object in 2d governing these deformations is a GKZ system of differential equations, representing 2d $tt^*$ structure \cite{Cecotti:1991me}, or the Picard-Fuchs equations for $X$ a Calabi--Yau manifold.\footnote{We refer to refs.~\cite{CK,MB} for background and references.} In 3d there is a new type of equations, which represents Ward identities satisfied by the partition function with insertions of line operators \cite{DGG14,KW13}. These shift the K\"ahler moduli $Q$ by finite amounts. We derive the system of $q$-difference equations from the 3d partition function for $X$ that replaces, and in the 2d limit reduces to, the differential GKZ system of the 2d theory. At the same time the 3d partition function satisfies differential equations in the mass parameters $t$, which also reduce to the differential GKZ system of the 2d theory. Mirror symmetry of 3d gauge theories acts on these 3d families in an interesting way.

The idea that the algebra of line operators in the 3d $\cx N=2$ theory should compute the quantum K-theory on the Higgs branch manifold was formulated in ref.~\cite{KW13}, in the context of a generalization of the relation between the Verlinde algebra and the quantum cohomology of Grassmannians. The present paper can be viewed as a realization of this idea for toric hypersurfaces. The connection between quantum K-theory and $q$-difference equations is central to the works \cite{OkL,Aganagic:2016jmx,Aganagic:2017gsx},\footnote{See also refs.~\cite{GK,PSZ,KPSZ}.} which study target spaces related to theories with twice the number of supersymmetries considered in this paper. The general differential equations for 3d $tt^*$ have been derived in ref.~\cite{CV13}.

\vskip3ex
\subsubsecb{Summary}
In sect.~\ref{sec:qdiffsys} we consider the  UV partition functions on $S^1\times S^2$ and $S^1\times D^2$ of 3d theories with a geometric Higgs phase corresponding to a K\"ahler manifold $X$ defined as a complete intersection hypersurface in a toric variety. We determine a system of $q$-difference operators annihilating these functions. These are 3d analogues of the GKZ (or Picard-Fuchs) differential operators prominent in 2d mirror symmetry and they reduce to them in the small radius limit. For a special choice of Chern--Simons terms, this system of difference equations matches those of the symmetrized version of Givental's permutation equivariant quantum K-theory \cite{Giv15all}. We propose that this theory gives the correct IR description for the 3d GLSM. In the small radius limit this 3d GLSM/quantum K-theory correspondence reduces to the well-known 2d GLSM/quantum cohomology correspondence.

Sect.~\ref{sec:def} describes families of 3d theories obtained by integrating in massive 3d particles. These depend on the new mass parameters, in addition to the FI parameters of the original theory. Insertions of massive field operators in the path integral are related to operator insertions of the permutation equivariant \cite{Giv15all} and ordinary  \cite{GivWDVV, Lee:2001mb} quantum K-theory (and to the operator insertions in quantum cohomology  in the small radius limit). We observe that the 3d partition function deformed by a large number of massive particles reproduces the topological string vertex of ref.~\cite{Aganagic:2003db} for $X$ a point. Remarkably these point vertices shared with the topological string can be glued in quantum K-theory for any dimension of $X$ and, moreover, applied to compact hypersurfaces by studying super-bundles \cite{Giv15all}.

In sect.~\ref{sec:LVlim} we study the geometric content of the partition function. In the large volume limit we obtain an interesting 3d generalization of the 2d central charge of a D-brane, related to an index on the loop space $LX$ of $X$. The 3d branes associated with the boundary conditions of the 3d theory carry charges in some (generalized) elliptic cohomology, related to K-theory on $LX$.  In the large volume limit we obtain integral $q$-series associated to a K\"ahler manifold $X$ with modular properties, which include the Witten genus under special conditions. We describe a basis of 3d branes in terms 3d matrix factorizations which give rise to a set of linearly independent solutions to the difference equations via a $q$-version of Mellin--Barnes type integrals.

 In sect.~\ref{sec:mirror} we consider the action of mirror symmetry for 3d gauge theories on the GLSM. The partition function for the gauge theoretic mirror theory $Y$ of $X$ generates a 3d version for a LG period integral, which reconstructs the Lagrangian cycles of the mirror geometry $Y$ within the Coulomb branch of the gauge theory $Y$. We show that these gauge theoretic mirrors include the K-theoretic mirrors presented in ref.~\cite{Giv15all}. 

In sect.~\ref{sec:EQK} we study the proposed IR theory, by computing explicitly the equivariant quantum K-theory invariants defined in ref.~\cite{Giv15all} at genus zero for a number of interesting examples. The $GL(N)$ equivariant quantum K-theory invariants are associated with Young tableaux and give a refinement of the ordinary quantum K-theory invariants, to which they reduce if representations are replaced by their dimensions in the symmetric group.

In sect.~\ref{sec:tft} we discuss the factorization properties of the 3d theory viewed as a topological field theory and propose a relation of the disk partition functions with insertions and the flat sections of 3d $tt^*$ equations of ref.~\cite{CV13}. By the 3d/quantum K-theory correspondence these satisfy a combined system of a differential connection in the mass parameters and a difference connection in the K\"ahler parameters. The connection matrices of the difference connection compute integral invariants associated with the entropy of defects created by line operators.

In sect.~\ref{sec:app} we study in some detail the case of Calabi--Yau $n$-folds with one K\"ahler parameter. We determine the general form of the $q$-period vector and the ring structure constants. We observe a universal relation between permutation equivariant quantum K-theory invariants and Gopakumar--Vafa invariants. We show that the 3d mirror map connecting the UV and the IR phases is integral and determined by the 3d BPS degeneracies. Taking the small radius limit gives a new proof of the integrality of the coefficients of the 2d mirror map.

In sect.~9 we discuss some open questions. Some details are collected in the appendices.

\section{$q$-difference systems for 3d $\cx N=2$ GLSMs} \label{sec:qdiffsys}
In this section we study certain quantities of a particular class of 3d $\cx N=2$ supersymmetric gauge theories, which will turn out to contain the information about the quantum product of the ring of 3d operators associated with the K-theory group $K(X)$ on a K\"ahler manifold $X$.\footnote{In this work $K(X)$ denotes the free part of the topological K-theory group $K^0(X)$, i.e., $K(X)=K^0(X)/K^0_\text{tor}(X)$ where $K^0_\text{tor}(X)$ is the torsion subgroup of $K^0(X)$.} These theories are 3d versions of the 2d $\mathcal{N}=(2,2)$ supersymmetric gauged linear sigma model (GLSM) on the type II string world-volume \cite{WitPhases,MP}, which has played a central role for 2d mirror symmetry. In the last years much progress has been made on the computation of $\mathcal{N}=2$ supersymmetric partition functions on curved spaces by localization.\footnote{See ref.~\cite{Pestun:2016zxk} for a review and references.} In the 2d case the partition function of the gauged linear sigma model (GLSM) on $S^2$ has been shown to compute the K\"ahler potential of the A-model \cite{Jockers:2012dk,Gomis:2012wy}. This gave a new way to compute the genus zero Gromov--Witten invariants for the manifold $X$ described by the GLSM. Since the computation of the partition function works also in higher dimension, it is then natural to follow a similar path for the lift to 3d.

In an attempt to set up a similar structure in one dimension higher, our starting point will be the 3d $\cx N=2$ supersymmetric gauge theory with the same field content as the 2d GLSM with K\"ahler target $X$. Additional data, such as 3d Chern--Simons terms, will be specified along the way. To follow the idea of a 3d lift of the GLSM/quantum cohomology correspondence, we study the 3d partition functions on a 3d membrane world-volume of topology $S^1\times C$, where $C$ is a disk or a sphere. In the small radius limit of $S^1$ the 3d theory reduces to a 2d theory on a Riemann surface $C$ and we expect to recover the results from quantum cohomology. 

Our starting point will be the works \cite{BDP,CV13} and in particular \cite{YS} on disk partition functions on $S^1\times_q D^2$.\footnote{See also ref.~\cite{SugTer}.}  The geometry is twisted, such that a loop around $S^1$ generates a $U(1)$ rotation corresponding to the combination $j_3+\Delta/2$ of the $R$-charge $\Delta$  and spin $j_3$ on $D^2$. $q$ is the weight for the twisting. For gauge group $U(1)$, the general form obtained by localization on the Coulomb branch is 
\be\label{Z3D}
Z(y_r,q)=\int_{|z|=1} \frac{dz}{2\pi i z} e^{-S_\text{class}}\prod_\alpha Z_\alpha\, ,
\ee
Here $z=e^{ih}$ is the $U(1)$ Wilson line on $S^1$ and $S_\text{class}$ the classical action. $Z_\alpha$ are the 1-loop determinants for matter fields of charge $q_\alpha$ and  R-charge $\Delta_\alpha$ \cite{YS,Gadde:2013wq}: 
\be\label{Zmatter}
Z^N_{\phi_\alpha} \sim \frac 1{(z^{q_\alpha}q^{\Delta_\alpha/2}y_r^{\gq_{\alpha r}},q)_\infty}\, , \qquad 
Z^D_{\phi_\alpha} \sim  (z^{-q_\alpha}q^{1-\Delta_\alpha/2}y_r^{-\gq_{\alpha r}},q)_\infty
\ .
\ee
Here $N$ ($D$) stands for a 3d chiral multiplet with Neumann (Dirichlet) boundary conditions and  $(x,q)_\infty$ denotes the $q$-Pochhammer symbol.\footnote{For $|q|<1$, $(x,q)_\infty=\prod_{n=0}^\infty (1-xq^n)$; see app.~\ref{app:qfunc} for more details.} For a given 3d field content, the partition function depends in addition on a choice of Chern--Simons couplings in the classical action $S_{class}$  and boundary conditions on $T^2=\p(D^2\times S^1)$. The $\sim$ denotes that an overall factor $q^{(...)}$ has been omitted for simplicity. The variables $y_r$ introduced above are chemical potentials for the global (flavor) symmetries, with $m_r=-\ln|y|$ representing real mass terms in 3d. We will often consider the case $y_r=1$ in the following and restore the $y_r$ dependence only when needed.   

With the appropriate normalization, the twisted partition function \eqref{Z3D} has an interpretation as an index of gauge invariant BPS states \cite{BDP,Gadde:2013wq,YS,Iqbal:2012xm}
\be\label{ind}
Z_{S^1\times_qD^2} \sim  \tr \left( (-1)^F e^{\beta H}q^{\Delta/2+j_3}y_r^{q_r} \right)\ .
\ee
It is natural to ask, how degeneracies of BPS states encode geometric information of the K\"ahler manifold $X$, such as the ``number'' of holomorphic curves. As mentioned above, both the UV phase with gauge fields included and the IR phases with gauge fields integrated out enjoy their own integral expansions. As will be discussed, these match Euler numbers on two different compactifications of the moduli space of maps adapted to the UV/IR regime. It is the IR phase, related to the quantum K-theory of ref.~\cite{Giv15all}, which is directly related to curve counting. These integral BPS sums in the 3d world-volume theory should be contrasted to the 2d case, where integrality properties of the genus zero Gromov--Witten invariants are related to BPS counting in target space \cite{GV98}.

\subsection{Projective space $\IP^{N-1}$ and degree $\ell$ hypersurfaces of $\IP^{N-1}$\label{subsec:3dpfpn} }
To illustrate the general structure, we first discuss the projective space $\IP^{N-1}$ and degree $\ell$ hypersurfaces $X$ therein. In our context, we can formally think of the projective space $\mathbb{P}^{N-1}$ as a degree zero hypersurface, such that we can uniformly treat both classes of examples as degree $\ell$ hypersurface $X$ in $\mathbb{P}^{N-1}$, where $\ell$ can be zero.\footnote{As discussed in sect.~\ref{subsec:IntTS}, the cases with and without a constraint are related by integrating in a 3d matter field.}  The generalization to toric hypersurfaces is given in the following section. The gauge group is $U(1)$, and the charges of the $N+1$ matter fields are 
\be\label{DefQ}
q_{0}=-\ell\ ,\qquad q_\alpha=1\ , \quad \alpha=1,\ldots,N\ .
\ee
A convergent series for $Z$ is obtained by summing the residues in- or outside the unit circle, depending on the value of the parameters in the action $S_{class}$. The details of the computation are collected in app.~\ref{app:3dpf}. The partition function depends on the complex FI parameter 
\be\label{FIp}
\tilde Q=e^{-2\pi\xi+i\theta}\ , 
\ee
which is the weight for the topological $U(1)_J$ symmetry dual to the gauge $U(1)$. 
For a special choice of Chern--Simons couplings,  the sum over the residues $z=q^{-(k-\eps)}$ for $k\geq 0$ takes the form 
\def\s{I}
\be\label{Z3D2}
Z=\ln(q)\, \int \frac{d\eps}{2\pi i}\ \FFd\cdot  \s(Q, q,\eps)\, ,
\ee
The $Q$ dependence is captured by the holomorphic series 
\be
\s(Q,q,\eps)=\sum_{k=0}^\infty Q^{k-\eps} a_k(q,\eps)\ ,
\qquad Q = \frac{\tilde Q}{(1-q)^{c_1}}\, , 
\ee
with $c_1=N-\ell$ the numerical coefficient of the first Chern class of $X$ and 
\be
a_k(q,e)=\frac{(-)^{c_1k}}{(q-1)^{c_1\eps }}
\frac{\Gamma_q(1-\eps)^N}{\Gamma_q(1-\ell\eps)}\frac{\Gamma_q(1+\ell(k-\eps))}{\Gamma_q(1+k-\eps)^N}\ . 
\ee
The $Q$-independent function is
\be\label{DeftFFs}
\FFd=\frac{q^R(q-1)^{c_1\eps }}{(-\eta(q))^{N-1+\delta_{\ell,0}}}\frac{1-q^{-\ell\eps}}{(1-q^{-\eps})^N}\
\frac{\Gamma_q(1+\eps)^N}{\Gamma_q(1+\ell\eps)}\ ,
\ee
where $\eta(q)$ is the Dedekind eta-function, $\Gamma_q$ the $q$-Gamma function\footnote{See app.~\ref{app:qfunc} for definitions and properties of $\Gamma_q$  and related functions.} and the exponent $R$ is determined by a choice of $R$ charges and the Chern classes of $X$. 

Since $Q$ is the weight of the topological $U(1)_J$, which is carried by vortices, the term $\sim Q^k$ in the sum $\s(Q,q,\eps)$ can be associated with the contribution of a vortex of charge $k$. The connection between the $D^2\times S^1$ partition function and vortex partition functions has been explored in refs.~\cite{BDP,DGH} for massive supersymmetric Higgs vacua. In this paper the main focus will be on massless case with a higher-dimensional Higgs branch corresponding to the $n$-dimensional K\"ahler manifold $X$. 

The 2d limit, or small radius limit, is defined by writing 
\be
q=e^{-\beta\hbar} \ ,
\ee
where $\beta$ is the radius of $S^1$ and $\hbar$ is the parameter for the $U(1)$ twist of the geometry. Then $\beta\to 0$ defines the 2d limit $q\to 1$. The 2d limit of the holomorphic series is 
\be\label{tmpper}
\lim_{\beta\to 0} \s(Q,q,\eps)=Q^{-\eps}\sum \hat Q^k 
\frac{\Gamma(1-\eps)^N}{\Gamma(1-\ell\eps)}
\frac{\Gamma(1+\ell(k-\eps))}{\Gamma(1+k-\eps)^N} \ ,
\ee
were we have used $\lim_{\beta\to0}\Gamma_q(x)=\Gamma(x)$. Moreover
$\hat Q$ is the renormalized FI parameter
\be\label{Qren}
\hat Q = Qe^{-c_1\ln(\hbar\beta)}\ ,
\ee
of the 2d theory. The generalized hypergeometric series \eqref{tmpper} is familiar from 2d mirror symmetry: for the Calabi--Yau case, i.e. $\ell=N$, the coefficients of an expansion in $\eps$ are linear combinations of the periods of the mirror manifold of $X$. To keep this parallel, we refer to the (coefficients of the) 3d vortex sum $\s(Q,q,\eps)$ also as the "$q$-periods''. 
\\

\subsubsec{Difference equations}
In the 2d theory, a concise way to describe the dependence of the series \eqref{tmpper} on the variable $Q$ is in terms of a system of differential equations. For $X$ a Calabi--Yau manifold, these are the well-known Picard--Fuchs equations, and their solutions are the periods of the mirror manifold \cite{CK,MB}. More generally, these equations reflect the flatness of the Gauss--Manin connection on the deformation space. We will now determine a system that is the 3d counterpart of the Picard--Fuchs operators. This system involves finite difference operators and has the $q$-periods as solutions. As explained in refs.~\cite{Dimofte:2011py,BDP,KW13}, difference equations arise in the 3d gauge theory as Ward identities of line operators. 

The difference equations for the $q$-periods arise from the recursion relation of the coefficients $a_k(q,\eps)$ going back to the basic identity of the $q$-Gamma function
\be\label{gammarec}
\Gamma_q(x+1)=\frac{1-q^x}{1-q\ }\ \Gamma_q(x)\, .
\ee
It implies
\be
a_{k+1}=(-(1-q))^{c_1}\frac{\prod_{i=1}^\ell(1-q^{\ell(k-\eps)+i})}{(1-q^{k+1-\eps})^N}\ a_k
\ .
\ee
Noting that the action of the difference operator $q^\theta =q^{Q\frac d {dQ}}$ on the summands produces factors
 \be
(1-q^{a\theta+b})Q^{k-\eps}=(1-q^{a(k-\eps)+b}) Q^{k-\eps}\ ,
\ee
we obtain the difference equation
\bea \label{DiffEq1}
\DL \ \s(Q,q,\eps)=0\ ,\qquad \DL =  
(1-q^{\theta})^N-(-)^{c_1}Q \prod_{j=1}^{\ell}(1-q^{\ell\theta+j}) \ .
\eea
For $\ell=0,1$, $\DL$ agrees with the Ward identity of ref.~\cite{KW13}  in the massless limit.\footnote{The generalization to non-zero mass terms corresponds to the $\mathbb{T^N}$-equivariant version with $y_r\neq 1$ and is straightforward; see eq.~\eqref{Sigmapnd}.}

In the above a term $(1-q^{-\eps})^N\sim \cx O(\eps^N)$ has been dropped for the following reason. The integral \eqref{Z3D2} picks out the residue of the product of the two factors. In the example, $\FFd$ has a pole of at most order $N$ (for $\ell=0)$ and only the first $N$ terms in the expansion
\be\label{seps}
\s(Q,q,\eps)=\omega_0(Q,q)+...+\eps^{N-1}\omega_{N-1}(Q,q)+...
\ee
of the holomorphic series contribute to residue.\footnote{Alternatively, with the replacement \eqref{epstoH} this represents the classical relation $H^N=0$ for the hyperplane class.} The coefficients $\omega_{i\leq N-1}(Q,q)$ then give $N$ independent solutions to the difference equation \eqref{DiffEq1}, see app.~\ref{app:sol}.

Similarly as in the 2d case, the set of difference equations can be interpreted as a set of equations, which expresses the flatness of a connection on the space parametrized by $Q$. The flat sections of this system will be identified with D-brane overlap functions  of ref.~\cite{CV13} in sect.~\ref{sec:tft}. The flat sections are linear combinations of the $q$-periods $\omega_{i\leq N-1}(Q,q)$ with coefficients in $(Q,q)$-dependent functions $f(Q,q)$ that are left invariant by the shift operator $Q\to Qq$, e.g. elliptic functions $e(x,\tau)$ with $x=\frac 1 {2\pi i }\ln (Q)$, $\tau=\frac 1 {2\pi i }\ln (q)$.\\[-3mm]

One can also consider the 2d limit on the difference operator to obtain a differential operator. Using
\be\label{DifftoDiff}
\lim_{\beta\to 0} \frac{1-q^\theta}{\! 1-q}\ = \ \theta\ .
\ee
 the leading term in the operator $\DL$ becomes
\be\label{DL2d}
\cL=\lim_{\beta \to 0 } \DL= \theta^N-(-)^{c_1}\hat Q\prod_{j=1}^{\ell}(\ell\theta+j) \,
\ .
\ee
$\cL$ is the quantum differential operator of quantum cohomology.\footnote{See chapter~10 of ref.~\cite{CK} for background material.} For $c_1(X)=0$ it is the well-known GKZ operator, which annihilates the periods of the mirror manifold of $X$ \cite{Bat,HKTY}. It reduces to the Picard-Fuchs operator upon an  additional factorization. The factorization is necessary, since the order of $\cL$ and $\DL$ is to high and they have too many solutions for $\ell>0$. In the example, they are of order $N=\dim(H^{2*}(\IP^{N-1}))$, while the reduced operators have order $N-1=\dim(H^{2*}(X))$. This factorization works similarly in 2d and 3d, see app.~\ref{app:sol}. \\[-4mm]

\subsubsec{Operator algebra} 
A new aspect of the 3d theory is, that the chiral ring is generated by line operators \cite{BDP,CV13}. The simplest line operator is a Wilson line wrapping the extra $S^1$. In the localized path integral \eqref{Z3D}, the insertion of a Wilson line operator of charge $m$  yields an extra factor of $z^{m}$. Passing to the vortex sum \eqref{Z3D2} at the residues $z^{-1}=q^{k-\eps}$, the insertion of a factor $z^{-m}=q^{m(k-\eps)}$ can be expressed on the series as
\be\label{WLshift}
z^{-m}:\ \s(Q,q,\eps)\ \to \ \s(Qq^m,q,\eps)=q^{m\theta} \s(Q,q,\eps) \ .
\ee
Thus the charge one Wilson line in the $U(1)_g$ theory acts on the vortex partition function as the shift operator $\hat p= q^\theta$. The operators defined above satisfy the commutation relations 
\be\label{comr}
\hat p\hat Q-\hat Q\hat p=(q-1)\hat Q\hat p\, ,\qquad [\hbar \theta,Q]=\hbar Q\ ,
\ee
where the second equation again represents the 2d limit defined as in \eqref{DifftoDiff}.
In the 2d theory it is known, that the small quantum cohomology algebra of $\IP^{N-1}$ is obtained as the quasi-classical limit of the differential operator $\cx L$ \eqref{DL2d}, after the replacement $\hbar\theta\to H$ \cite{GivH1,GivEGW}
\be\label{Lqc}
\cx L \xrightarrow{\hbar\to 0} H^{N}=Q\, .
\ee
A similar replacement of the operators $(\hat p,\hat Q)$ by commuting variables $(p,Q)$ in the classical limit yields 
\be
(1-p)^N=Q\ ,
\ee
which is the small quantum K-theory ring of $X=\IP^{N-1}$ \cite{GivLee}. This is a first hint that the 3d partition function on the Higgs branch $X$ with Wilson line insertions computes more generally a certain quantum K-theory ring on $X$, and the classical limit of the difference operator characterizes this ring at special moduli. Similarly, we obtain
\be
  (1-p)^N = (-)^{c_1} Q (1-p^\ell)^\ell \ ,
\ee
as a prediction for the small quantum K-theory ring of a degree $\ell$ hypersurface $X$ in $\IP^{N-1}$.

The commutation relations \eqref{comr} and the relation \eqref{Lqc} have been obtained in 1994 from a heuristic construction of $S^1$ equivariant Floer co-homology on the universal cover $\tLX$ of the free loop space $LX$ of $X$ \cite{GivH1} . It was also noticed that the results following from this ansatz take a form which can be interpreted as some sort of path integral. The above observations indicate, that the relevant path integral is that of the 3d gauge theory considered in this paper. Relations between 3d vortex sums and ordinary quantum K-theory have been noticed for special examples before, e.g. in refs.~\cite{BDP, KW13,Bonelli:2013mma,BKK}. As will be explained below, the 3d path integral really computes the permutation invariant version of quantum K-theory constructed more recently in ref.~\cite{Giv15all}.

\subsection{Difference systems for toric hypersurfaces}
The previous discussion can be generalized to other gauge groups and matter content. Here we discuss the case of toric complete intersections. For abelian gauge group $G=U(1)^k$, the K\"ahler manifold $X$ is defined as an intersection of hypersurfaces in a toric variety 
\be
W=\IC^N/\!\!/(\IC^*)^k\ .
\ee
A phase of the model determines a fixed basis $\{q^a\}$ of charge vectors in the K\"ahler cone. The entries 
\be \label{defqai} 
q^{a}_\alpha\in\IZ\ , \qquad a=1,\ldots,k\  ,
\ee 
are the charges of the chiral matter fields $\varphi_\alpha$ under the $a$-th $U(1)$ factor. The target space $X$ is a Calabi--Yau manifold if $\sum_\alpha q^a_\alpha=0$ for all $a$ \cite{WitPhases}.

It will be useful to know the 3d partition function for the sphere and a disk times the circle. The two partition functions are expected to be related by a factorization, which reflects the insertion of a complete basis of 3d branes, as in \cite{BDP,CV13}. \\

\subsubsec{Partition function on $S^1\times_q S^2$}
The 3d partition function on $S^1\times_q S^2$ has been studied in refs.~\cite{Imamura:2011su,Dimofte:2011py,Kapustin:2011jm}. Consider a $U(1)$ theory with $N+1$ charged matter fields $\varphi_\alpha$ of general charges. The fields with index $\alpha>0$ parameterize the toric variety $W$, a weighted projective space, and the index $\alpha=0$ is reserved for the field that imposes the hypersurface condition. The details of the computation are relegated to app.~\ref{app:3dpar}, where the following expression is derived:
\be
\label{ResExS}
Z_{S^2\times_q S^1}=\ln(q)\ \oint_0 \frac{d\eps}{2\pi i } \s(\bar Q,\bar q,\eps)\, \FFs\, \s(Q,q,\eps)\ .
\ee 
Here $\bar q = q^{-1}$ and the bar on $Q$ means ordinary complex conjugation. The $Q$-dependence of the partition function is again captured by a holomorphic function $\s(Q,q,\eps)$ and its conjugate. It is given by a generalized $q$-hypergeometric series 
\begin{multline} \label{DefOmq}
\s(Q,q,\eps)=\sum_{n\geq 0} (-)^{n(c_1+q_0)}q^{d(n,\eps)}\left(\frac Q{(1-q)^{c_1}}\right)^{n-\eps} \\
\times \frac{\Gamma_q(1-q_0(n-\eps))}{\Gamma_q(1+q_0\eps)} 
\prod_{\alpha>0}\frac{\Gamma_q(1-q_\alpha\eps)}{\Gamma_q(1+q_\alpha(n-\eps))}\ .
\end{multline}
Here $c_1=\sum_{\alpha=0}^N q_\alpha$ is again the numerical coefficient of the  first Chern class of $X$.  For simplicity we show the expression for canonical choice of $R$-charges, $\Delta_0=2$ and $\Delta_{\alpha>0}=0$. The exponent $d(n,\eps)$ depends on the 3d Chern--Simons (CS) couplings. It is shown in the appendix, that it can be set to zero by a judicious choice of CS terms in the classical action. The "folding factor" for the square $|I(Q,q,\eps)|^2$ in the residue integral is
\be\label{FFEx}
\FFs=
\frac{(1-q^{q_0\eps})}{\prod_{\alpha>0}(1-q^{-q_\alpha\eps})}\
\frac{\prod_{\alpha>0}\Gamma_q(1+q_\alpha\eps)}{\Gamma_q(1-q_0\eps)}\
\frac{\Gamma_q(1+q_0\eps)}{\prod_{\alpha>0}\Gamma_q(1-q_\alpha\eps)} \ .
\ee
\\[-1mm]

\subsubsec{Partition function on $S^1\times_q D^2$}
Instead of factorizing the partition function on $S^1\times_q S^2$ into a holomorphic and anti-holomorphic series in $Q$ consider the partition function on $S^1\times_q D^2$. This computation fixes a normalization factor which can not be obtained unambiguously from the factorization and is needed to determine the 3d analogue of the D-brane central charge. The general computation is given in app.~\ref{app:3dpf}; for the $U(1)$ theory the result is 
\be\label{ResExD}
Z_{S^1\times_q D^2} = \ln(q)\, \oint_0 \frac{d\eps}{2\pi i}  \FFd \cdot \s(Q,q,\eps)\ ,
\ee
with
\begin{multline}\label{DefOmqD}
\s(Q,q,\eps)=\sum_{n\geq0} \left(\frac{Q}{(1-q)^{c_1}}\right)^{n-\eps}q^{d(n,\eps)}
(-)^{c_1n}\prod_{\alpha\in D} \frac{\Gamma_q(1-q_\alpha(n-\eps))}{\Gamma_q(1+q_\alpha\eps)} \\
\times \prod_{\alpha\in N} \frac{\Gamma_q(1-q_\alpha\eps)}{\Gamma_q(1+q_\alpha(n-\eps))}\ .
\end{multline}
Here $N\ (D)$ refers to 3d chirals with  Neumann (Dirichlet) conditions at the boundary. The $D$ fields will be taken to represent the sections for the hypersurface constraints. The result above is shown for the canonical choice of $R$-charges $\Delta_\alpha=0\ (2)$ for $N$ ($D)$. For the example of a degree $\ell$ hypersurface in $\IP^{N-1}$ we have $N$ fields with Neumann conditions of charge 1 and one field with Dirichlet conditions of charge $-\ell$. The folding factor in \eqref{ResExD} is
\be\label{DeftFF}
\FFd=(-\eta(q))^{\tD-\tN}\, q^R(1-q)^S\, \frac{\prod_D(1-q^{q_\alpha\eps})}{\prod_N(1-q^{-q_\alpha\eps})}\
\frac{\prod_N\Gamma_q(1+q_\alpha\eps)}{\prod_D\Gamma_q(1-q_\alpha\eps)}\ ,
\ee
where $\eta(q)=q^{\frac 1 {24}}\prod_{n=1}^\infty(1-q^n)$ and $\tD\, (\tN)$ is the number of Dirichlet (Neumann) fields, respectively. For the canonical choice of $R$-charges, the exponents are $S=0$ and $R=-\operatorname{ch}_2 \eps^2-\frac 12 c_1 \eps$, where $\operatorname{ch}_2$ and $c_1$ are the numerical coefficients of the second and first Chern characters of $X$, respectively.
\\
 
\subsubsec{Comparison and 2d limit}
The holomorphic data  $\s(Q,q,\eps)$ appearing in the two partition functions agree up to a minus sign that can be absorbed in the definition of the classical action. The form of the $S^2$ partition function then indicates that it can be factorized into two $S^1\times_q D^2$ partition functions, similarly as in \cite{Pas}. The factorizability of the $\cx N=2$ supersymmetric 3d partition function is expected on general grounds \cite{Dimofte:2011py,CV13}.
For $c_1=0$ one can again take the naive 2d limit $q\to 1$ in \eqref{DefOmq} and replace the $q$-Gamma with ordinary Gamma functions. In this limit, the series $\s(Q,q,\eps)$ reduces to the generalized hypergeometric series prominent in 2d mirror symmetry \cite{Bat,HKTY}. It represents the building blocks of the periods of the toric Calabi--Yau complete intersection~$X$.
\\

\subsubsec{Systems of difference equations}
The residue formulas for the partition functions have a straightforward generalization to the $U(1)^n$ gauge theory with matter fields of charges $q_\alpha^i$, $\alpha=0,...,N$, $i=1,...,n$. To describe a complete intersection $X$ in a toric variety, we consider $\tN$ chiral fields with Neumann boundary conditions and $\tD$ chiral fields with Dirichlet boundary condition, as defined in ref.~\cite{YS}. A field $\varphi_\alpha$ with Dirichlet boundary conditions  and negative $U(1)^n$ charges $q_\alpha^i$ implements a hypersurface constraint of degree $|q_\alpha^i|$. The charge vectors $q^a$ are defined up to linear transformations. To obtain a vortex expansion at large values of the FI parameters, we choose a basis $\{q^a\}$ that corresponds to a large volume phase.\footnote{See refs. \cite{WitPhases,MP,CK} for a discussion of phases in the 2d GLSM.}

The general expressions for the partition function, the vortex sum and the folding factor are given in eqs.~\eqref{Zgen}, \eqref{OmD2}, and \eqref{ResD2} in app.~\ref{app:3dpf}. We allow for generic CS terms, contributing a factor 
\be
q^{d(k,\eps)}\ ,\qquad d(k,\eps)=\frac 12 A_{ij}(k_i-\eps_i)(k_j-\eps_j)+B_i(k_i-\eps_i) \ ,
\ee
in the vortex sum. Here $k_i$ is the vortex number and $\eps_i$ the Wilson line integration variable for  $U(1)_i$.

The derivation of the recursion relation for the coefficients of the vortex sum $\s(Q_a,q,\eps_a)$ does not depend on the details of the integration contour, assuming a convergent contour exists.\footnote{See ref.~\cite{Gerhardus:2015sla} for a discussion of integration contours in the 2d case.} Proceeding as before, one obtains the following set of $n$ difference operators annihilating the vortex sum \eqref{OmD2} 
\be\label{DiffEqgen}
 \DL_a =  
\prod_{\substack{\alpha\in N\\q^a_\alpha>0}}\prod_{j=0}^{q^a_\alpha-1}(1-q^{\vartheta_\alpha-j})-Q_a 
q^{\sum_i A_{ai}\theta_i}\prod_{\alpha\in D}\prod_{j=1}^{|q^a_\alpha|}(1-q^{-\vartheta_\alpha+j}) \,
\prod_{\substack{\alpha\in N\\q^a_\alpha<0}}\prod_{j=0}^{|q^a_\alpha|-1}(1-q^{\vartheta_\alpha-j})\ .
\ee
Here $a=1,...,n$ and \\[-5mm]
\be\label{defvarth}
\vartheta_\alpha = \sum_a q_\alpha^a \theta_a\ ,\qquad \theta_a = Q_a\frac{\p}{\p Q_a}\ .
\ee
\ \\[-5mm]
In the above we have absorbed some constants by the redefinition  
\be Q_a \to Q_a(-)^{c_a}q^{-\frac 12 A_{aa}-B_a}\ .\ee 

The difference operators $\DL_a$ represent the Ward identities for the line operators in the 3d theory with gauge group $U(1)^n$ and with general matter charges. In the 2d limit, the Ward identities reduce to the familiar differential operators central to 2d mirror symmetry. E.g. for a hypersurface one obtains
\be\label{DefcL}
\cL_a=
\prod_{\alpha, q^a_\alpha>0}\prod_{j=0}^{q^a_\alpha-1}(\vartheta_\alpha-j)-\frac{Q_a}{(\beta\hbar)^{c_1}} 
\prod_{j=1}^{|q^a_0|}(-\vartheta_0+j) \,
\prod_{\alpha,q^a_\alpha<0}\prod_{j=0}^{|q^a_\alpha|-1}(\vartheta_\alpha-j)
\ ,
\ee
and these are for $c_1(X)=0$ again the well-known GKZ operators  of refs.~\cite{Bat,HKTY} that annihilate the periods of the mirror manifold of $X$.\\

\subsubsec{Comparison with equivariant quantum K-theory}
So far, we have considered the UV phase of the 3d gauge theories with a Higgs branch representing a K\"ahler manifold $X$. We have found that the vortex sum of the 3d GLSM, and thus the partition function, is annihilated by the system of difference operators \eqref{DiffEqgen}. We are now ready to identify the topological theory associated with the IR phase of the 3d gauge theory, i.e., the theory that replaces quantum cohomology in the 3d generalization of the 2d GLSM/quantum cohomology correspondence. 

In ref.~\cite{Giv15all} Givental constructs a $GL(N)$ equivariant quantum K-theory with an action of the permutation symmetry $S_n$ on a correlator with $n$ insertions. In the simplest case $N=1$, the permutation symmetric theory, only the totally symmetric representation of $S_n$ appears. In ref.~\cite{Giv15all} the so-called $I$-function is computed for the symmetric quantum K-theory for a super-bundle $\Pi E$ over a toric space $W$.  This $I$-function $I_{\Pi E}$ satisfies a set of difference equations, which are reproduced below for convenience:
\begin{multline}\label{diffeqgiv}
\prod_{j:m_{ij}>0}\prod_{r=0}^{m_{ij}-1}(1-q^{-r}q^{m_{kj}\theta_k})\cx I_{\Pi E}\\=
Q_i\prod_{j:m_{ij}<0}\prod_{r=0}^{-m_{ij}-1}(1-q^{-r}q^{m_{kj}\theta_k})
\prod_a\prod_{r=1}^{l_{ia}}(1-q^{r}q^{l_{ka}\theta_k}) \cx I_{\Pi E}\ .
\end{multline}
Here $m_{ij}$ are the defining vectors of $W$ and $l_{ia}$ the degrees of the hypersurface constraints generalizing the single hypersurface considered above; in our notation $q^i_j=m_{ij}$ and $q^i_0=-l_{i1}$. Eq.~\eqref{diffeqgiv} contains minor corrections to the formula in ref.~\cite{Giv15all} (4th on page 5 of p.VI), which, however, follow straightforwardly from the derivation given there. After this modifications and setting the effective Chern--Simons terms in \eqref{DiffEqgen} to zero the operators defined by the equations \eqref{diffeqgiv}  agree with the $\DL_a$. The general case with non-zero Chern--Simons terms relates to the level structure of quantum K-theory described in ref.~\cite{RZ18}.

In the large volume phase, i.e., small $Q_a$, the (reduced) system of linear difference equations \eqref{DiffEqgen} has $\dim(H^{2*}(X))$ independent solutions, reproducing the solutions of the differential equations \eqref{DefcL} in the 2d limit. The agreement of the equations \eqref{DiffEqgen} and \eqref{diffeqgiv} implies that the $S^1\times_q D^2$ partition function of the 3d GLSM computes, up to linear combination with coefficients in $q$-shift invariant functions, (a certain value of) the $I$-function of the symmetric quantum K-theory.\footnote{More details on the definitions of the permutation equivariant quantum K-theory of ref.~\cite{Giv15all} will be given in sect.~\ref{sec:EQK}, where we study the quintic and other examples.}

\subsection{Period matrix and monopole expansion\label{subsec:mon}}
The difference operators \eqref{DiffEqgen} acting on the 3d partition function have the general form $\DL_a I =0$ with 
\be
\DL_a = \DL_a^+-Q_a  \DL_a ^- \ , 
\ee
In this form the difference equations represent Ward identities for line operators, that generalize those of ref.~\cite{KW13} to 3d theories associated with toric complete intersection hypersurfaces.

On the other hand, defining $\tilde\DL_a ^-  = \DL_a^- |_{\theta_a  \to \theta_a -1}$, the difference equations can be rewritten as an eigenvalue problem 
\be\label{defmon}
\DLb_a  I =  Q_a I\ ,\qquad \DLb_a = (\tilde\DL_a^-)^{-1}\DL_a^+\ .
\ee
More generally, we observe that the vortex sums $I_{\vec b} := (\prod_i q^{\theta_{b_i}})\,  I$ associated with insertions of Wilson lines $\prod_i z_{b_i}^{-1}$ in the Coulomb integral represent eigenfunctions of $\DLb_a$ with different eigenvalues 
\be\label{eveq}
\DLb_a  I_{\vec b} =  \zeta_{a,\vec b} I_{\vec b} \ ,\qquad \zeta_{a,\vec b} =q^{\sum_i \delta_{a,b_i}}Q_a\  .
\ee
The linearly independent solutions to these equations are the building blocks for the 3d generalization of what is called the period matrix in the context of 2d mirror symmetry. E.g., for $X=\IP^{N-1}$ there are $N$ independent solutions for an eigenvalue $Qq^k$, represented by the expansion of $I_k$ as in \eqref{seps}, and the $I_k$ are linearly independent for $k=0,...,N-1$. This gives an $N\times N$ matrix $\cx T$ of solutions which comprises an operator/state correspondence between chiral operators and boundary states in the 3d theory, as will be discussed in sect.~7.

We observe that the difference equations can be used to resum the vortex sums as a partition function for monopole operators,\footnote{See refs.~\cite{AHISS,KapB,IS,Ken}.} which carry the topological $U(1)_J$ charge with weight $Q_a$. For illustration we take again the example of the $U(1)$ theory with $N$ fundamentals, corresponding to the target $X=\IP^{N-1}$. The vortex sum \eqref{DefOmqD} for Chern--Simons level $\kappa$ is 
\be
I(\IP^{N-1}) = \sum_{n=0}^\infty \frac{Q^nq^{\kappa \frac{n(n-1)}{2}}}{\prod_{r=1}^n(1-Pq^n)^N}\ ,
\ee
where we introduced the notation $P=q^{-\eps}$ and dropped the overall factor $Q^{-\eps}$. The omission of the leading term $Q^{-\eps}$ requires the replacement $q^\theta\to Pq^\theta$ in the difference operator (which is eq.~\eqref{DiffEq1} with an extra factor $q^{\kappa\theta}$ in the second term as in eq.~\eqref{DiffEqgen}). For zero Chern--Simons level $\kappa=0$, the above expression agrees with the K-theoretic $J$-function for $\IP^{N-1}$ at zero input \cite{GivLee}, with $P$ interpreted as the Chern character of the line bundle $\cx O(-1)$, fulfilling the relation $(1-P)^N=0$. In the derivation of the 3d partition function this constraint arises from the residue integral on the Coulomb branch, as explained below eq.~\eqref{DiffEq1}. 

On the other hand, viewing $I$ as an index, which counts states of different electric charges weighted by $P$, before taking the integral, we should not impose the constraint $(1-P)^N=0$.\footnote{The definition of the vortex sum as a character on the moduli space of vortices along the lines of ref.~\cite{NekABCD} will be discussed in sect.~\ref{subsec:defent}.} The exact difference equation fulfilled by this counting function is not eq.~\eqref{DiffEq1} but the {\it inhomogeneous} equation
\be\label{deqmod}
(1-Pq^\theta)^N \ \tilde{\cx I} = Q \, (Pq^\theta)^\kappa\, \tilde{\cx I} +(1-P)^N\ .
\ee
Here we use $\tilde{\cx I}$ to distinguish the counting function without the constraint on $P$ from $I$. This modified difference equation can be used to resum $\tilde{\cx I}$ as a power series in $P$ with exact coefficients  in the $U(1)_J$ weight  $Q$. Defining\footnote{The extra factor $(1-P)^N$ takes into account the contribution from spin zero fields in the counting function, which had been included in the gluing factor $f$ in eq.~\eqref{DeftFFs} before.}
\be
\cx I=(1-P)^{-N}\, \tilde{\cx I} =\sum_{\ell=0}^\infty  P^\ell\, \cx I_\ell(Q,q) \  ,
\ee
eq.~\eqref{deqmod} gives a recursion relation for the coefficients $\cx I_\ell(Q,q)$. For $\kappa=0$ the solution has the form 
\be\label{Pexp}
\cx  I(\IP^{N-1})=\sum_{n=0}^\infty \frac {c^N_n(Q,q) P^n}{\prod_{r=0}^n (1-q^rQ)}=
\frac1{(1-Q)}+\cx O(P)\ ,
\ee
where $c^N_n(Q,q)$ is a polynomial of degree $<n$. The leading term $\sim P^0$ is independent of the target space $X$ and is the partition function for the electrically neutral monopole operator of $U(1)_J$ charge one and spin zero. The subleading terms count charged operators composed of monopoles and charged matter fields and these depend on the target $X$. E.g., for $N=1$ one obtains $c^{N=1}_n(Q,q)=1$ for all $n$,  whereas for $N=2$, i.e.  $X=\IP^1$,
\be
\cx  I(\IP^1)=\frac{1}{1-Q}+P\, \frac{2}{(Q,q)_2}+P^2\, \frac{3+q Q}{(Q,q)_3}
+ P^3\, \frac{4+2qQ(1 +q)}{(Q,q)_4}+\ldots
\ee
with $(Q,q)_n=\prod_{r=0}^{n-1}(1-Qq^r)$. In the sector of $U(1)$ charge one, the monopole operator comes with two spin states of weights $\sim q^0,q^1$, and it is dressed by a single mode of one of the two matter fields. In the higher charge sectors, there are corrections to the naive counting from $Q$-dependent terms in the polynomials $c_{n>1}$, which would be interesting to understand from the field theory point of view. A similar expansion as in eq.~\eqref{Pexp} exists for non-zero $\kappa>0$ with $Q$ replaced by $QP^\kappa$. This is the weight of the neutral operator made from charged matter fields and the monopole operator of non-zero charge induced by the Chern--Simons term \cite{IS}.

Equations of the form $\eqref{eveq}$ have appeared in the context of 3d $\cx N=4$ supersymmetric theories in ref.~\cite{Verma} and connected to the action of the monopole operators \cite{AHISS,KapB,IS,Ken} of the theory on the Higgs branch defined by quantization of the theory on $\IR^2 \times \IR_t$ and with $\mathcal{N}=(2,2)$ boundary conditions imposed on a plane $\IR^2$ at fixed $t$. The present set up describes a lift of the discussion of ref.~\cite{Verma} to the $S^1$ compactification of a 4d theory. That is to say the pair of a 3d $\cx N=4$ bulk theory with a 2d $\mathcal{N}=(2,2)$ boundary theory of ref.~\cite{Verma} is lifted to a pair of a 4d $\cx N=2$ bulk theory with a 3d $\cx N=2$ boundary theory compactified on an additional $S^1$. The $\cx N=2$ 3d partition functions discussed in this paper are the lifts of the $\mathcal{N}=(2,2)$ 2d partition functions of the boundary theory. The action of the monopole operator of the 4d $\cx N=2$ theory on the vortex moduli spaces is described by the K-theoretic lift of the cohomological operations in ref.~\cite{Verma}. The eigenvectors for the eigenvalue problem \eqref{defmon} should represent generalized Whittaker vectors of the $S^1$-compactified 4d Coulomb branch algebra. It would be interesting to apply the methods of ref.~\cite{Verma} to the present setup. A connection between quasi-map moduli spaces and Whittaker functions for $\mathfrak{gl}_{\ell+1}$ has been described in ref.~\cite{Gerasimov}.

\subsection{Spectral manifolds}
A certain phase of the 3d GLSM associates to a K\"ahler manifold $X$ the system of difference equations \eqref{DiffEqgen}. The shift operators $\hat P_a=q^{\theta_a}$ and the K\"ahler moduli $Q^a$ satisfy the commutation relations $\hat P_a\hat Q_b=q^{\delta_{ab}}\hat Q_b\hat P_a$ generalizing  \eqref{comr}. In the commuting limit $q=1$, the equations obtained by replacing operators $(\hat P_a,\hat Q_a)$ with commuting variables $(p_a,Q_a)$ in the difference operators assign to a hypersurface $X\subset W$ with $n$ K\"ahler moduli a spectral surface $\Sigma(X)$ of $\dim_\IC = n$: \footnote{Despite of the notation, the difference operators $\DL_a$, and therefore $\Sigma(X)$, depend on the embedding of $X$ as a hypersurface in $W$ and on the phase for the 3d GLSM.}
\be
X \ \leadsto\  \DL_a(X) \ \leadsto \ \Sigma(X):\  \cap_a\{ f_a(p_b,Q_b)=0\} \subset (\IC^*\times \IC^*)^n\ .
\ee
Even for the simplest theories with $X=\IP^{N-1}$ one obtains an interesting series of spectral curves, which is related to known type II string compactifications of the form 
\be\label{Sigmapn}
f=(1-p)^N-Q+xz\ ,\qquad \Sigma(\IP^{N-1})=\{f=0\}\cap \{z=0\}\ .
\ee
These geometries are mirror to a 3-fold fibration $Y$ of an $A_{N-1}$ singularity and have been related in ref.~\cite{AVm} to M-theory compactifications on local manifolds $S^3\times R^4/Z_N$. Moreover, the equation for $\Sigma(X)$ was shown to be equivalent to the condition $d\cx W =0$ for a supersymmetric vacuum in a dual type IIA theory with D6 branes and disk superpotential $\cx W$.\footnote{See also ref.~\cite{CV13} for a discussion in terms $(p,q)$-webs of fivebranes.} The quantum K-theory for $\IP^{N-1}$ should thus be closely related to the topological string on $Y$; we come back to this issue in sect.~\ref{sec:topvert}. Turning on real mass terms $y_i$, $i=1,\ldots,N$, for the fields, which corresponds to studying the $\mathbb{T}^{N-1}$ equivariant K-theory of $\IP^{N-1}$, describes a blow up of the $A_{N-1}$ singularity with difference operator and the spectral surface 
\be\label{Sigmapnd}
\DL = \prod_{i=1}^N (1-y_iq^\theta) -Q\quad \leadsto \quad \Sigma:\ \prod_{i=1}^N (1-y_ip)-Q=0\ .
\ee
More generally, the M-theory compactifications with D and E groups of \cite{AVm} give the spectral curves for 3d theories related to weighted projective spaces $W\IP(a_i)$, with $a_i$ the Dynkin numbers for the respective group. 

The spectral manifold is related to the twisted superpotential $W $ of the $S^1$ compactified 2d theory including the contribution from the Kaluza--Klein modes \cite{AHKT,NekSha,NekSha2}. In the $q\to 1 $ limit, the 3d partition function behaves as 
\be\label{spsu}
Z\sim \int \frac{dz_i}{2\pi i z_i}e^{\frac 1 {\ln q} \cx W(z,Q)},\qquad 
\ee
where $\cx W(z,Q)$ is regular. An expansion of the difference equation around $q=1$ yields the equation for the spectral curve $\Sigma(X)$ with $p_a=\exp(Q_a\p_{Q_a} \cx W)$. Moreover $\Sigma(X)$ is Lagrangian w.r.t. to the holomorphic symplectic form $\sum_a\frac{dQ_a}{Q_a}\wedge \frac{dp_a}{p_a}$ and describes the manifold of supersymmetric vacua of the 3d theory coupled to a 4d bulk by gauging the global symmetry \cite{DGG14}.

In the above example $X=\IP^{N-1}$, the integrand in the 3d partition function can be written as $e^{W(z,Q,q)}$ with 
\be
W(z,Q,q) = N\sum_{k>0} \frac{z^k}{k(1-q^k)} -\ln Q\ln z /\ln q -A \ln^2 z/2\ln q\ ,
\ee
where we added a Chern--Simons term with coefficient $A$. The spectral curve \eqref{Sigmapnd} arises in the limit $q\to 1$, where $W\to \cx W/\ln q +...$. There are other semi-classical limits of the integrand at $q^m=1$ for $m\in\mathbb{N}$,  as noticed before in ref.~\cite{Giv15all}. In an expansion around $q^m=1$, the leading terms come from the summands with $k=m\cdot n$, and one obtains the spectral curves
\be
\Sigma(X)_m: f_m=(1-p^m)^N-(p^AQ)^m=0\ ,
\ee
which describe orbifolds of the spectral curve for $m=1$.

In the context of open topological string, going back from the spectral curve $\Sigma$ to the difference operator $\DL$ has been interpreted in ref.~\cite{Aganagic:2003qj} as a quantization of the mirror curve $\Sigma$ with the Hamiltonian $H=\DL$. Upon adding a hypersurface constraint, the operators depend explicitely on the quantization parameter $q$ as in \eqref{DiffEq1}. This is expected on general grounds \cite{Aganagic:2011mi}, and it would be interesting to study these operators from the point of quantization.

\section{Deformations} \label{sec:def}
\subsection{Integrating in massive bulk fields \label{subsec:Def1}}
We consider now the modification of a given 3d theory by integrating in new massive matter fields. For concreteness we consider the $U(1)$ partition function for $X=\IP^{M-1}$ as a starting point. We assume that the Chern--Simons couplings are initially chosen to cancel exponentials in $q$ that depend on the vortex number~$k$. We have seen that the sum over the poles $z=q^{-k+\eps}$ produces the vortex sum\footnote{\label{fn14}In this section we define the vortex sum without the overall factor $Q^{-\eps}$ for convenience. Dropping this factor has to be compensated by the replacement $q^\theta\to q^{\theta-\eps}=Pq^\theta$ in the difference operators.} 
\be\label{vortexsum}
I(Q,q) = \sum_{k=0}^\infty I_k Q^k\ ,\qquad I_k = \frac{1}{\prod_{\ell=1}^{k}(1-Pq^{\ell})^M}\ .
\ee
Here we introduced the notation $P=q^{-\eps}$ for later use. We now consider the effect of integrating in a new massive matter field of $U(1)$ charge $a$, R-charge $\Delta$ and twisted mass parameter $y= e^{-m}$. To keep the effective Chern--Simons term of the theory fixed in the limit  of infinite mass $y\to0$, and to cancel potential 3d anomalies in the fermion measure, we define the integrating in procedure to include compensating Chern--Simons background couplings specified in the following table
\be
e^{-S_{CS}}= e^{\frac{1}{2\ln q}k_{ij}\ln x_i \ln x_j}\ ,\qquad 
\begin{tabular}{c|ccc}
&$\ln z$&$\ln y$&$\frac 12 \ln q$\\
\hline
$\ln z$&$\frac {a^2}2$&$\frac a 2 $&$a\frac {\Delta-1}2$\\
$\ln y$&..&$\frac 1 2 $&$\frac {\Delta-1}2$\\
$\frac 12 \ln q$&..&..&$\frac {(\Delta-1)^2}2$\\
\end{tabular}
\ee
where $x_i\in\{z,y,q^{1/2}\}$.
For $a=1$, $\Delta=0$ this reduces to the choice made in ref.~\cite{BDP}.

Integrating in a new massive particle with Neumann boundary conditions, charge  $-a<0$ and mass $y$ together with the specified CS couplings gives a new factor\footnote{This is for $R$-charge 0; the general case is obtained by the replacement $y\to yq^{\frac \Delta 2}$.} 
\be\label{simplefac}
\frac 1 {(z^{-a}y ,q)_{\infty}}
\ee 
 in the integrand. It does not contribute new poles inside the integration contour chosen before. After passing to the sum over residues, the effect of the new field is a multiplicative factor in the $k$-th vortex sector:
\be
I_k \to I_k(y)=\frac 1 {(y P^aq^{ka},q)_\infty}\cdot I_k\ .
\ee
On the vortex sum, 
  this transformation can be represented by the action of a difference operator, namely\be\label{int1}
I(Q,q) \to I(Q,q,y)=\frac 1 {(y P^aq^{a\theta},q)_\infty}\ I(Q,q)\ .
\ee
The partition function of the theory with massive deformations fulfills a deformed difference equation.  From the commutation relation \eqref{comr}, we obtain the deformed Ward identity  $\DL' I(Q,q,y)=0$ with 
\be\label{DLdef}
\DL'= (1-Pq^\theta)^M-Q\prod_{\ell=0}^{a-1}\left(1-y(Pq^{\theta})^aq^\ell \right) \ .
\ee 
The transformation of the vortex sum defined by the integrating in operation is interesting in two ways: For large mass it defines a deformation family of the original theory. Upon interpolation to zero mass one obtains a theory with different target space $X$. 

\subsection{Perturbative expansion and quantum K-theory \label{subsec:pertexp}}
For large mass, i.e., small $y=e^{-m}$, we can view the result of the above operation as a small deformation of the given theory with target $X$. Using the relation~\eqref{qpoch}, the difference operator \eqref{int1} can be expanded as 
\be\label{IIop}
\frac 1 {(yP^aq^{a\theta},q)_\infty}=\exp\left(\sum_{r>0}\frac{(yP^aq^{a\theta})^r}{r(1-q^r)}\right)\ .
\ee
Similarly, integrating in $N$ fields with masses $y_i,i=1,...,N$, results in the difference operator 
\be\label{IIopmt}
\exp\left(\sum_{r>0}\frac{\tr V^r  (P^aq^{a\theta})^r}{r(1-q^r)}\right)\ ,
\ee
where $V=\textrm{diag }(y_1,...,y_N)$.
Recall that the operator $(Pq^\theta)^a$ was obtained from the insertion of a charge $a$ Wilson line $z^a$ in the dynamical $U(1)$ gauge field, and the variables $y_i$ contain the Wilson line in the background gauge field associated with the real mass deformation. Writing $\tr U=z^a$, the $q^0$ term of the operator \eqref{IIopmt} takes the familar form
\be\label{csmtr}
Z_\text{MT}(U,V)= \exp \left( \sum_r\frac 1 r \tr U^{r} \tr V^r \right)\ .
\ee
This operator has played an important role in the duality between the topological string and 3d CS theory \cite{Labastida:2000yw}. Adding this factor in the path integral computes CS correlation functions with insertions of multi-traces (MT) of Wilson line operators in the dynamical gauge field $U$ and the background gauge field $V$. The standard correlators with single-trace (ST) insertions coupled to a background fields are generated by the $r=1$ term 
\be\label{csstr}
Z_\text{ST}(U,V)=\exp(   \tr U \tr V \big)\ .
\ee

The 3d index \eqref{ind} counts the number of gauge invariant BPS operators. The deformed theory includes operators dressed by the massive modes. The weights of theses modes are given by a ``complexification'' of the Wilson line background $V$ compared to eq.~\eqref{csmtr} and in addition there is an infinite tower of modes with different spins for each field, weighted by the $q$-variable. For an interesting interpretation of the multi-traces, $V$ should be viewed as an $U(N)$ connection. The 3d indices with standard insertions of single-traces, restricting to the $r=1$ term of \eqref{IIopmt}, are generated by the difference operator 
\be\label{IIopst}
\exp\left(\frac{y(P^aq^{a\theta})}{(1-q)}\right)\ .
\ee

\subsubsec{Comparison to quantum K-theory}
The difference operator \eqref{IIopmt} for the multi-trace insertions is identical to the deformation operator of the $GL(N)$-equivariant quantum K-theory defined in part~I of ref.~\cite{Giv15all}. This theory computes quantum K-theory correlators with $n$ insertions, equivariant under the action of the permutation group $S_n$ on the $n$ insertions. The operator for the single trace insertions \eqref{IIopst} is the deformation operator in the ordinary quantum K-theory \cite{GivER,GivTon}. It is also possible to consider both types of insertions at the same time in the mixed quantum K-theory of, see part~VII of ref.~\cite{Giv15all}.

The above agreement identifies the massive deformations of the UV gauge theory with the deformations of mixed quantum K-theory, and more precisely, insertions of massive field operators in the path integral with the insertions in the correlators of the associated quantum K-theory.\\[2mm]

\subsubsec{Comparison to the 2d $A$-model}
As will be discussed in detail in sect.~\ref{sec:LVlim}, the factor $P$ in the 3d partition function represents the Chern character of the bundle $\mathcal{O}(-1)$ over $\IP^{M-1}$ in the target space geometry. The integrating in of a massive particle of charge $-a$ has produced an operator involving the (classical) K-theory element $P^a\in K(X)$. For $X=\IP^{M-1}$, $(1-P)^M=0$ and there are $M$ independent directions. The vector space is spanned by, say, matter fields of charge $-a=(0,1,\ldots,N-1)$.\footnote{To express a deformation of the theory by a field outside this charge window in terms of this basis, one has to use the Ward identity for the Wilson lines, i.e., the deformed difference equation \eqref{DLdef}, not the classical relation.} In general, there will be $\dim(K(X))$ parameters $\tau_\ell$ associated with a basis of $K(X)$, times the number of species of such sets.

In addition, the 3d GLSM with gauge group $U(1)^k$ depends on $k=\dim H^2(X)$ FI parameters, or vortex weights, $Q_a$. The values of those can be deformed with the help of the chiral Wilson line operators, see \eqref{WLshift}. In total, the 3d PF depends on the twist $q$ and the parameters
\be \label{ampar}
\begin{aligned}
\textrm{massive particles: }&&&\tau_k,\ k=0,\ldots,\dim(K(X))-1\ , \\ 
\textrm{Wilson lines: }&&&Q_a,\ a=1,\ldots,\dim(H^2(X))\ .
\end{aligned}
\ee
Here $\tau_k\approx -\ln(y_k)$ is a complexified mass parameter for a single species.

How do the above operators and deformations match to the 2d $A$-model in the small radius limit? In the $A$-model one has $\dim H^{2k}(X)=\dim(K(X))$ independent cohomologial operators \cite{Witten:1988xj}. A cohomological basis is obtained from the K-theory basis in the small radius limit $\beta\to 0$ via the Chern isomorphism. For $\IP^{M-1}$ one may choose
\be\label{pbtmp}
\Phi_k = (1-P)^k \ \to \ (\beta H)^k\ ,\qquad P=e^{-\beta H}\ ,\ k=0,...,N-1\ .
\ee
The deformations of the $A$ model in $H^2(X)$ are associated with the complexified K\"ahler parameters $t_a, a=1,...,\dim(H^2(X))$. However, we have obtained two types of parameters for each element of $H^2(X)$, the mass parameters $\tau_a$ and the Novikov variables $Q_a$. This curious doubling of the parameters for $H^2(X)$ exists already in quantum cohomology \cite{GivEGW}. In 2d, the parameters are redundant in the IR theory in the following sense: after a reparametrization of the deformations, the partition function depends only on the combinations $Q_ae^{t_a}$, where $t_a=t_a(\tau,Q)$ are the flat coordinates on the deformation space.

In the 3d theory, the parameters $\tau_a$ and $Q_a$ parametrize different directions in the deformation space, and there is no a priori reason to expect them to lead to equivalent deformations. The 2d behavior of the deformations can be studied from the integrating in operator \eqref{IIop}. After a linear reparametrization of the parameters $y$ adapted to the basis \eqref{pbtmp} for $K(X)$, it takes the form
\be
\exp\left(\sum_{a=0}^{N-1}\sum_{r>0}\frac{y_a^r(1-(Pq^k)^r)^a}{r(1-q^r)}\right)\ .
\ee
For fixed $a$, the small radius limit of the exponent is, with $P=e^{-\beta H}$ and $q=e^{-\beta\hbar}$,
\be
\sum_r\frac{y^r_a(1-(Pq^k)^r)^a}{r(1-q^r)}  =  
\sum_r\frac{y^r_a\beta^{a-1}}{r^{2-a}}\frac{(H+k\hbar)^a}\hbar = t_a \frac{(H+k\hbar)^a}{\hbar}\ ,
\ee
with $t_a = \beta^{a-1}\sum_r y_a^r/r^{2-a}$. The scalar term $\sim H^0$, multiplying the weight $Q^k$, is 
\be
\exp(\, t_a (k\hbar)^a/\hbar\, )\ .
\ee
Only in the 2d limit and only for $a=1$ it can be absorbed in the vortex weight by the redefinition $Q\to Qe^{t_1}$. The shift is
\be
\textrm{ MT: } 
t_1 = -\ln(1-y_1)
,\  \qquad \textrm{ ST: }t'_1 = y_1\
 ,
\ee
for the multi-trace \eqref{IIopmt} and single-trace \eqref{IIopst} perturbations, respectively.

In quantum cohomology, the dependence on the combinations $Q_ae^{t_a(Q,\tau)}$  follows from the divisor equation.\footnote{See ref.~\cite{CK} for a review and references.} There is no divisor equation in quantum K-theory. The $Q$ and $\tau$ deformations are still related in a more general way: a change of parameters $(Q,\tau)\to (Q',\tau)$ leads to a theory in the deformation family of the original one, with deformation parameters $(Q,\tau'(Q))$ \cite{GivTon}.

\subsection{Equivariant quantum K-theory and topological string vertex}  \label{sec:topvert}
In p.~II of ref.~\cite{Giv15all} Givental reconstructs the equivariant quantum K-theory for toric $X$ from $\mathbb{T}^n$ equivariant fixed point localization on $X$, by gluing the vertices associated with fixed points along fixed curves.  The point vertices are obtained by assigning a special input $V$ to the operator \eqref{IIopmt} for the target $X=pt$. We first observe that these vertices are in fact equal to the topological vertex
\be\label{topvert}
I(pt)=\exp\left(\sum_{r>0}\frac{\tr V^r  }{r(1-q^r)}\right)=\sum C_{00\nu}(q^{-1})s_\nu(x)=Z_{\text{top.vert.}}(\mathbb{C}^3)\ , 
\ee
with $x_i=-q^{-\frac 12}y_i$. The expression on the r.h.s. is the topological string vertex in the Schur representation for a stack of branes on a single leg of the toric Calabi--Yau 3-fold $\mathbb{C}^3$ \cite{Okounkov:2003sp}.\footnote{Relations between vortex sums and topological string vertex have been exploited earlier in ref.~\cite{DGH}.} In this context, $q=e^{ig_s}$, with $g_s$ the string coupling constant, $C_{00\nu}$ is the value of topological vertex for a holomorphic disk with a boundary labeled by a 2d partition $\nu$  and $s_\nu(x)$ the associated Schur function.

The coincidence of the $GL(\infty)$ equivariant quantum K-theory  for a point and the topological string vertex raises interesting questions. Firstly, the relation $q=e^{ig_s}$  combined with the small radius limit to 2d/quantum cohomology shows, that the 3d theory gives a resumation of the expansion in the string coupling; a simple illustration will be given in sect.~\ref{sec:outlook}.

Secondly the gluing of the point vertices \eqref{topvert} along fixed curves in \cite{Giv15all} reminds of the gluing of topological string vertices to obtain the partition function for a toric Calabi--Yau 3-fold $X$ \cite{Aganagic:2003db}. A noteworthy difference is that the gluing formalism of ref.~\cite{Giv15all} works for any number of fixed curves connected to the vertex \eqref{topvert} and can be applied to compact hypersurfaces $X$ by studying super-bundles. The gluing rule of ref.~\cite{Giv15all} sums up the contributions from $N$ fixed curves connected to a point vertex into a single input $V$. As explained below, this amounts to using an effective vertex with global $SU(N)$ structure from the point of the topological string. 

For the $U(1)$ theory with $N$ matter fields of charge one, corresponding to $X=\IP^{N-1}$, $SU(N)$ is a global symmetry at zero mass. As noted around \eqref{Sigmapn}, the spectral curve associated with the difference operator agrees with the mirror curve for an $A_{N-1}$ singularity studied in ref.~\cite{AVm}. For a single chiral $N=1$ one obtains $X=pt$, or more precisely the stack $X=\IC/\!/\IC^*$, including the degenerate orbit. The spectral curve $\Sigma$ is the curve for the mirror of $\IC^3$ \cite{Aganagic:2003db}. For $N=2$, the curve is a singular version of the mirror curve for $\mathcal{O}(-2)_{\IP^1}\oplus \mathcal{O}(0)_{\IP^1}$, at zero volume of the $\IP^1$. Using results of ref.~\cite{IK}, it has been already observed in refs.~\cite{Pas,BDP}, that the 3d vortex sum for $N=2$ with non-zero real masses coincides with the open string partition function for a brane moving  on  $\mathcal{O}(-2)_{\IP^1}\oplus \mathcal{O}(0)_{\IP^1}$. 

The above generalizes to $N>2$,  where $\Sigma$ describes the singular limit of the mirror curve for a chain of $N-1$ $\IP^1$'s of zero size, with intersections corresponding to the $A_{N-1}$ Dynkin diagram. Non-zero volume corresponds to introducing real masses, which represent equivariant parameters for the $\mathbb{T}^{N-1}$ action, leading to the deformed equations \eqref{Sigmapnd}. The vortex sum solving the deformed difference operator coincides with the $\mathbb{T}^{N-1}$ equivariant I-function of ref.~\cite{GivLee}
\be
I= (1-q)\sum \frac{Q^d}{\prod_{k=1}^d\prod_{i=1}^N(1-q^ky_iP)}\ .
\ee
It is shown in part II of \cite{Giv15all}, how to rewrite the evaluation $I^{(i)}=I(y_i^{-1})$ at the $\mathbb{T}^{N-1}$ fixed point $P=y_i^{-1}$  in terms of the point vertex \eqref{topvert}, with a special input $V$ determined by a recursion relation summing up the pole contributions from fixed curves connected to the fixed point. On the other hand, $I^{(i)}$ coincides with the effective topological string vertex for the $A_{N-1}$ geometry, called the half $SU(N)$ vertex in ref.~\cite{IK}. The representations at the external legs are trivial, except for a fundamental at the $i$-th leg. The precise match is, up to a reparametrization,  parallel to the discussion in ref.~\cite{Pas}, where the half $SU(N)$ vertex with these representations has been discussed in detail  in the context of factorization of the 3d partition function on $S^3$.

The above discussion generalizes further to a degree $\ell$ compact hypersurface in $\IP^{N-1}$, with the spectral curve $\Sigma$ associated to the commuting limit of the difference operator \eqref{DiffEq1} for $\ell\neq 0$ and a modified effective $N$-vertex obtained by adding the weight factor from the hypersurface constraint associated with the field of charge $-\ell$. We conclude that the sewing rules of ref.~\cite{Giv15all} can be interpreted as gluing effective $SU(N)$ vertices associated with the topological string vertex.  It will be interesting to compare the gluing rules of ref.~\cite{Giv15all} and ref.~\cite{Aganagic:2003db} in more detail for toric 3-folds. A proposal for the computation of the {\it all genus} topological string partition function on compact Calabi--Yau 3-fold by gluing effective vertices has been made recently in ref.~\cite{Vafatalk}.

\subsection{Change of target space \label{subsec:IntTS}}
Another rewriting of eq.~\eqref{int1} is 
\be
\frac 1 {(yq^{\Delta/2}P^aq^{ka},q)_\infty}=\frac{\prod_{\ell=0}^{ak-1}(1-yPq^{\ell+\Delta/2})}{(yq^{\Delta/2}P^a,q)_\infty}\ ,
\ee
where we have restored the $R$-charge.
For $\Delta=2$, the interpolation to zero mass $y=1$ gives the transformed vortex sum
\be
I'(\IP^{M-1}, Q,q,y=1)= \frac{1}{(qP^a,q)_\infty} \cdot \, \sum Q^{k-\eps}\frac{\prod_{\ell=1}^{ak}(1-P^aq^{\ell})}{\prod_{\ell=1}^{k}(1-Pq^{\ell})^{M}} \ , 
\ee
which is the vortex sum for a degree $|a|$ hypersurface $X\subset \IP^{M-1}$, times the $k$-independent factor. Accordingly the deformed difference equation \eqref{DLdef} reduces to \eqref{DiffEq1} in the massless limit.

The pre-factor modifies the folding factor $f_{D^2}$ of the theory on the new target  $X$.\footnote{The  inverse of the pre-factor has an interpretation as a twisting class interpolating between untwisted quantum K-theory and the twisted version of quantum K-theory described in ref.~\cite{TonTwist} and part~XI of ref.~\cite{Giv15all}.} For $|a|=1$, i.e. a degree one hypersurface in $\IP^{M-1}$, one expects to obtain the integrand for $\IP^{M-2}$, but one finds
\be
f_{D^2}(\IP^{M-1})\ \frac{1}{(qP,q)_\infty}\ I'(\IP^{M-1}, Q,q,y=1)=
\frac 1 {\theta(P^{-1})}\cdot \big( f_{D^2}(\IP^{M-2}) I(\IP^{M-2},Q,q)\big) \ .
\ee
The r.h.s. differs from the integrand for $\IP^{M-2}$ because of the $\theta$-function. One can get rid of this factor by integrating in a Dirichlet field of opposite charge 
\be
\frac{\theta(yqz^{-a},q)}{(yqz^{-a},q)_\infty} = (y^{-1}z^a,q)_\infty\ .
\ee
together with compensating CS terms. $\theta$-functions arise as the one-loop determinant of fields living on the $T^2$ boundary of $D^2\times_q S^1$ \cite{YS,Gadde:2013wq}. The need of additional CS terms can be seen from the fact, that the $\theta$-function is not invariant under shifts $x\to xq$, i.e., it would change the difference equation. An invariant combination is 
\be\label{thetainv}
\theta(x,q)\cdot e^{\ln (-x)^2/2\ln q-\ln (-x)/2} \ .
\ee
More general factors of this type are used in sect.~\ref{subsec:branef} to construct a complete basis of solutions to the difference equation.

Note that by similar steps but in the reverse direction, one can use mass deformations and integrating in to move up in dimension from $\IP^{M-1}$ to $\IP^M$, and more generally to create general toric spaces $W$ starting from the trivial vortex sum 
\be
I(Q) = \sum_k Q^k =\frac1 {1-Q}\ .
\ee

\section{Geometric indices and three-dimensional E-branes\label{sec:LVlim}}
In this section we study the geometric content of the 3d partition functions, starting from the expansion around the limit of large K\"ahler moduli. We discuss some modifications that arise in the step from 2d to 3d related to the 3d lift of D-brane boundary conditions, such as a new type of K-theory charge and linearly independent bases of $q$-Mellin--Barnes integrals. Moreover we discuss new genera associated to a K\"ahler manifold $X$ by the 3d theory.

\subsection{Large volume limits and index theorems\label{subsec:LVlimD}}
To obtain a better geometric understanding of the large volume limit, it will be useful to discuss first the relation of the 2d disk partition function to classical index formulas. The large volume limit of the 2d disk partition reproduces the perturbative central charge of a D-brane \cite{HoriRomo,HonOk} 
\be\label{Zlv2d}
Z^{LV}_{D^2}(E_\alpha) \sim \int_X e^{-J}\hG \ch(E_\alpha)\ .
\ee
Here $E_\alpha$ is a sheaf that defines a $B$-type boundary condition at $S^1=\p D^1$, $J$ is a K\"ahler class on $X$ and $\hG$ the so-called Gamma class \cite{Libgober,Iritani2,Katzarkov:2008hs}, a certain square root of the Todd class
\be\label{hG}
\hG=\sqrt{\td(X)}e^{i\Lambda_X} , \qquad \hG^*\hG = \hat A (X) = e^{-c_1/2}\td (X)\ .
\ee
$\hG^*$ denotes the dual of $\hG$ defined by the reflection $x_\alpha\to -x_\alpha$ on the Chern roots $x_\alpha$. The expression without the factor $e^{i\Lambda_X}$ had been derived from anomaly inflow arguments in refs.~\cite{Cheung:1997az,MM97}. The additive class $\Lambda_X$ governs the perturbative corrections to the K\"ahler metric of the GLSM on $X$ \cite{Halverson:2013qca}. Explicit expressions in terms of the Chern classes of $X$ will appear below. The Gamma class intertwines between the tensor product of sheaves and the wedge product on the Chern characters
\be \label{eq:2dprod}
\left\langle\ch(E_\alpha^*)(\hG)^* e^{c_1(X)/2},\ch (E_\beta) \hG\right\rangle_X=\int_X \td (X) \ch(E_\alpha^*\otimes E_\beta) \ .
\ee
Here $\langle a,b\rangle_X=\int_X a\wedge b$ for $c_1(X)=0$. The right hand side is the Witten index for the open string  stretched between the two D-branes defined by $E_\alpha$ and $E_\beta$  \cite{HIV}
\be\label{HIVind}
\textrm{ind}_{\bar \p_{E}}=\sum_k (-1)^k\dim \textrm{ Ext}^k(E_\alpha,E_\beta) \buildrel HRR \over = \int_X \td (X) \ \ch(E_\alpha^*\otimes E_\beta) \ .
\ee
The Hirzebruch--Riemann--Roch index theorem used in the last step  has a simple derivation from supersymmetric quantum mechanics on $S^1$ \cite{AG,FW}. The Todd class comes from the path integral over the bosons of the sigma model, and the Chern character from  fermions on $S^1$ coupled to the connection on $E$. 

In the 2d partition function \eqref{Zlv2d}, the boundary $S^1$ is filled by a disk. Only half of the bosonic modes on $S^1=\p D^2$ can be extended smoothly to the interior; one can choose coordinates such that these are positive modes defined on a holomorphic disk. The bosonic determinant for these modes is a certain square root $\hG$ of the full determinant $\td (X)$ on $S^1$. The precise form can be obtained as  the $S^1$ equivariant Euler class for the normal bundle to the positive energy modes on the loop space $LX$ of $X$ \cite{GGIritani, Lu}:
\be\label{es12d}
\frac1{e_{S^1}(N_+)}
\sim (\hbar/2\pi)^{n}\hbar^{c_1(X)/\hbar}\hG\ ,
\ee
where $\hbar$ the generator of the $S^1$ action rotating the loops.
Schematically, this "half-index'' on the boundary $S^1=\p D^2$ can be obtained by removing the contribution from the negative modes in the full index
\be\label{2dres}
\hat A(X) \to \hat A(X) \cdot e_{S^1}(N_-) \sim \hat A (X)/\hG^* = \hG \ .
\ee
Adding the contribution of the boundary fermions one obtains the $J$-independent terms of the large volume limit  \eqref{Zlv2d} of the disk parition function. To summarize, the large volume limit of the 2d disk partition functions is the half-index computed by the sigma model on the boundary $S^1$ with target space $X$.

We now turn to the 3d case, where the boundary is the 2d torus $\p(S^1\times_q D^2)\simeq T^2$. The indices computed by the 2d supersymmetric sigma models on $T^2$ have been first studied in refs.~\cite{WitLoop, Alvarez87} in the context of the supersymmetric strings. The relevant differential operator is the Dirac-Ramond operator associated with the loop space $LX$ of $X$. Correspondingly, one expects that the large volume limit of the 3d partition functions computes similar indices as the one discussed above in 2d,  with $X$ replaced by $LX$.

To this end we consider the large volume limit of \eqref{Z3D} defined by taking a generic direction in the K\"ahler class $J$ where all $Q_a\to 0$. The leading term comes from setting $n=0$  in the series \eqref{DefOmqD}. For simplicity we describe the one modulus case and write  $J=tH$, with $H$ the hyperplane class. After the formal replacements
\be\label{epstoH}
\eps\to -H/\hbar\ , \qquad q_\alpha\eps\to -D_\alpha/\hbar \ ,\qquad D_\alpha=q_\alpha H\ ,
\ee
the integral \eqref{ResExD} can be viewed as an integral over $X$
\be \label{RestoVolInt}
\int \frac{d\eps}{2\pi i} \frac{1}{e(X,\eps)} \mu(\eps)=\int_X \mu(H)\ ,
\ee
where the integrand $\mu(H)$ is a class in rational cohomology on $X$ and $e(X)$ is the rational function associated to the Euler class of $X$, see Table~\ref{tab:charclasses}. 

Using the expressions given in app.~\ref{app:3dpf}, we find for the large volume limit of the disk partition function \eqref{DefOmqD}, with a choice of CS terms that sets  $d(k,\eps)=0$: 
\be\label{Zlv3d}
Z_{S^1\times_q D^2}^{LV}(\mathcal{O}_{LX})=\frac{\ln q}{(-\eta)^{\dim(X)}}\int_X e^{-J} \ e^{c_1^\beta(X)/2}\frac{\hAq (X)}{\hGq^*} \cdot e^{-\ch^\beta_2(X)/\ln q }\ .
\ee
This expression corresponds to the trivial brane $\mathcal{O}_{LX}$ with Neumann boundary conditions. The K\"ahler class $J$ is defined as
\be
J=\beta \hat t_i J_i\ ,\qquad \hat t_i = \ln(Q_i/(1-q)^{c_{1i}})/\ln q+\frac 12 c_{1i}\ .
\ee
It is normalized with an extra factor of $\beta$ relative to the K\"ahler classes on $X$.  Similarly, the superscript $\beta$ on $\ch_2^\beta(X)$ and $c_1^\beta(X)$ denotes that these classes are defined in the 3d normalization, e.g. $c_1^\beta(X)=\sum_i c_{1i}\, \beta J_i=\beta c_1(X)$. 

The remaining cohomology classes in \eqref{Zlv3d} are multiplicative and can be characterized by a function $f(x)$ in a single variable $x$ with $f(0)=1$. Given $f(x)$, we define the class $C(f,X)$  for the 3d GLSM with target $X$ using the splitting principle as
\be\label{Cfx}
C(f,X)=\frac{\prod_N f(x_\alpha)}{\prod_D f(-x_\alpha)}\ ,
\ee
where $N$ and $D$ denote again fields with Neumann/Dirichlet boundary conditions. The characteristic functions for the classes appearing in the above formulas are listed in Table~\ref{tab:charclasses}. 
\begin{table}[t]
\begin{center}
\mbox{
\hbox{
\vbox{\offinterlineskip
\halign{\vrule width 1.2pt\strut~~$#$~~~\hfil&~$#$~~~\hfil\vrule&~~$#$~~~\hfil&~$#$~~~\hfil\vrule width 1.2pt\cr
\noalign{\hrule height 1.2pt}
\text{2d} &  x_\alpha=D_\alpha & \text{3d} & x_\alpha=\beta D_\alpha \cr
\noalign{\hrule}
e(X) & f(x)=x & e(X) & f(x)=x \cr
c(X) & f(x)=1+x & c(X) & f(x)=1+x \cr
\td (X) & f(x)=\frac{x}{1-e^{-x}} & \hAq  (X) & f(x)=e^{-x/2}\frac{x(q)_{\infty}^2}{\theta(e^{-x},q)} \cr
\hG & f(x)=\Gamma(1-x/\hbar) & \hGq & f(x)=\Gamma_q(1+x/\ln q) \cr
\noalign{\hrule height 1.2pt}
}}}}
\end{center}
\caption{Displayed are the defining functions for the characteristic classes appearing in the discussed 2d and 3d partition functions, where $(q)_\infty=(q,q)_\infty$ and $ \theta(y,q)=(y,q)_\infty(q/y,q)_\infty$. Upon evaluating with respect to the Chern roots $x_\alpha$ of the holomorphic tangent bundle $T_X^{1,0}$, we obtain the corresponding characteristic classes of the space $X$. In this table only the Euler class $e(X)$ is not multiplicative, but nevertheless obeys $e(E\oplus F)=e(E)e(F)$ because it is identified with the top Chern class $c_{\dim(X)}(X)\equiv e(X)$ of the total Chern class $c(X)$, which is again multiplicative.} \label{tab:charclasses}
\end{table}

Comparing with the previously discussed 2d case, the large volume expression \eqref{Zlv3d} has the expected form for the trivial brane with $\chq(E)=1$. The full index on the loop space $LX$ can be informally written as\footnote{As discussed around eq.~\eqref{G2exp} below, this expression is $SL(2,\IZ)$ invariant, and thus well-defined as an index, only for $\ch_2(E)-\ch_2(X)=0$.}
\be\label{3dfullind}
\textrm{ind}_{\bar D_E} \simeq \int_X \hAq (X) \ \chq(E)\ .
\ee
Here $X$ represents the fixed locus $X\subset LX$ of the $S^1$ action rotating the loops. The $S^1$ equivariant characteristic classes $\hAq$ and $\chq$ are defined on the restriction of bundles to the fixed point set. The class $\hGq$ satisfies an identity analogous to \eqref{hG}:
\be\label{AGGrel}
\hAq  (X) = \hGq\ \hGq^* \ \hat A^\beta (X) \ .
 \ee
The class $\hGq$ , which we call the $q$-Gamma-class, represents the Chern character of the K-theoretic Euler class for the normal bundle $N_+$ of positive loops (written for a simple factor in \eqref{Cfx})
\be\label{ENK}
\frac 1 {e^K_{S^1,\alpha}(N_+)}=\frac{1}{\prod_{k=1}^\infty (1-q^ke^{x_\alpha})} = \frac{(1-q)^{x_\alpha/\ln q}}{\prod_{k=1}^\infty(1-q^k)^n}\Gamma_q\left(1+\frac{x_\alpha} {\ln q}\right).
\ee
The expression on the r.h.s. reduces in the 2d limit to \eqref{es12d} using zeta-function regularization. The large volume limit \eqref{Zlv3d} of the disk partition function on $S^1\times_q D^2$ is then related to the full index \eqref{3dfullind} by the loop space analogue of eq.~\eqref{2dres}. The last factor in eq.~\eqref{Zlv3d} originates from the Chern--Simons couplings of the theory.

Heuristically speaking, the Dirac operator on $X$ is to K-theory what the Dirac operator on $LX$ is to elliptic cohomology \cite{Witgen}. We conclude that 3d-brane charges take their values in a certain generalization of elliptic cohomology $\EC(X)$.  This suggests that 3d-branes are represented by objects in a derived category associated with $\EC(X)$. In the following we also refer to these objects as elliptic branes or short ``E-branes''. In lack of a better understanding of $\EC(X)$, we view the E-branes as the analogues of D-branes in $S^1$-equivariant K-theory on the loop space $LX$.  In sect.~\ref{subsec:branef} we will construct a basis of linearly independent K-theory charges and show that eq.~\eqref{Zlv3d} has the generalization
\be\label{Zlv3dg}
Z_{S^1\times_q D^2}^{LV}(E)\sim \frac{1}{\eta^{\dim(X)}} \int_X e^{-J} \ e^{c_1^\beta(X)/2} \frac{\hAq( X)}{\hGq^*}\ \chq(E)\  e^{-(\ch^\beta_2(X)-\ch^\beta_2(E))/\ln q }\ .
\ee
We then tentatively assign the data displayed in Table~\ref{tab:tableind} to the 3d GLSM.
\begin{table}[t]
\begin{center}
\mbox{
\hbox{
\vbox{\offinterlineskip
\halign{\strut\vrule width 1.2pt~~#~\hfil\vrule width 1.2pt&~#~~~\hfil\vrule&~#~~~\hfil&#\vrule width 1.2pt\cr
\noalign{\hrule height 1.2pt}
&\multispan{2}{\hfil~~~~\bf 3d~theory~~~~~~$\xrightarrow{\ \beta\to0\ }$~~~~\bf 2d theory\hfil}&\cr
\noalign{\hrule height 1.2pt}
Full index&$\displaystyle\textrm{ind}_{\bar D_E}=\int_X \hAq (X) \chq(E)$&$\displaystyle\textrm{ind}_{\bar \p_E}=\int_X \td X \ch(E)$&\cr
Gamma class& $\displaystyle\hGq\sim \frac{1}{e^K_{S^1 }(N_+)} $&$\displaystyle\hG\sim \frac{1}{e_{S^1 }(N_+)}$&\cr
Half-index&$\displaystyle Z^{LV}_{S^1\times D^2 }\sim $ eq.\eqref{Zlv3dg}&$\displaystyle Z^{LV}_{D^2}\sim \int_X e^{J}\hG \ch(E)$ 
&\cr
\noalign{\hrule}
Boundary theory &E-branes&D-branes$\displaystyle\phantom{\sum^a}$&\cr
Brane charge&$K_{S^1}(LX)$&$K(X)$$\displaystyle\phantom{\sum^a_a}$&\cr
\noalign{\hrule height 1.2pt}
}}}}
\end{center}
\caption{Displayed are various indices and the boundary data of the 3d GLSM together their dimensional reduction to the 2d GLSM given in terms of the limit $\beta \to 0$.} \label{tab:tableind}
\end{table}
In the small radius limit, the 3d quantities on the l.h.s. of Table~\ref{tab:tableind} should reduce in a well-defined sense to those on the r.h.s.  In particular the 2d boundary conditions with K-theory charge  in $K(X)$ corresponding to D-branes descend from E-branes with K-theory charge in $K_{S^1}(LX)$, associated with the  boundary conditions for the 3d world-volumes.

In the following we address some simple issues related to the l.h.s. of Table~\ref{tab:tableind}. There are many interesting questions concerning the 3d lift to which we do not know the answers, such as the emergence of a generalized elliptic cohomology from the boundary SCFT,\footnote{See, however, ref.~\cite{Aganagic:2016jmx}.} the anomaly inflow mechanisms and an analysis of the category of boundary conditions along the lines of ref.~\cite{HHP}. We hope to come back to these questions in the future.

\subsection{Small radius limit and an $SL(2,\mathbb{Z})$ anomaly}
The 3d disk partition function \eqref{Zlv3d} is naturally defined as an expansion in small $|q|$. To obtain an expansion in the small radius limit  $|q|\to 1$, one needs to use an  $SL(2,\mathbb{Z})$ transformation $S: \tau \to -1/\tau$ on the complex structure of the boundary $T^2$. The relevant $J$-independent factor of the integrand corresponds to the characteristic function
\be\label{fx2d}
\tilde f_{D^2}(x)=e^{-\frac{x^2}{2\ln q }}\cdot e^{-\frac x 2}\ \frac{x\,(q)_\infty^2}{\theta(e^{-x},q)}\cdot \frac1{\Gamma_q(1-\frac x {\ln q})  }\ .
\ee
The $S$-transform of $\tilde f_{D^2}(x)$ is
\be\label{stf}
e^{-\frac{x^2}{2\ln q }}\cdot e^{-\frac x 2}\ \frac{x\, (q)_\infty^2 }{\theta(e^{-x})} =e^{-\frac{x_{2d}}2} \frac{x_{2d}(q')_\infty^2}{\theta(e^{-x_{2d}},q')}=: t(x_{2d})\ ,
\quad \tau'=-\frac1\tau\ ,\ x_{2d}=-\frac x \tau\ .
\ee
Taking $q'\to 0$ on the r.h.s., with $x_{2d}=-\frac{x}\tau$ fixed, gives 
\be
e^{-x_{2d}/2}\frac{x_{2d}}{1-e^{-x_{2d}}}\ ,
\ee
which is minus the characteristic function for the $\hat A$-genus in the 2d frame. Noting that the small radius of the $q$-Gamma function is 
\be
\Gamma_q\left(1+\frac x {\ln q}\right) \ \xrightarrow{\ q \to 1\ } \ \Gamma\left(1-\frac{x_{2d}}{2\pi i}\right)\ ,\qquad q=e^{2\pi i \tau}\ ,
\ee
we recover the 2d result  from refs.~\cite{HoriRomo,HonOk}.

Eq.~\eqref{stf} shows that the factor $e^{-\frac{\ch^\beta_2(X)}{2\ln q }}$ in eq.~\eqref{Zlv3d}  arises from the  failure of modular invariance of $t(x)$. The latter has a series expansion in $x$ in terms of the Eisenstein functions \cite{Schellekens:1986xh,Zagier86}
\be\label{G2exp}
t(x)= e^{-x/2} \frac{x(q)_\infty^2}{\theta(e^{-x},q)}=\exp\left(\sum_{k=1}\frac2{2k!}G_{2k}(\tau)x^{2k}\right)\ .
\ee
$t(x)$ is the characteristic function for the Witten genus, except for the term from  $k=1$. This term is multiplied $\ch_2(X)$ and the vanishing of this class is the condition for the twisted Dirac operator on the loop space to be well-defined \cite{WitLoop}. For non-trivial Chern character the condition is $\ch^\beta_2(X)-\ch_2^\beta(E)=0$, which is the coefficient of the corresponding term in eq.~\eqref{Zlv3dg}.

We emphasize that the original index \eqref{ind} is well-defined and gives an integral series regardless of the condition $\ch^\beta_2(X)-\ch_2^\beta(E)=0 $. In the following we assume that the $SL(2,\mathbb{Z})$ anomaly can be tolerated, or canceled, once the 2d boundary theory is coupled to the 3d bulk. The 2d formula \eqref{Zlv2d} was first obtained by an independent anomaly inflow argument on the D-brane boundary of the string \cite{Cheung:1997az,MM97}, including the necessary correction to make sense of the index on submanifolds without spin structure. Here we would need some sort of anomaly inflow for a membrane ending on an E-brane, that cancels an anomaly in the string structure.\footnote{An anomaly cancellation for M-theory membranes was discussed in ref.~\cite{WitE8}.}

On a technical level, a standard way to achieve $SL(2,\IZ)$ invariance is to replace the Eisenstein function $G_2(\tau)$ in eq.~\eqref{G2exp} by its $SL(2,\IZ)$ covariant cousin $\hat G_2(\tau)=G_2(\tau)+\frac{1}{8\pi \tau_2}$. This amounts to the replacement of $\hAq  (X)$ in eq.~\eqref{AGGrel} by 
\be\label{sl2zinv}
\td^\beta(X)\, \hGq\Gamma_{X,\bar q}\, e^{\varrho(q)\, \ch_2^\beta(X) } \quad \text{with} \quad
\varrho(q)=\tfrac{1}{\ln q}-\tfrac 1 {\ln (q) - \ln(q^*)} \ ,
\ee
combined with a similar shift of $\chq(E)$ to $\chq(E) e^{-\varrho(q) \ch_2^\beta(E)}$ for a non-trivial E-brane $E$ in eq.~\eqref{3dfullind}.  It would be interesting to understand this modification in terms of an obstruction to the holomorphic factorization of the sphere partition function due to unpaired zero modes on the boundary.

\subsection{BPS indices associated to K\"ahler manifolds} \label{LVBPSinv}
The $S^1\times_qD^2$ partition function computes the  index \eqref{ind} and a similar relation also holds for the $S^1\times_q S^2$ partition functions \cite{Imamura:2011su,Kratten,Kapustin:2011jm,Dimofte:2011py}. The BPS indices have series expansion with integral coefficients in the fugacities $(q,y_r)$, or  more specifically $(q,Q_i)$ in the case of the unperturbed theory associated with a K\"ahler manifold $X$. For small $|q|$ and $|Q|$ one expects them to be power series in $q$ and $Q_i$, starting with one in an appropriate normalization. We obtain the prediction that the 3d UV partition functions assign to the K\"ahler manifold $X$ an integral power series $\Ii_X(Q,q)$ with certain modular transformation properties. In the large volume limit, it reduces  to an integral $q$-series $\Ii_X(q)$ 
\be
X \ \xrightarrow{\ Z_{S^1\times C}\ } \  \Ii_X(Q,q) \ \xrightarrow{\text{\ LV limit\ }} \  \Ii_X(q)\ , 
\ee
where $C$ is either $D^2$ or $S^2$. As can be seen from eq.~\eqref{G2exp} and its relation to the Witten genus for ch$_2(X)=0$, the integral series $\Ii_X(q)$ are relatives of known cobordism invariants associated with $q$-Gamma functions.\\[1mm]
  
\subsubsec{Sphere index for $X$}
Let us first consider the sphere partition function, which is somewhat simpler due to the absence of anomalous terms and boundary factors. Repeating steps similar to the one around eq.~\eqref{Zlv3d}, one obtains for the large volume limit of the sphere partition function 
\be\label{ZSlv3d}
Z_{S^1\times_q S^2}^{LV}\sim \int_X e^{-J-\bar J} \ \td^\beta(X)\ \frac{\hGq}{\hGq^*}\ (1-q)^{2c_1(x/\ln q)}\ ,
\ee
with $J=\beta J_i \ln Q_i/\ln q$.
The $J$-independent terms correspond to the characteristic function 
\be\label{fS2func}
f_{S^2}(x)=\frac{x}{1-e^{-x}}\cdot \frac{(qe^{-x},q)_\infty}{(qe^{x},q)_\infty}\ .
\ee
For $c_1(X)=0$, the characteristic class $C_{S^2}(X)\equiv C(f_{S^2},X)$ has the expansion \footnote{\label{fnbeta} On the r.h.s. we drop the superscript $\beta$ on the 3d normalized Chern classes, c.f., Table~\ref{tab:charclasses}.}
\begin{multline}
C_{S^2}(X)= 1+\frac{1}{12} c_2+\frac{c_3 \psi _q(2,1)}{\ln^3(q)} 
   +\frac{1}{720} \left(3 c_2^2-c_4\right) \\
   +\frac{(c_5-c_2 c_3) \psi _q(4,1)+ c_2c_3 \ln^2(q)\psi_q(2,1)}{12 \ln ^5(q)} + \ldots \ .
\end{multline}
By integrating this class over $X$ we obtain the series  $\Ii_X(q)$ with an integral $q$-expansion as can be seen by applying the Hirzebruch--Riemann--Roch index theorem on $X$. The second factor in eq.~\eqref{fS2func} corresponds to a multiplicative characteristic class, which can be rewritten with formula~\eqref{qpoch} as
\be
\frac{(qe^{-x},q)_\infty}{(qe^{x},q)_\infty} = \exp\left[ -{\sum_{k=1}^{+\infty} \frac{q^k(e^{-kx} - e^{kx})}{k(1-q^k)} } \right] \ .
\ee
Applying now the splitting principle the class $C_{S^2}(X)$ takes the form 
\be
\begin{aligned}
C_{S^2}(X) &= \td(X) \exp\left[ -{\sum_{k=1}^{+\infty} \frac{q^k}{k(1-q^k)} \left[ \ch(\Psi_k(T^{0,1}_X)) - \ch(\Psi_k(T^{1,0}_X))\right]}\right] \\
& = \td(X) \sum_{\nu,\mu} f_{\nu,\mu}(q) \, \ch( S_\nu(T^{0,1}_X) \otimes S_\mu(T^{1,0}_X) )\ ,
\end{aligned}
\ee
with the Adams operator $\Psi_k$, $k=0,1,2,\ldots$, acting on the anti-holomorphic and holomorphic tangent bundles $T^{0,1}_X$ and $T^{1,0}_X$.\footnote{For a complex bundle $E$ we have the isomorphism $\overline{E} \simeq E^*$, which implies on the level of Chern classes $c_k(\overline{E}) = c_k(E^*) = (-1)^k c_k(E)$. In particular we have $T_X^{0,1} \simeq T_X^{1,0\,*}$, which allows us to write the characteristic class purely in terms of the Chern classes $c_k(X)\equiv c_k(T_X^{1,0})$.} In the second line the bundles $S_\nu(T^{0,1}_X)$ (resp. $S_\mu(T^{1,0}_X)$) denote the subbundles of $T^{0,1\,{\otimes |\nu|}}_X$ (resp. $T^{1,0\,{\otimes |\mu|}}_X$) associated to the representation of the symmetric group $S_{|\nu|}$ (resp. $S_{|\mu|}$) of the Young tableau $\nu$ (resp. $\mu$) with $|\nu|$ (resp. $|\mu|$) boxes. With the Schur representation of the topological vertex~\eqref{topvert}, we can explicitly spell out the coefficient functions $f_{\nu,\mu}(q)$ (labeled by a pair of Young tableaux) according to\footnote{Here we apply the splitting principal by replacing a complex vector bundle $E$ in terms of a direct sum of line bundles $L_1\oplus\ldots\oplus L_{\operatorname{rk}(E)}$, which is equivalent to $E$ on the level of characteristic classes. Using the identities $\ch(\Psi_k(E)) = \ch(\Psi_k(L_1))+ \ldots + \ch(\Psi_k(L_{\operatorname{rk}(E)}))$ and $\ch(S_\nu(E))=s_\nu(\ch(L_1),\ldots,\ch(L_{\operatorname{rk}(E)}))$, we arrive together with eq.~\eqref{topvert} at the explicit form of the functions $f_{\nu,\mu}(q)$.}
\be
f_{\nu,\mu}(q) = \left(-q^{1/2}\right)^{|\nu|+|\mu|} C_{00\nu}(q)C_{00\mu}(q^{-1}) \ .
\ee 
As a consequence of the algebraic properties of the vertex $C_{00\nu}(q)$ (c.f., ref.~\cite{Aganagic:2003db}), the coefficient functions $f_{\nu,\mu}$ are rational functions in $q$ with integral power series expansions. Thus, the series
\be
\Ii_X(q) = \sum f_{\nu,\mu}(q) \chi(X, S_\nu(T^{0,1}_X) \otimes S_\mu(T^{1,0}_X) )\ ,
\ee
becomes a sum of holomorphic Euler characteristics of the bundles $S_\nu(T^{0,1}_X) \otimes S_\mu(T^{1,0}_X)$ on $X$ with an integral $q$-expansion, as expected from the relation of the 3d partition function to an index of BPS states.\footnote{Alternatively, one would like to apply a suitable index theorem on the loop space~$LX$ \cite{Witgen}. Following the approach of ref.~\cite{Hirzebruch}, we can directly argue for integrality by identifying the second factor in eq.~\eqref{fS2func} with the Chern character of the bundle $\bigotimes_{n=1}^{\infty} \Lambda_{-q^n}(T^{0,1}_X)\otimes \bigotimes_{n=1}^{\infty}S_{q^n}(T^{1,0}_X)$, where $\Lambda_{t}(T^{0,1}_X) = \sum_{k=0}^{+\infty} t^k (\Lambda^k T^{0,1}_X)$ and $S_{t}(T^{1,0}_X) = \sum_{k=0}^{+\infty} t^k (S^k T^{1,0}_X)$ are the generating functions of the skew-symmetric and totally-symmetric tensor products of the bundles $T^{0,1}_X$ and $T^{1,0}_X$, respectively. Thus, $\Ii_X(q)$ furnishes a generating function in $q$ of particular sums of holomorphic Euler characteristic of the above tensor products of bundles.} 

The 2d limit $x/\ln q \to -x_{2d}/2\pi i$ of the class $C_{S^2}(X)$ is  
\be
C^{2d}_{S^2}(X)= 1-2c_3\,\zeta (3)+2 (c_2c_3-c_5)\zeta   (5)+ \cdots \ .
\ee
This  is the characteristic class that determines the perturbative corrections to the K\"ahler potential of the 2d theory \cite{Halverson:2013qca}. It is obviously non-integral due to the irrational coefficients proportional to $\zeta(n)$. The first correction term integrating to $-2\zeta(3)\frac{\chi(X)}{(2\pi i)^3}$ is well-known from mirror symmetry and represents a four-loop correction to the sigma model. The transcendental $\zeta(3)$ is obtained in ref.~\cite{Candelas:1990rm} by analytic continuation of the periods over the moduli space, or central charges of D-branes in modern language. Its 3d ancestor is 
\be\label{zeta3}
\psi_q(2,1)=\ln^3  (q)\cdot q\frac d {dq} \ln M(q)\ \xrightarrow{\ \text{2d limit} \ } \ -2 \zeta(3)\ ,
\ee
where $M(q)$ is the MacMahon function \eqref{MacM}, the generating function of 3d partitions. This suggests that the irrational coeffcients in the connection matrix for the analytic continuation of the periods of 2d mirror symmetry arise as the limiting values of integral BPS counting functions of the 3d theory, which appear in the connection matrix of the analytic continuation of central charges of E-branes. \\

\subsubsec{Disk index for $X$}
Similarly, we can now analyze the multiplicative characteristic class $C_{D^2}(X)$ based on the function~\eqref{fx2d}. Restricting to the $J$-independent terms
the disk partition function at degree zero yields the multiplicative characteristic class based on the function
\be
\hat{f}_{D^2}(x)=\frac{x}{1-e^{-x}} \cdot \frac{(q)_\infty}{(qe^x,q)_\infty} \ .
\ee
Here the hat `$\ \hat{}\ $' indicates that the anomalous contribution of the second Chern class to the modular symmetry $SL(2,\mathbb{Z})$ is removed. For $c_1(X)=0$, the resulting multiplicative characteristic class $C_{D^2}(X) \equiv C(\hat f_{D^2},X)$ yields the expansion\footnote{See footnote~\ref{fnbeta}.}
\bea
\hskip-2cm 
C_{D^2}(X) &=&
1-\frac{c_2 \left(12\psi_1-\ln^2(q)\right)}{12 \ln^2(q)}
+\frac{c_3 \psi_2}{2 \ln^3(q)} +\frac{1}{720 \ln ^4(q)}
\big(360 c_2^2 \psi_1^2\\&&
+ 60(c_2^2 - 2 c_4)\psi_3
-60 c_2^2 \ln^2(q) \psi_1 +(3c_2^2-c_4)\ln^4(q)
\big) +\ldots \ , \nonumber
\eea
where $\psi_k=\psi_q(k,1)$.
It is again a multiplicative characteristic class with an integral $q$-expansion. The integrality can again be argued for with the help of the Hirzebruch--Riemann--Roch index theorem. Namely, $C_{D^2}$ becomes a sum over Young tableaux~$\nu$ of the form
\be
C_{D^2}(X) = \td(X) \, \left( \frac{\eta(q)}{q^{1/24}} \right)^{\dim(X)}  \sum_{\nu} (-q^{1/2})^{|\nu|} C_{00\nu}(q^{-1}) \, \ch(S_\nu(T^{1,0}_X)) \ .
\ee
As the factors $(-q^{1/2})^{|\nu|} C_{00\nu}(q^{-1})$ enjoy an integral $q$-expansion, the large volume limit of the disk partition function~\eqref{Zlv3d} yields the series $\Ii_X(q)=\eta(q)^{-\dim(X)}\int_XC_{D_2}(X)$, which realizes integral sums of holomorphic Euler characteristics 
\be
\Ii_X(q)=\sum_{\nu} (-q^{1/2})^{|\nu|} C_{00\nu}(q^{-1}) \, \chi(S_\nu(T^{1,0}_X))\ ,
\ee
with integral $q$-coefficients, where we removed a constant factor to normalize the leading term to one.

\subsection{BPS indices beyond the large volume limit}
Analogously to the derivation of the large volume limit of the disk partition function~\eqref{Zlv3dg}, the entire 3d disk partition function can be written with eq.~\eqref{RestoVolInt} in the geometric form 
\be\label{Zlv3dggen}
Z_{D^2\times_q S^1}(E)\sim \!\!\!\! \sum_{\gamma \in H_2(X,\mathbb{Z})}  e^{-t\cdot \gamma} \int_X  e^{-J} \ e^{c_1^\beta(X)/2} \frac{\hAq( X)}{\widehat{\Gamma}^*(\gamma)}\ \chq(E)\  e^{-\frac{\ch^\beta_2(X)-\ch^\beta_2(E)}{\ln q} }\ ,
\ee
where $t_i=\hat t_i-\frac{c_i}2$ and $t \cdot \gamma=\sum t_i \gamma_i$. The sum over $\gamma$ runs over the non-negative curve classes in $X$, which label the topological sectors of the discussed vortex configurations. The class $\widehat{\Gamma}^*(\gamma)$ is defined in terms of the Chern roots~$x_\alpha$ associated to the chiral fields with charges $q_\alpha$ and with Neumann and Dirichlet boundary conditions as
\be \label{wGclass}
\widehat{\Gamma}^*(\gamma)= \frac{\prod_{\alpha\in N} \Gamma_q(1-\frac {x_\alpha} {\ln q} +q_\alpha \cdot \gamma)}{\prod_{\alpha\in D} \Gamma_q(1+\frac {x_\alpha} {\ln q} - q_\alpha \cdot \gamma)} \ .
\ee
For $\gamma=0$ the class $\widehat{\Gamma}^*$ simplifies to the multiplicative characteristic class $\hGq^*$, and we recover the large volume disk partition function~\eqref{Zlv3dg}.

Using eq.~\eqref{defgammaq} we can argue that the class~\eqref{wGclass} takes the general form
\be \label{wGclassexp}
\widehat{\Gamma}^*(\gamma) = g^\gamma(q,\ch(\mathcal{L}_\alpha)) \cdot \hGq^* \ ,
\ee
in terms of the line bundles $\mathcal{L}_\alpha$ with $c_1(\mathcal{L}_\alpha)=x_\alpha$. By construction the functions $g^\gamma(q,\ch(\mathcal{L}_\alpha))$ have again an integral $q$-expansion. Therefore, repeating the arguments of sect.~\ref{LVBPSinv}, we explicitly find that the 3d partition function~\eqref{Zlv3dggen} yields an integral q-series $\Ii_X(Q,q)$ in all topological vortex sector labeled by $Q$. While the integrality property is again expected from the interpretation of $\Ii_X(Q,q)$ as a generating function of BPS indexes, the expression~\eqref{wGclassexp} offers a geometric interpretation of the BPS indexes in terms of holomorphic Euler characteristics of complex vector bundles built from the line bundles $\mathcal{L}_\alpha$.

\subsection{3d brane factors and Mellin--Barnes integrals \label{subsec:branef}}
The aim of this section is to describe integral bases of E-branes which generate a basis of K-theory charges and give rise to a set of  linearly independent partition functions with a large volume limit \eqref{Zlv3dg}.

\subsubsection{Integral solutions of Mellin--Barnes type} 
The reduced system of $q$-difference operators \eqref{DiffEqgen} has $k=\dim(K(X))$ linearly independent solutions, the $q$-periods in eq.~\eqref{seps}.\footnote{Here, linear dependence is defined with coefficients in $q$-dependent functions, i.e., different elements in $K_{S^1}(LX)$ are considered equivalent if they correspond to the same local solution up to an overall $q$-dependent factor.} The reduced difference equations and the $q$-period vector for the degree $N$ hypersurface in $\IP^{N-1}$ are given in app.~\ref{app:sol} and will serve as an example. The boundary condition considered so far selects one particular linear combination of the $q$-periods. To obtain more general solutions to the $q$-difference system we consider insertions of extra "brane factors" in the residue integral
\be\label{ZE}
Z(E)=\ln(q)\, \int \frac{d\eps}{2\pi i}\ \FFd\cdot  \s(Q, q,\eps)\cdot \ff_E(q,q^\eps)\, .
\ee
So far  $\ff_E(q,q^\eps)=1$, which by eq.~\eqref{epstoH} corresponds to the brane on $X$ with $\chq(E)=1$, i.e., full Neumann boundary conditions.

We seek a set $\{\ff_{E_\alpha}\}$ of brane factors, such that $i)$ the partition functions with insertions of $\ff_{E_\alpha}$ give a complete basis of solutions to the original $q$-difference system and $ii)$ the basis is integral in the sense that the the large volume limit generalizes the index \eqref{Zlv3d} to an $S^1$-equivariant bundle $E$ on $LX$ as in \eqref{Zlv3dg}. More generally one may add the factor $\ff_E(z,q)$ in the original Coulomb integral \eqref{Z3D}
\be\label{ZEg}
Z(E)=\int \frac{dz}{2\pi z} \left(e^{-S_\text{class}}\prod_\alpha Z_\alpha\right) \cdot \ff_E(z,q)\ ,
\ee
such that it reduces to \eqref{ZE} upon evaluation at the poles. Eq.~\eqref{ZEg} may serve as the starting point for an analytic continuation of $Z(E)$ over the deformation space by contour deformation. Integral solutions to differential or difference equations of the above type are referred to as Mellin--Barnes integrals.\footnote{See refs.~\cite{Aganagic:2016jmx,Aganagic:2017gsx} for a discussion in the context of $\cx N=4$ supersymmetry and ref.~\cite{Knapp} for a recent discussion in the context of the 2d GLSM.}

In order that the integral with an insertion of $\ff_E$ fulfills the same $q$-difference equation as the original integrand with $\ff=1$, the factor $\ff_E(z,q)$ has to be invariant under shifts of $z$
\be
\ff_E(z,q)=\ff_E(zq,q)\ .
\ee
Indeed the derivation of the $q$-difference equation around \eqref{gammarec} can be lifted straightforwardly to the integrand before summing over poles, if one assumes that the integration contour does not pass poles under a shift $z\to qz$. Three simple shift invariant functions that may serve as building blocks are 
\be \label{bfcs}
\ff_1(z,q)= e^\frac{2\pi i \ln z}{\ln q}\ , \quad
\ff_2(z,q)= \theta(z,q)e^{-\frac{\ln^2(-zq^{-1/2})}{2\ln q}}\ , \quad
\ff_3(z,q)= \prod_i \frac{\theta(zx_i,q)}{\theta(zy_i,q)}\ ,
\ee
recalling that $\theta(zq,q)=-z^{-1}\theta(z,q)$. The functions $\ff_1,\ff_2$ are invariant only under $z\to qz$, but not $z\to e^{2\pi i}z$, while $\ff_3$ is invariant under both shifts, i.e., elliptic,  if the arguments $x_i,y_j$ satisfy $\prod_i  x_i = \prod_i y_i$.\footnote{More generally, one may replace $z$ by powers of $z$ in \eqref{bfcs} with an appropriately modified condition.} Factors of the type $\ff_3$ are rational in factors of type $\ff_2$  and have been used in ref.~\cite{Meijerq} to define $q$-analogues of Meijer functions.\footnote{Ref.~\cite{Meijerq} considers also non-elliptic factors $\ff_3$, but these lead to functions satisfying different difference equations than the original solution. For a relation between elliptic ratios and Chern--Simons interactions see ref.~\cite{BDP}.} 
In addition to shift invariance, the factor $\ff$ has to have appropriate convergence properties on the integration contour used in \eqref{ZEg}. The three factors have a simple physical interpretation in the 3d partition function: $\ff_1$ represents an integral mixed CS term for the $U(1)$-$R$-symmetry and is generated by the monodromy in the FI~term
\be\label{FImon}
e^{{\ln z \ln Q}/{\ln q}} \ \xrightarrow{\  Q\to e^{2\pi i } Q \ } \  e^{{\ln z \ln Q}/{\ln q}}\ e^{{2\pi i \ln z}/{\ln q}}\ .
\ee
The choice $\ff_2$ is related to integrating in $\mathcal{N}=(0,2)$ boundary fermions and will be discussed in detail below. An elliptic factor $\ff_3$ describes an anomaly free combination of boundary fields.

Up to minor modifications, $\ff_2$ is the 1-loop determinant of a $\mathcal{N}=(0,2)$ fermi multiplet on $T^2=\p(S_1 \times_q D^2)$ computed in ref.~\cite{YS}
\be
Z_{fermi}(v)=e^{\frac 1 {2\ln q}\ln^2(v q^{-1/2})}q^{-1/24} \  \theta(v)\ , \qquad  v=z^{q_\alpha}q^{\Delta_\alpha/2}y_r^{\gq_{\alpha r}},
\ee
using the same notation as around \eqref{Zmatter}. The 1-loop determinant of a fermi multiplet on $T^2$ had been computed earlier in the context of string theory \cite{AGMV,Schellekens:1986xh} and in the derivation of the loop space index theorems in ref.~\cite{WitLoop,Alvarez87}. The result differs in the prefactor of the theta function, which depends on a choice of regularization for the infinite products in the determinants. The regularization obtained from the $S^1\times_q D^2$ partition function in ref.~\cite{YS} produces  the shift invariant factor $\ff_2$, up to the change of sign in the exponent of $\ff_2$, which is necessary for shift invariance.

\subsubsection{Integrating in boundary fields and Dirichlet directions \label{sss:dir}}
We now describe a simple basis of branes that can be obtained by a boundary version of integrating in massive particles and relate it to Dirichlet boundary conditions on $X$ and $LX$. Let us again discuss the 2d case first.\footnote{Constructions of D-brane boundary conditions using 1d boundary fermions have been discussed in refs.~\cite{Mayr:2000as,Govindarajan:2001kr,HHP,HoriLin}.} Integrating in a periodic boundary fermion $\eta$ on $S^1$ charged with respect to a gauge symmetry $U(1)$ contributes a determinant factor
\be \label{fdet2d}
\ff=\det(1-e^{\frac{iF}{2\pi}}) = 1-ye^{-x_\eta}\ ,
\ee
where $2\pi i x_\eta$ is the eigenvalue of the $U(1)$ field strength $F$ in the representation of $\eta$ and $y$ is a weight representing the non-zero mass for $y\neq 1$; it corresponds to the $S^1$-equivariant version of the index theorem \cite{AG}. The two-dimensional $\mathbb{C}$ module generated by the fermion zero mode is spaned by $\ket 0$ and $\eta\ket 0=\ket 1$, where $\ket q$ has $U(1)$ charge $q$. $\ff$ is the Chern character of the alternating bundle $E=\sum_{i=0}^1 (-)^i \wedge^i L$, where $L$ is the bundle associated to the fermion $\eta$:
\be\label{cheta}
\ch(\, \ket 0 \buildrel \eta \over  \to \ket 1\, )=1-ye^{-x_\eta} \ . 
\ee
Starting with the ordinary $\bar\p$ index on a weighted projective space $X=W\IP^{n-1}$  and integrating in $\eta$ produces the integrand 
\be
\td(X)\cdot  \ch(E) = \prod_{\alpha=1}^{n}\frac{x_\alpha}{1-e^{-x_\alpha}}\cdot (1-ye^{-x_\eta})\ .
\ee
If $x_\eta= x_\alpha$ for some $\alpha$, say $\alpha=1$, taking the massless limit $y\to 1$ cancels a bosonic determinant factor in $\td(X)$, giving 
\be\label{td2dm}
\prod_{\alpha=2}^{n}\frac{x_\alpha}{1-e^{-x_\alpha}}\cdot x_1 = \td(H) \cdot c_1(N_H)\ .
\ee
Here $H\subset X$ is the hypersurface with normal bundle $N_H$ defined by setting the bosonic field (homogeneous coordinate) in $\varphi_1$ to zero. Integrating in $\eta$ has created a Dirichlet boundary condition $\varphi_1=0$.

We will now use a similar idea to describe Dirichlet conditions on $LX$.
Consider integrating in boundary fermions in the 3d partition function with determinant 
\be\label{deff2}
\ff_2= \theta(ye^{-x})c(ye^{-x})\ , \qquad c(x)=\exp\left(\frac{1}{2\ln q }\ln^2 (-xq^{-1/2})\right)\ .
\ee
Using an $S$-transformation as in \eqref{stf} one may check that the 3d brane factor $\ff_2$ reduces to the 2d Chern character \eqref{fdet2d} in the small radius limit. 
Repeating the argument around \eqref{td2dm} gives, in the massless limit $y\to 1$,\footnote{Eq.~\eqref{td2dm} is a special case of the Grothendieck--Riemann--Roch formula for $X$. The following equation should represent a special case of a Grothendieck--Riemann--Roch formula for $LX$.}
\be\label{td3dm}
\hAq (X) \cdot \ff_2 = \hAq (H) \cdot x_1\cdot 
[iC^{-3}e^{i\pi x_1/\ln q}e^{x_1^2/2\ln q} (q)_\infty^2]\ ,
\ee
where $C=q^{-\frac 1 {24}}q'^{\frac 1 {24}}$.
The r.h.s. is related to the charge for a 3d brane associated to the Dirichlet condition $\varphi_1=0$. The factor in the square bracket comes from the regularization of the bulk theory coupled to the 2d boundary theory; thus the above manipulation should be considered on the integrand of the half-index:
\
\be
\frac{1}{(-\eta)^{d}}\frac{\hAq(X)}{\hGq^*} e^{-\ch^\beta_2(X)/\ln q}\cdot \ff_2 = 
\frac{e^{i\pi x/\ln q}}{(-\eta)^{d-1}}\frac{\hAq(H)}{\Gamma_{H,q}^*} e^{-(\ch^\beta_2(X)-\ch^\beta_2(E))/\ln q}\cdot c^K_1(N_H) 
\ee
where $\ch_2^\beta(E)=x_1^2/2$ and the factor $e^{i\pi x/\ln q}$ is a half-integral contribution to the FI term equal to $\ff_1(e^{x/2})$. Moreover
\be
c^K_1(N_H)=x_1\frac{iC^{-3}q^{-1/24}(q)_\infty}{\Gamma_q(1-x_1/\ln q)}=
x_1\prod_{n=1}^\infty (1-e^{-x}q^n)\cdot [-iC^{-3}q^{-1/24}(1-q)^{-x_1/\ln q }]\ ,
\ee
can be interpreted as an Euler class of the normal bundle including the contribution from negative loops, cpw. \eqref{ENK}. 

The connection of the fermion determinant to the $S^1$-equivariant Chern character for a bundle on $LX$ can be illustrated treating the 2d fermion $\eta$ on $T^2$ as a 1d fermion on $S^1$ with infinitely many Fourier modes $\eta_k$, $k\in \mathbb{Z}$, weighted by $q^k$. Restricting for the moment to the non-negative modes $k\geq 0$, the Fock space generated by these modes of the single 2d fermion is the infinite sequence 
\be
\ket 0 \to \sum_{0\leq k_1} \eta_{k_1} \ket 0 \to \sum_{0\leq k_1<k_2} \eta_{k_1} \eta_{k_2} \ket 0 \to ...
\ee
corresponding to an alternating bundle $E_+=\sum_{i=0}^\infty (-)^i \wedge^i L$, where the subscript means restriction to $k\geq 0$. The equivariant character generalizing the 2d expression \eqref{cheta} is
\bea
&&1-\sum_{0\leq k_1} q^{k_1} y e^{-x} + \sum_{0\leq k_1<k_2} q^{k_1+k_2} y^2 e^{-2x} - \sum_{0\leq k_1<k_2<k_3} q^{k_1+k_2+k_3} y^3  e^{-3x} + ...  \nonumber \\
&=& 
1-\frac{ye^{-x}}{1-q}+\frac{y^2e^{-2x}q}{(1-q)(1-q^2)}-\frac{y^3e^{-3x}q^3}{(1-q)(1-q^2)(1-q^3)}+...\nonumber \\
&=&\sum_{k=0}^\infty \frac{(-ye^{-x})^k q^{k(k-1)/2}}{(q)_k}=(ye^{-x},q)_\infty\ ,
\eea
where \eqref{qpsum} has been used in the last step.
By an appropriate choice of the vacuum $\ket 0$, the modes for negative $k$ can be treated as another set of modes with positive $k$ but opposite $U(1)$ charge. Multiplying the two contributions gives for the total Chern character
\be
\chq(E)=(ye^{-x},q)_\infty(qy^{-1}e^{+x},q)_\infty=\theta(ye^{-x},q)\ ,
\ee
which is the brane factor $\ff_2$, up to the prefactor from the regularization. 

\subsubsection{3d matrix factorizations}
The new boundary degrees of freedom added in the last step have to be coupled to the rest of the theory in a supersymmetric way. Boundary conditions for the 2d theory with $B$-type supersymmetry can be defined by matrix factorizations $W=E\cdot J$ of the superpotential \cite{KapLi}, and a similar description exists for boundaries of the 3d theory with $\mathcal{N}=(0,2)$ supersymmetry \cite{GukovMF,YS}. The factors $E$ and $J$ determine the supersymmetric couplings of the boundary fermions in 1d or 2d, respectively. The action of  $\mathcal{N}=(0,2)$ Fermi multiplets has been thoroughly studied in the context of linear sigma models for heterotic strings, starting with refs.~\cite{WitPhases,Distler:1993mk}.

The Chern characters considered above are related to simple matrix factorizations of Koszul type described as follows. In the 2d theory one considers $r$ fermionic annihilation and creation operators $\eta_i$ and $\bar \eta_i $ with anti-commutators $\{\eta_i,\bar \eta_j\}=\delta_{ij}$ and $\{\eta_i,\eta_j\}= 0 = \{\bar \eta_i,\bar \eta_j\}$ acting on a vacuum $\ket 0$ with $\eta_i\ket 0 =0$. The Fock space obtained by acting with the $\bar \eta_i$ on $\ket 0$ is a sum of graded vector spaces. The Koszul type complex is defined by a fermionic map $Q=\sum x_i \eta_i$ connecting consecutive vector spaces $\bigoplus_{i_1<i_2\ldots<i_k}\bar \eta_{i_1}\ldots\bar \eta_{i_k}\ket 0$ and $\bigoplus_{i_1<i_2\ldots<i_{k-1}}\bar \eta_{i_1}\ldots\bar \eta_{i_{k-1}}\ket 0$ of fixed fermion number. For $r$ fermions and bosonic maps $x_i$ of equal charge $q$ one obtains a complex of vector bundles
\be\label{oseq}
\cx O(q_0) \xrightarrow{\sum x_i\eta_i} \cx O(q_0+q)^{\oplus r} 
 \xrightarrow{\sum x_i\eta_i} \cx O(q_0+2q)^{\oplus \binom{r}2}  \xrightarrow{\sum x_i\eta_i} \, \cdots \, 
 \xrightarrow{\sum x_i\eta_i} \mathcal{O}(q_0+rq)\ ,
\ee
where $q_0+rq$ is the charge of the vacuum. The constructions of 2d boundary conditions using more general complexes of fermions has been given, e.g., in refs.~\cite{Mayr:2000as,Govindarajan:2001kr}. These complexes can be associated to 2d matrix factorizations by specifying in addition the action of the $U(1)$ $R$-symmetry group on the vector spaces \cite{HHP}.

For the degree $N$ hypersurface in $\IP^{N-1}$ one considers factorizations of the superpotential\footnote{In this section $\varphi_i$ denotes a superfield  and $x_i$ its lowest components; we use also $p=x_0$.}
\be
W=\varphi_0\, g_N(\varphi_i)\ .
\ee
The GLSM has two phases, a large volume phase where $p=0$ and the equation $g_N(x_i)=0$ cuts out a hypersurface in $\IP^{N-1}$ parametrized by $x_i$, and a Landau-Ginzburg phase where $p\neq 0$ and $x_i$ parametrize $\IC^N/\mathbb{Z}_N$ \cite{WitPhases}. Two special Koszul complexes associated with the fields $\varphi_0$ and $\varphi_{i>0}$ considered in ref.~\cite{HHP} are
\be\label{mfs}
\begin{aligned}
a)\quad &\cx O(q_0) \overset{g_N}{\underset{p}{\rightleftarrows}} \cx O(q_0+N) \ ,&&Q=g_N\eta_0+p\bar\eta_0\ ,\\
b)\quad &\cx O(q_0)\overset{x_i}{\underset{W_i}{\rightleftarrows}}
\cx O(q_0+1)\overset{x_i}{\underset{W_i}{\rightleftarrows}}\, \cdots \, 
\overset{x_i}{\underset{W_i}{\rightleftarrows}}
\cx O(q_0+N)\ ,
&\quad&Q=\sum_{i=1}^N x_i\eta_i+W_i\bar\eta_i\ ,
\end{aligned}
\ee
corresponding to factorizations $W=p\cdot g_N(x_i)$ and $W=\sum_i x_i \cdot W_i$ with $W_i=\p_{x_i}W$, respectively. The factorization $a)$ represents a trivial configuration near the LG point $Q^{-1}=0$, where $p\neq 0$ and the boundary potential is strictly positive. Similarly the factorization $b)$ is trivial near large volume, where the set $x_i=0$ for all $i$ is excluded.

We can use the {\it same} sequences to define the couplings of the 2d boundary fermions for a 3d matrix factorization associated to the boundary $\p(S^1\times_q D^2)$. The difference lies in the different contribution of the higher-dimensional fields to the path integral. The fermion zero mode is replaced by a chiral fermion $\psi(z)$. 

The simplest quantity to consider is the graded sum of cohomologies, which computes the Chern character for a bundle on $X$ and $LX$  for boundary fermions in 1d and Fermi multiplets in 2d, respectively. For the complexes in \eqref{mfs} these are, up to normalization factors, the Chern characters computed in sect.~\ref{sss:dir}
\be\label{chmf}
\begin{tabular}{ccc}
 &  2d  & \ \ \   3d \\
$a)\quad$  &$1-e^{-Nx}$ & \qquad $\ff_2(e^{-Nx})$\\
$b) \quad$  &$(1-e^{-x})^N $& \qquad $\ff_2(e^{-x})^N$
\end{tabular}
\ee
By positivity of the boundary potential, the 3d matrix factorizations associated with the sequences $a)$ and $b)$ should correspond to trivial E-branes in the IR near the LG point and large volume point, respectively. This is consistent with the fact that an insertion of the brane factor in the Mellin--Barnes integral considered below makes the integrand of the residue integral regular in the respective regime.

Similarly, the boundary conditions corresponding to $k$ Dirichlet directions on $LX$ considered in the previous sections represent another set of simple 3d matrix factorizations with 
\be
c)\ \ \ Q=g_n\eta_0+(x_1\eta_1+...x_k\eta_k)+\bar \eta_0 p,\qquad \chq(E)\sim\ff_2(e^{-Nx})\ff_2(e^{-x})^k\ .
\ee
To describe more general cases, one needs to understand the equivalence relations between E-branes, i.e., the analogue of tachyon condensation for D-branes studied in ref.~\cite{KapLi,HHP}. This an important open problem. In the 3d theory, the objects in the  sequence \eqref{oseq} do not represent $\IC$-modules  associated with 1d fermionic zero modes, but the non-trivial $\bar Q_+$ cohomology underlying the elliptic genus \cite{Witgen}. These spaces are modules of a chiral algebra generated by 2d chiral fermions $\psi(z)$ and these have to be matched in a 3d generalization of subtracting "trivial" branes.\footnote{Modules of chiral algebras appear also in the context of triangulations of 4-manifolds \cite{Feigin:2018bkf}, where distinct triangulations are proposed to yield equivalence relations among chiral algebras. It would be interesting to see, if such equivalences are meaningful in the context of E-branes as well.}

\subsubsection{Bases of solutions via  Mellin-Barnes integrals}
The boundary conditions described above allow to construct bases of linearly independent integral solutions. Here we consider again the degree $N$ hypersurface in $\IP^{N-1}$ for simplicity. The partition function with spectrum \eqref{DefQ} can be written as the Mellin--Barnes type integral 
\be\label{mb1}
Z(E^{LV}_0)\sim \int Q^{\sigma} e^{-CS}\ \frac{\Gamma_q(-\sigma)^N}{\Gamma_q(-N\sigma)}  d\sigma \ .
\ee
Here $\sigma=-\ln z/\ln q $ and $E_0^{LV}$ stands for the brane on $X$ with Neumann boundary conditions in the large volume phase.\footnote{We will not be careful about the normalization and the Chern--Simons terms hidden in $e^{-CS}$, which are fixed as in sect.~\ref{subsec:3dpfpn} such that the $q$-difference system is given by eq.~\eqref{DiffEq1}.} The contour is initially chosen to sum up the poles $\sigma=n$ for $n\geq 0$ and gives \eqref{Z3D2}.

In \eqref{mb1}, the 3d chiral  $\varphi_0$ of charge $-N$ has Dirichlet boundary condition, which sets the superpotential at the boundary to zero and is supersymmetric without introduction of boundary terms. Using the identity of the $q$-Gamma function
\be
\Gamma(\sigma)\Gamma(1-\sigma)= \frac{(q)_\infty^2(1-q)}{\theta(q^\sigma)}\ ,
\ee
the above integral can be rewritten as 
\be\label{mb2}
Z(E_0^{LV})\sim \int Q^{\sigma} e^{-CS}\ \Gamma_q(-\sigma)^N \Gamma_q(1+N\sigma)\  \ff_2(q^{-N\sigma}) \ d\sigma \ .
\ee
This expression describes a $\varphi_0$ field of charge $+N$ with Neumann boundary conditions. In this case there is a non-vanishing boundary variation for supersymmetry transformations, which needs to be cancelled by  coupling to a $\mathcal{N}=(0,2)$ boundary Fermi multiplet via 3d matrix factorization \cite{YS,GukovMF}. The Fermi multiplet contributes a factor $\ff_2(q^{-N\sigma})$ as in \eqref{chmf}.

Starting from either \eqref{mb1} or \eqref{mb2}, a basis of solutions is obtained by integrating in boundary fermions with brane factors $\ff_2(z)$
\bea\label{lvb}
Z(E^{LV}_a) &\sim& \int Q^{\sigma} e^{-CS}\ \frac{\Gamma_q(-\sigma)^N}{\Gamma_q(-N\sigma)}  \cdot \ff_2(q^{-\sigma})^a d\sigma , \quad a=0,...,N-2\ .\nonumber
\eea
The above E-branes constitute a $\dim(K(X))$-dimensional basis $\{E^{LV}_a\}$ of linearly independent integral K-theory charges, and the Mellin--Barnes integrals give a basis of linearly independent solutions to the $q$-difference system \eqref{DiffEq1} near small $|Q|$. However they do not give global solutions, as the integrand does not have poles at $\sigma<0$. The regularity of the integrand in this regime is due to the factor $\ff_2(q^{-N\sigma})$ in eq.~\eqref{mb2} and is consistent with the claimed triviality of the  factorization $a)$ in the three-dimensional theory. 

At the Landau--Ginzburg point $|Q|$ is small and $p\neq 0$ \cite{WitPhases}. This excludes Dirichlet boundary conditions for $\varphi_0$. Imposing Dirichlet conditions on all $\varphi_{i>0}$ and Neumann conditions on $\varphi_0$, gives the integral 
\bea\label{lgb}
Z(E^{LG}_a) &\sim& \int Q^{\sigma} e^{-CS}\ \frac{\Gamma_q(1+N\sigma)}{\Gamma_q(1+\sigma)^N} \ \ff_1(q^{\sigma})^a d\sigma 
\eea
with $a=0$.
Summing over the poles at $\sigma=-k/N$ one obtains the solution~\eqref{solLG} as a series in $Q^{-k/N}$. The complete basis of solutions \eqref{solLGg} is generated by phase rotations $Q\to Qe^{2\pi i}$ \eqref{FImon}, adding powers of brane factors $\ff_1$.  The integrand does not have poles in the regime $\sigma>0$, as it is obtained from the one in eq.~\eqref{mb1} by multiplication with $\ff_2(z)^N/\ff_2(z^N)$. This is proportional to the Chern character of the matrix factorization $b)$ and the regularity of the integrand for $\sigma>0$ confirms the triviality of the 3d matrix factorization at large volume. 

Due to the absence of poles in the opposite regime, neither of above integrals defines a global solution to the $q$-difference equation. It is straightforward to introduce brane factors that reduce to the local solutions above and have residues in both regimes. E.g., allowing for rational factors in $\ff_1$ gives the integral
\bea\label{lgb2}
Z&\sim& \int Q^{\sigma} e^{-CS}\ \frac{\Gamma_q(1+N\sigma)}{\Gamma_q(1+\sigma)^N} \ \frac{1}{1-e^{2\pi i \sigma}}d\sigma \ .
\eea
which at large volume gives $Z\sim Z(E_0^{LV})$,  while $Z\sim -\frac 1 N \sum_{a=1}^{N-1} a Z(E_a^{LG})$ at the Landau-Ginzburg point.  Integrals of this type are expected to arise from more general 3d matrix factorizations, which are obtained from the above studied complexes of Koszul type by using equivalences between E-branes.

\section{Mirror symmetry} \label{sec:mirror}
$\cx N=2$ supersymmetric 3d gauge theories have a symmetry that is called mirror symmetry \cite{ISa,AHISS, Oz,KapStr}. It maps the Higgs branch of one theory to the Coulomb branch of the dual theory and vortices of the former to bound states of electrons and monopoles in the latter. Since the 3d partition function computes the vortex sum $I(Q,q)$, one may expect a nice action of 3d mirror symmetry on this quantity. It has been shown in ref.~\cite{AHKT} that 3d mirror symmetry may be related to the 2d Hori--Vafa mirrors \cite{HV} in the small radius limit of an $S^1$ compactification. Combining this with the IR flow to equivariant quantum-K-theory  and quantum cohomology, respectively, one may hope to learn something new about certain aspects of 3d/2d mirror symmetry. 

In the following we relate the vortex sum of the original partition function for $X$ to the partition function of the gauge theoretic mirror, called $Y$. The latter takes the form of a 3d version of Landau--Ginzburg type overlap integrals, giving a $q$-generalization of the 2d expressions derived in ref.~\cite{HIV}. The same type of integrals appears in the definition of K-theoretic mirrors of ref.~\cite{Giv15all}, showing that these are special cases of 3d gauge theoretic mirrors.

\subsection{Partition functions for gauge theoretic mirrors}
The $\cx N=2$ mirror pairs relevant to the class of 3d GLSM considered in this paper have been described in refs.~\cite{DT,AHKT}. For a theory $X$ of the type considered in the previous section, its Higgs branch is mirror-dual to a theory $Y$ in its Coulomb branch. Such a mirror pair $(X,Y)$ of $\mathcal{N}=2$ supersymmetric 3d theories is given by the gauge theory data displayed in Table~\ref{tab:Mpair}.
\begin{table}[t]
\begin{center}
\mbox{
\hbox{
\vbox{\offinterlineskip
\halign{\vrule width 1.2pt\strut~#~\hfil&\vrule~#~\hfil\vrule width 1.2pt\cr
\noalign{\hrule height 1.2pt}
3d theory $X$ --- Higgs branch:&3d theory $Y$ --- Coulomb branch:\cr
\noalign{\hrule}
gauge group $U(1)^k$ & gauge group $U(1)^{N-k}$ \cr
$N$ chiral multiplets $\varphi_\alpha$ of charge $q_\alpha^a$ & $N$ chiral multiplets $\hat\varphi_\alpha$ of charge $\hat q_\alpha^r$ \cr
FI parameters $\zeta^a$, masses $m_\alpha$ & FI parameters $\hat\zeta^a$, masses $\hat m_\alpha$ \cr
($\alpha=1,\ldots,N$; $a=1,\ldots,k$) & ($\alpha=1,\ldots,N$; $r=1,\ldots,N-k$) \cr
\noalign{\hrule height 1.2pt}
}}}}
\end{center}
\caption{The table exhibits the gauge theory data of a pair of mirror dual 3d theories. Namely, the Higgs branch of the 3d theory $X$ in the left column and the Coulomb branch of the 3d theory $Y$ in the right column are dual to each other.} \label{tab:Mpair}
\end{table}
The $R$ charges can be chosen as in \eqref{Rchcan}. The charges have to fulfill the condition 
\be\label{qorth}
\sum_\alpha q^a_\alpha\hat q_\alpha^r=0\ , \qquad \begin{array}{l}a=1,\ldots,k\ ,\\ r=1,\ldots,N-k\ .\end{array}
\ee
Moreover, the effective FI terms and masses on the two sides are related by 
\be\label{MiMaTong}
\zeta^a=\sum_\alpha q_\alpha^a \hat m_\alpha\ ,\qquad
\hat \zeta^r=\sum_\alpha \hat q_\alpha^r  m_\alpha\ .
\ee
For concreteness we consider an example from the previous section, the $\IP^{M-1}$ theory perturbed by a massive particle of $U(1)$ charge $-\ell$ (with $\ell \le M$). The charges and masses of the theory are 
\be
(q^{a}_\alpha) = (-\ell,1,\ldots,1)\ , \qquad  (m_\alpha) = (m_0,0,\ldots,0)\ ,\ \qquad \alpha = 0,\ldots,M\ ,
\ee
where the first entry is for the massive particle with fugacity $y= e^{-m_0}$.  The mirror theory $Y$  is an  $U(1)^{M}$-theory with $M+1$ matter fields. A choice of charges satisfying \eqref{qorth} is 
\def\mo{\!\!$-$1}
\def\sp{\phantom{-}}
\be\label{qhat}
(\hat q^r_\alpha) = \begin{pmatrix}
\hat q^1_0&\sp0&\sp0&\sp0&\sp\cdots&\sp0\\
\hat q^1_1&\sp1&\sp0&\sp0&\sp\cdots&\sp0\\
\hat q^1_2&\sp0&\sp1&\sp0&\sp\cdots&\sp0\\
\vdots&\sp\vdots&\sp\vdots&\sp\vdots&\sp\ddots&\sp\vdots\\
\hat q^1_{M-1}&\sp0&\sp0&\sp0&\sp\cdots&\sp1\\
\hat q^1_M&-1&-1&-1&\sp\cdots& -1
\end{pmatrix}\ ,\qquad 
\begin{array}{l} 
\hat q^0_\alpha= \begin{cases} 1 & 0\le \alpha\le\ell \ , \\ 0 & \ell < \alpha \le M \ , \end{cases}\\[6ex]
r=0,\ldots,M-1\ , \\ \alpha = 0,\ldots,M \ . \end{array}
\ee
The constraints \eqref{MiMaTong} read
\be
\zeta^1 = \sum_{\alpha>0}\hat m_\alpha-\ell \hat m_0\ ,\qquad
\hat\zeta^r = \begin{cases} m_0 & r=0 \ , \\ 0 & \text{else} \ . \end{cases}
\ee
With $Q=e^{-\zeta^1}$, $y_\alpha = e^{-\hat m_\alpha}$, the first equation becomes
\be\label{Qrel}
\prod_{\alpha>0}y_\alpha = Q y_0^\ell\ .
\ee
The difference operator for the theory $X$  is (see eqs. \eqref{DiffEqgen},\eqref{DLdef})
\be\label{deexp}
\DL = \prod_{\alpha>0}(1-q^{\vartheta_\alpha})-Q\prod_{j=1}^\ell(1-yq^{-\vartheta_0+j}) \ .
\ee
In view of the general mirror map  $Q_a=\prod_\alpha y_\alpha^{q^a_\alpha}$ and the definition $\vartheta_\alpha = \sum_a q_\alpha^a \theta_a$ (see eqs.~\eqref{MiMaTong},\eqref{defvarth}), the shift operators act on the mirror side $Y$ by shifts of the mass parameters 
\be\label{varthY}
q^{\vartheta_\alpha} y_\beta =  q^{\delta_{\alpha\beta}}y_\beta\ .
\ee

To write down the disk partition function for $Y$ one needs to know the map between boundary conditions under mirror symmetry. This question has been recently studied in ref.~\cite{DGP} for a class of examples on a case by case basis, with the answers depending on the details. For the theories consider here we will make some choices motivated below  and then check their consistency. A hint comes from the relevant composite operators for theory $Y$, which are of the form \cite{AHKT}
\be\label{Xfields}
X(n_\alpha)=\prod_{\alpha\in I}\hat\varphi_\alpha^{n_\alpha} \prod_{\alpha\notin I}(\hat\varphi_\alpha^\dagger)^{-n_\alpha}\ ,\qquad  I=\{\alpha:n_\alpha\geq 0\}\ .
\ee
Gauge invariance requires $\sum_\alpha n_\alpha\hat q^r_\alpha=0$, which is solved by $n_\alpha=\sum_a p_a q^a_\alpha$. These operators are dual to vortices with windings $p_a$ in $U(1)^k$. In the above example, positive winding $p$ in $U(1)$ gives positive $n_{\alpha>0}$ and negative $n_0$, i.e., the BPS operators involve the modes of the chiral fields $\hat \varphi_{\alpha>0}$ and the anti-chiral field  $\hat \varphi_0^\dagger$. The vortices have bosonic zero modes for $\alpha\in I$, but not for $\alpha\notin I$ \cite{WitPhases}, and fermionic zero modes for all fields of non-zero charge \cite{IS,Ken}. These match the modes of bulk fields restricted to the boundary, if one takes  Neumann (Dirichlet)  boundary conditions for $\alpha\in I$  ($\alpha\notin I$).

Starting from eq.~\eqref{Zmatter}, the partition function with these boundary conditions is
\be \label{Zmirror}
\begin{aligned}
Z_Y&\sim  \int \prod_{r=0}^{M-1}d\ln z_r\, 
e^{\frac{-m_0 \ln z_0}{\ln q}}z_0 \frac{(z_0y_0q,q)_\infty }{\prod_{\alpha=1}^{M-1} (z_0^{\hat q^0_\alpha}z_\alpha y_\alpha,q)_\infty\left(\frac{z_0^{\hat q^0_M}y_M}{\prod_{r=1}^{M-1}z_r},q\right)_\infty}\\ 
&\sim \int \prod_{r=0}^{M-1}d\ln x_r \ e^{-\frac{m_0 \ln x_0}{\ln q} }\, x_0\, \frac{(x_0q,q)_\infty}{\prod_{\alpha=1}^M (x_\alpha,q)_\infty}\ ,
\end{aligned}
\ee
where we neglect overall constants. The second expression is obtained by a change of integration variables, with the $x_\alpha$ satisfying the same equation \eqref{Qrel} as the $y_\alpha$. The non-trivial choices made in the above ansatz concern the $R$ charges and gauge charge for the $\alpha=0$ direction. The extra factor $z_0$ in the integrand is generated by a shift $m_0\to m_0-\ln q$, which accounts for the non-zero $R$-charge of the field $\varphi_0$. The other modification is the weight $z_0$ in the determinant in the numerator, which is the weight of the anti-chiral  with gauge charge $-1$ appearing in \eqref{Xfields}. With this choice $Z_Y$ is annihilated by the difference operator \eqref{deexp} of the theory $X$, using eqs.~\eqref{varthY},\eqref{phdeq}, as should be the case for dual boundary conditions. Using the sum formula \eqref{qpoch} for the $q$-Pochhammer symbols, $Z_Y$ can be rewritten as a LG type of integral 
\be\label{LGper}
Z_Y(\alpha)\sim \int_{\Gamma_\alpha} \prod\frac{dx_i}{x_i} e^{W},\qquad W=\frac{\ln(y)\ln(x_0)}{\ln(q)}+W(x_i,q)\ ,
\ee
with 
\be\label{3dW}
W(x_i,q)=\sum_{k>0}\frac{w(x^k,q^k)}{k(1-q^k)},\qquad 
w(x_i,q)=\sum_{\alpha>0} x_\alpha-qx_0\ .
\ee
The linear function $w(x_i)=w(x_i,q=1)$ is the superpotential of the 2d mirror derived in ref.~\cite{HV}. The expression \eqref{LGper} has the form of the Landau--Ginzburg period of the 2d theory, with $w(x_i)$ replaced by $W(x_i,q)$. Integrals of the type \eqref{LGper} have been studied by Givental for the massless case in ref.~\cite{Giv15all}, where they were introduced from scratch as solutions to a given system of difference equations for symmetric quantum K-theory and used to define a concept of K-theoretic mirrors. This identifies {\it Givental mirrors} as special cases of known 3d gauge theoretic mirrors. 

In the 2d theory with $c_1=0$, the Landau--Ginzburg period can be rewritten as an integral over a Lagrangian cycle of the mirror Calabi--Yau manifold~$Y$ of $X$. In the 3d gauge theory, the integral \eqref{LGper} arises as the integral over zero modes of the gauge fields, i.e., Wilson line moduli of the 3d theory. It would be interesting to understand in more detail, how the Calabi--Yau geometry emerges from the gauge theory moduli space. 

\subsection{Direct integration of 3d Landau--Ginzburg integrals}
The partition functions of two dual gauge theories $X$ and $Y$ should be equal for a mirror pair of boundary conditions. In the following we identify integration contours $\Gamma_\alpha$ for the Landau-Ginzburg integrals $Z_Y(\alpha)$ which reproduce the partition functions $Z_X(E_a^{LG})$ in eq.~\eqref{lgb} upon direct integration.

Convergent integration contours $\Gamma_\alpha$ for the Landau--Ginzburg integrals can be constructed as gradient flows of the real part $\operatorname{Re} W$ of the superpotential, starting from the critical points of $W$, see refs.~\cite{HIV,BDP,CV13,Giv15all}. A detailed analysis of gradient flows for the superpotential of basic 3d gauge theories has been made for several examples in ref.~\cite{BDP}. The result is that the flows depend on the values of the parameters $(Q,q,y)$, but at the end, the partition functions of a mirror pair match in all regimes of parameters for dual boundary conditions , possibly up to monodromy.

We consider a class of integration cycles for small $Q^{-1}$ and $q$ which are 3d lifts of the integration cycles used in the direct integration of 2d LG integrals \cite{Berglund:1993ax}. To this end, we write $W = W_0(x_{\alpha>0}) + \delta W(x_0)$, and treat the second term as a perturbation, using the constraint \eqref{Qrel}:
\be
\delta W(x_0) = \frac{\ln(y)\ln(x_0)}{\ln(q)}-\sum_{k>0}\frac{q^kx_0^k}{k(1-q^k)},\qquad x_0 = \psi \prod_{\alpha>0}x_\alpha^{1/\ell},\quad \psi = Q^{-1/\ell}.
\ee
Expanding the exponential for small $\psi$ gives
\be
x_0 e^{\delta W(x_0)} = \sum_{k=1}^\infty \frac{\psi^\kh}{(\bar q)_{k-1}}\prod_{\alpha>0}x_\alpha^{\kh/\ell}\ ,\qquad \kh = k+\delta\ ,\quad \delta = \frac{\ln y}{\ln q} \ .
\ee
Inserting this expansion in $Z_Y$, the integral factorizes as
\be
Z_Y = \sum_{k>1} \frac{\psi^{\kh}}{(\bar q)_{k-1}}\ \prod_{\alpha>0} \int dx_\alpha \frac{x_\alpha^{\kh/\ell-1}}{(x_\alpha,q)_\infty}\ .
\ee
The basic integrals evaluate to 
\be
\int_C \frac{dy}{y} \frac{y^{-\zeta}}{(y,q)_\infty} = \frac{(1-q)^{-\zeta}}{\Gamma_q(1+\zeta)} \ ,
\ee
where $C$ is a contour that sums up the poles of the denominator. This integral is a 3d lift of the Hankel type integrals \cite{WW} for the ordinary Gamma function and reduces to it in the 2d limit $\beta\to 0$ after the variable change $y=\hbar\beta \hat y$, $q=e^{-\hbar\beta}$. Collecting all factors one obtains $Z_Y = -(1-q)^{1+\delta}\omega^{LG}_0(\psi)$ with 
\be
\omega^{LG}_0(\psi)=  \sum_{k>1} \left(\frac{\psi}{(1-q)^{(\ell-N)/\ell}}\right)^\kh\frac{(-)^kq^{k(k-1)/2}}{\Gamma_q(k)\Gamma_q(1-\frac \kh \ell)^N}\ .
\ee
The series $\omega^{LG}_0(\psi)$ converges for small $|\psi|$ and $|q|$ and is annihilated by the difference operator \eqref{deexp}, as it should (cpw. app.~\ref{app:sol}). For $\ell=N$ the result agrees with the partition function $Z_X(E_0^{LG})$ in \eqref{lgb} describing the E-brane with full Dirichlet conditions for the theory $X$. To obtain the mirror of the other branes $E^{LG}_{a>0}$ for $X$ one notes that the solution of the constraint \eqref{Qrel} involved the choice of a root for the factors $x_\alpha^{1/\ell}$. Different roots can be absorbed into redefinitions $\psi \to \eta^i \psi$ with $\eta^\ell=1$. These choices gives further solutions
\be
\omega^{LG}_i(\psi) = \omega^{LG}_0(\eta^i \psi)\ ,
\ee
that match to the other boundary conditions in  eq.~\eqref{lgb}.

\section{Computation of quantum K-theory invariants \label{sec:EQK}}
In this section we explicitly compute permutation equivariant quantum K-theory invariants by using Givental's reconstruction theorems applied to three-dimen\-sio\-nal partition functions. 

The genus zero quantum K-theory invariants of a K\"ahler manifold $X$ are holomorphic Euler characteristics over the moduli space of stable maps $\overline{\mathcal{M}}_{0,m}(X,\beta)$ with $m$ marked points into the class $\beta \in H_2(X,\mathbb{Z})$ of the form
\begin{equation}
   \left\langle t_1(q) ,  \ldots, t_m(q) \right\rangle_{0,m,\beta} \,=\, 
   \chi_{\overline{\mathcal{M}}_{0,n}(X,\beta)}(\operatorname{ev}_1^*t_1(L_1)\otimes\ldots\otimes\operatorname{ev}_m^*t_n(L_m)\otimes\mathcal{O}^\text{vir})  \ .
\end{equation}   
Here the inputs $t_i(q)$ take values in 
\begin{equation}
   t_i(q) \in K(X)[q,q^{-1}] \ ,
\end{equation}
where the Laurent polynomials $t_i(q)$ in $q$ with coefficients in $K(X)$ get evaluated with the universal cotangent line bundles over $(C,x_1,\ldots,x_m,f)\in\overline{\mathcal{M}}_{0,m}(X,\beta) $ at the marked point~$x_i$, respectively, and $ev_i: \overline{\mathcal{M}}_{0,m}(X,\beta) \to X$ is the evaluation map at the marked point~$x_i$. Finally, $\mathcal{O}^\text{vir}$ is the virtual structure sheaf of the moduli space $\overline{\mathcal{M}}_{0,m}(X,\beta)$ constructed in ref.~\cite{Lee:2001mb}. 

Givental's permutation equivariant quantum K-theory refines the ordinary quantum K-theory invariants with respect to the symmetric group $S_n$ (for $n\le m$) acting as automorphisms on the moduli space of stable maps $\overline{M}_{0,m}(X,\beta)$ by permuting the last $n$ marked points. Then the holomorphic Euler characteristics with $n$ identical inputs $t(q)\equiv t_{m-n+1}(q) = \ldots = t_m(q)$ are equivariantly refined to
\begin{multline}
   \left\langle t_1(q) ,  \ldots, t_{m-n}(q);\, t(q), \ldots, t(q) \right\rangle_{0,m,\beta}^{S_n} \,=\, \\
   \sum_{\nu \in \operatorname{Irrep}(S_n)} \chi^{S_n,\nu}_\beta(t_1(q) ,  \ldots, t_{m-n}(q);\, t(q)) \cdot \nu \ .
\end{multline}
The sum runs over all irreducible representations $\nu$ of the symmetric group $S_n$, and $\chi^{S_n,\nu}_\beta$ are the equivariant Euler characteristics of the irreducible representation $\nu$ of the symmetric group $S_n$.

In particular, the equivariant quantum K-theory invariants associated to the one-dimensional symmetric representations $\text{sym}=\tableau{2}\!\!\cdot\!\!\cdot\!\!\cdot\!\!\tableau{1}$ read
\begin{equation}
     \left\langle t_1(q) ,  \ldots, t_{m-n}(q);\, t(q), \ldots, t(q) \right\rangle_{0,m,\beta}^{S_n,\text{sym}}\,=\, \chi^{S_n,\text{sym}}_\beta(t_1(q) ,  \ldots, t_{m-n}(q);\, t(q)) \ .
\end{equation}    
They are referred to as the symmetric quantum K-theory invariants. The unrefined ordinary quantum K-theory invariants are recovered from equivariant invariants as
\begin{multline} \label{eq:RelOrdKTheory}
    \left\langle t_1(q) ,  \ldots,  t_{m-n}(q), t(q), \ldots, t(q) \right\rangle_{0,m,\beta} \,=\, \\
    \sum_{\nu \in \operatorname{Irrep}(S_n)} \chi^{S_n,\nu}_\beta(t_1(q) ,  \ldots, t_{m-n}(q);\, t(q)) \cdot \dim \nu \ ,
\end{multline}
in terms of the dimensions of the irreducible representations $\nu$. 

Analogously to the cohomological Gromov--Witten invariants, the quantum K-theoretic invariants are conveniently encoded in the K-theoretic Givental $J$-functions. They enjoy for the ordinary quantum K-theory, the equivariant quantum K-theory and the symmetric quantum K-theory the expansions \cite{Giv15all}
\begin{equation}\label{Jexp}
\begin{aligned}
   J_K(t) \,&=\, (1 - q) + t(q) + \sum_{\beta\ge 0}\sum_{n\ge 0} \sum_i \frac{\Phi^i}{n!} \left\langle \frac{\Phi_i}{1-qL},t(q),\ldots,t(q) \right\rangle_{0,n+1,\beta} Q^\beta   \ , \\
   J_K^{\text{eq}}(t) \,&=\, (1 - q) + t(q)\cdot s_{\tableau{1}} + \sum_{\beta\ge 0}\sum_{n\ge 0} \sum_{i,\nu}  s_\nu\cdot\Phi^i\cdot\chi^{S_n,\nu}_\beta\left( \frac{\Phi_i}{1-qL};\,t(q) \right)  Q^\beta\ , \\
   J_K^{\text{sym}}(t) \,&=\, (1 - q) + t(q) + \sum_{\beta\ge 0}\sum_{n\ge 0} \sum_i \Phi^i\cdot\chi^{S_n,\text{sym}}_\beta\left( \frac{\Phi_i}{1-qL};\,t(q)\right)  Q^\beta  \ .
\end{aligned}
\end{equation}
Here the first term is called the dilaton shift and the second term is referred to as the input of the $J$-function. $\Phi^i$ and $\Phi_i$ denote a basis and a dual basis of $K(X)$, and $s_\nu$ are the Schur polynomials of the Young tableaus of the irreducible representations~$\nu$ in the Novikov ring $\Lambda=\mathbb{Q}[[N_1,N_2,\ldots]]$ of Newton polynomials $N_r=x_1^r+x_2^r+\ldots$. In particular, we have $s_{\tableau{1}} = N_1$. These K-theoretic $J$-functions and their inputs respectively take values in the formal rings
\begin{equation}
\begin{aligned}
  J_K(t) \, &\in \,  \mathcal{K}  &\text{with}&& t\,&\in\, \mathcal{K_+}  \ , \\
  J_K^{\text{eq}}(t) \, &\in \,  \mathcal{K}\otimes\Lambda  &\text{with}&& t\cdot s_{\tableau{1}}\,&\in\, \mathcal{K_+}\otimes\Lambda  \ , \\
  J_K^{\text{sym}}(t) \, &\in \,  \mathcal{K} &\text{with}&& t\,&\in\, \mathcal{K_+}  \ , \\
\end{aligned}
\end{equation}
with \cite{GivER}
\begin{equation}\label{Kdef}
\begin{aligned}
 \mathcal{K} \,&=\,  K(X) \otimes \mathbb{C}(q,q^{-1})\otimes\mathbb{C}[[Q]] \ , \\
  \mathcal{K}_+\,&=\,K(X) \otimes \mathbb{C}[q,q^{-1}]\otimes\mathbb{C}[[Q]] \ , \\
  \mathcal{K}_- \,&=\,K(X) \otimes 
     \left\{\,r(q) \in R(q) \,\middle|\, \text{$r(0)\ne\infty$ and $r(\infty)=0$} \right\} \otimes\mathbb{C}[[Q]] \ ,
\end{aligned}   
\end{equation}
such that $\mathcal{K} = \mathcal{K}_+ \oplus \mathcal{K}_-$ and where $R(q)$ denotes the field of rational functions in the variable $q$.\footnote{$\mathcal{K}_\pm$ are Lagrangian subspaces of $\mathcal{K}$ with respect to the symplectic pairing $\Omega(f,g) = (\operatorname{Res}_{q=0} + \operatorname{Res}_{q=\infty}) \frac{dq}{q} \left(f(q),g(q^{-1})\right)_K$ with the product $\left( \mathcal{E}, \mathcal{F} \right)_K = \chi(X,\mathcal{E}\otimes \mathcal{F})$ on $K(X)$ \cite{GivER}.} Note that the K-theoretic invariants of the ordinary/symmetric and permutation equivariant $J$-functions lie in the subspace $\mathcal{K}_- \subset \mathcal{K}$ and $\mathcal{K}_- \otimes \Lambda\subset \mathcal{K}\otimes\Lambda$, respectively. All three K-theoretic $J$-functions are canonically identified for vanishing input, namely
\begin{equation}
  J_K(0) = J_K^{\text{eq}}(0) = J_K^{\text{sym}}(0) = (1-q) + \sum_{\beta\ge 0}\sum_i \Phi^i \left\langle \frac{\Phi_i}{1-qL} \right\rangle_{0,1,\beta} Q^\beta \ .
\end{equation}
For a detailed discussion on equivariant quantum K-theory, we refer the reader to original refs.~\cite{Giv15all}.

\subsection{The point}
As discussed in sect.~\ref{sec:topvert}, the 3d vortex sum for the target $X=\text{pt}$ coincides with the topological vertex for a stack of branes on a single leg of $\IC^3$. To set the stage for the forthcoming computations, we briefly review the K-theoretic Givental $J$-functions for this case (see p.I of \cite{Giv15all}):
\begin{equation}
   J_K^\text{eq}(t)= (1-q) e^{\sum_{k=1}^{+\infty} \frac{t\cdot N_k}{k(1-q^k)}} 
\end{equation}
Here $N_r = x_1^r + x_2^r + \ldots$ are the Newton polynomials. Expressed in terms of the topological vertex according to eq.~\eqref{topvert}, the $J$-function takes the form
\be
J_K^\text{eq}(t)= (1-q) \left[ \sum_\nu   (-q^{-1/2})^{|\nu|} \sum_\nu C_{00\nu}(q^{-1}) s_\nu(x) \right]^t \ .
\ee
For $t=1$, it becomes (up to normalizations) the generating function of the topological vertex~$C_{00\nu}(q^{-1})$.\footnote{After the replacement $C_{00\nu} \to (-1)^{|\nu|} C_{00\nu}$, the obtained expressions agree with the topological vertex in the canonical framing as normalized in ref.~\cite{Aganagic:2003db}.} Expanding in the Schur polynomials $s_\nu$ we arrive for the first few leading orders in marked points at
\begin{equation} \label{eq:CorEqPt}
\begin{aligned}
   \left\langle \frac{1}{1-qL};\, 1, 1 \right\rangle_{0,3}^{S_2} &= 
     \frac{1}{(1-q^2)}s_{\tableau{2}} +
      \frac{q}{(1-q^2)}s_{\tableau{1 1}} \ , \\
   \left\langle \frac{1}{1-qL};\, 1, 1, 1 \right\rangle_{0,4}^{S_3} &=
      \frac{1}{(1-q^2)(1-q^3)}s_{\tableau{3}} +
      \frac{q}{(1-q)(1-q^3)}s_{\tableau{2 1}} \\
      &\qquad+\frac{q^3}{(1-q^2)(1-q^3)}s_{\tableau{1 1 1}} \ , \\
    \left\langle \frac{1}{1-qL};\, 1, 1, 1 , 1\right\rangle_{0,5}^{S_4} &=  
      \frac{1}{(1 - q^2) (1 - q^3) (1 - q^4) }s_{\tableau{4}} \\
      &\qquad +\frac{q}{(1 - q) (1 - q^2) (1 - q^4)}s_{\tableau{3 1}} \\
       &\qquad+ \frac{q^2}{(1 - q^2)^2 (1 - q^3)}s_{\tableau{2 2}}\\
       &\qquad +\frac{q^3}{(1 - q) (1 - q^2) (1 - q^4)}s_{\tableau{2 1 1}} \\
       &\qquad +\frac{q^6}{(1 - q^2) (1 - q^3) (1 - q^4)}s_{\tableau{1 1 1 1}} \ .
\end{aligned}
\end{equation}
Here the Schur functions $s_{\nu}$ are labeled by the Young tableau of the irreducible representation $\nu$ of the symmetric group,\footnote{The monomials of the Schur functions $s_\nu$ in $\Lambda$ are given by the associated semi-standard Young tableaus with entries in the positive integers.}
which they obey the ring relations
\begin{equation}
   s_\nu \cdot s_\mu \,=\, \sum_{\rho \in \operatorname{Irreps}(\nu\otimes\mu)} s_{\rho} \ .
\end{equation}
In terms of the Newton polynomials they are for instance given by
\begin{equation}
\begin{aligned}
  &s_{\tableau{1}} = N_1 \ , \quad 
  s_{\tableau{1 1}} =  \frac12 (N_1^2 - N_2) \ , \quad
  s_{\tableau{2}} = \frac12 (N_1^2 + N_2) \ , \\
  &s_{\tableau{1 1 1}} = \frac16 (N_1^3 - 3 N_1 N_2 +2 N_3) \ , \quad
  s_{\tableau{2 1}} = \frac13 (N_1^3 - N_3) \ , \\
  &s_{\tableau{3}} =  \frac16 (N_1^3 + 3 N_1 N_2 +2 N_3)\ .
\end{aligned}
\end{equation}

By projecting on the symmetric representations of the permutation equivariant invariants~\eqref{eq:CorEqPt}, we readily obtain with eq.~\eqref{qpoch} the permutation symmetric quantum K-invariants
\be
\begin{aligned}
\left\langle \frac{1}{1-qL};\, 1, \ldots, 1 \right\rangle_{0,n+1}^{S_n,\text{sym}}
&= (1-q) \operatorname{Coeff}(e^{\sum_{k>0}\frac{x^k}{k(1-q^k)}},x^n) \\
&= \frac{1}{\prod_{i=2}^n(1-q^i)} \quad\text{for}\quad n\ge 2 \ ,
\end{aligned}
\ee
which are in agreement with the holomorphic Euler characteristics directly obtained from the permutation symmetric $J$-function $J^\text{sym}_K$. Furthermore, by employing the relation~\eqref{eq:RelOrdKTheory} we recover from the invariants~\eqref{eq:CorEqPt} together with the dimensions of the representations of $S_n$
\begin{equation} \label{eq:SnReps}
\begin{aligned}
  &\dim \tableau{2} \,=\, \dim \tableau{1 1} \,=\, 1 \ , \quad
  \dim \tableau{3} \,=\, \dim \tableau{1 1 1} \,=\, 1 \ , \quad
  \dim \tableau{2 1} \,=\, 2 \ , \\
  &\dim \tableau{4} \,=\, \dim \tableau{1 1 1 1} \,=\, 1 \ , \quad
  \dim \tableau{2 2} \,=\, 2 \ , \quad
  \dim \tableau{3 1} \,=\, \dim \tableau{2 1 1} \,=\, 3 \ ,
\end{aligned}  
\end{equation}
the ordinary quantum K-invariants
\begin{equation}
    \left\langle \frac{1}{1-qL};\, 1, \ldots, 1 \right\rangle_{0,n+1}\,=\,\frac{1}{(1-q)^{n-1}} \quad\text{for}\quad n\ge 2 \ .
\end{equation}
This is in agreement with the ordinary quantum K-theoretic $J$-function $J_K$, as directly given by K-theoretic string equation \cite{Lee:2001mb}.

\subsection{The projective surface} \label{sec:P2}
The projective surface $\mathbb{P}^2$ is our next example. Its classical K-theory ring $K(\mathbb{P}^2)$ is generated by $\Phi_k = (1-P)^k$, $k=0,1,2$, with the tautological line bundle $P\equiv\mathcal{O}(-1)$ of $\mathbb{P}^2$, and its intersection pairing for these generators of $K(\mathbb{P}^2)$ reads
\begin{equation}
  \left( \Phi_k , \Phi_\ell \right) \,=\, \int_{\mathbb{P}^2} \operatorname{td}(\mathbb{P}^2) \operatorname{ch}(\Phi_k \otimes \Phi_l) \,=\, 
  \begin{pmatrix}
   \ 1 & \ 1 & \ 1 \\ \ 1 & \ 1 & \ 0 \\ \ 1 &\ 0 &\ 0
  \end{pmatrix} \ .
\end{equation}  
The $J$-function $J_K$ with vanishing input \cite{Giv15all}(p. II) 
\begin{equation}\label{eq:JKP2zero}
  J_K(0) \,=\, J_K^\text{eq}(0) \,=\, J_K^{\text{sym}}(0) \,=\, (1-q) \sum_{d=0}^{+\infty} \frac{1}{\prod_{i=1}^{d}{(1-q^iP)^3}} Q^d \ ,
\end{equation}
coincides with the vortex sum \eqref{vortexsum} obtained from the partition function, up to the normalization factor $(1-q)$.

Let us now focus on the permutation equivariant quantum K-theoretic $J$-function with non-vanishing input. Using Givental's reconstruction theorem \cite{Giv15all}(p. VIII)  for the permutation equivariant K-theoretic $J$-function, we can generate a non-trivial input as follows 
\begin{equation} \label{eq:ReconCP2}
   J_K^\text{eq}(t(\epsilon)) \,=\, e^{\sum_{r=1}^{+\infty}\frac{\sum_{\ell} \Psi_r(\epsilon_\ell) P^{\ell r} q^{\ell r Q\partial_Q}}{r(1-q^r)}} J_K(0) \ .
\end{equation}
The operator acting on $ J_K(0)$ is of the form \eqref{IIopmt} obtained in sect.~\ref{subsec:pertexp} by integrating in new massive modes in the partition function. 
The mass parameters are described by  $\epsilon=\sum_\ell \epsilon_\ell P^\ell$, which is a formal series in the Newton polynomials $N_r$ and the variable $Q$ of the Novikov ring $\Lambda\otimes\mathbb{C}[[Q]]$ with coefficients in the polynomial ring $K(\mathbb{P}^2) \otimes \mathbb{C}[q,q^{-1}]$. Furthermore, $\Psi_r$ denotes the Adams operator, which acts on variables $Q$, $q$, and the ring $\Lambda$ of Newton polynomials $N_r$ as
\begin{equation}
  \Psi_r(Q) \,=\, Q^r \ ,\quad
  \Psi_r(q) \,=\, q^r \ , \quad
  \Psi_r(N_k) \,=\, N_{r k} \ . 
\end{equation}
In order to generate with formula~\eqref{eq:ReconCP2} the permutation equivariant $J$-function $J_K^\text{eq}(t(\epsilon))$ with input
\begin{equation} \label{eq:InputCP2}
   t(\epsilon) \,=\, a \Phi_1 + b \Phi_2 \ ,
\end{equation}
we arrive to leading order in $Q$ and to leading order in the degree of the Schur polynomials $s_\nu$ at
\begin{multline} \label{eq:epsP2}
  \epsilon = (a \Phi_1 + b \Phi_2)\cdot s_{\tableau{1}} 
  + \frac{Q}2 \Big( b(1-b)(\Phi_0 - \Phi_1) \cdot s_{\tableau{2} } \\
      - b(1+b)(\Phi_0 - \Phi_1)\cdot s_{\tableau{1 1}} + \ldots \Big) + \ldots \ .
\end{multline}
Note that the coefficients of the elements of the Novikov ring $\Lambda\otimes\mathbb{C}[[Q]]$ are in the polynomial ring $K(\mathbb{P}^2) \otimes \mathbb{C}[q,q^{-1}]$. 

From the $J$-function $J^\text{eq}_K(t)$ with input~\eqref{eq:InputCP2} we for instance determine the permutation equivariant invariants at degree $Q$ for two marked points
\begin{equation}
\begin{aligned}
   \sum_{k=0}^2 \Phi_k \left\langle \tfrac{\Phi^k}{1-qL};\, \Phi_1\right\rangle_{0,2,1} &= 
      \left( \tfrac{1-4q+6q^2}{(1-q)^4} \Phi_0  +  \tfrac{1-3q}{(1-q)^3} \Phi_1 + \tfrac{1}{(1-q)^2} \Phi_2  \right) s_{\tableau{1}} \ , \\
    \sum_{k=0}^2 \Phi_k \left\langle \tfrac{\Phi^k}{1-qL};\, \Phi_2 \right\rangle_{0,2,1} &= 
     \left( \tfrac{1-3q+3q^2}{(1-q)^3} \Phi_0  +  \tfrac{1-2q}{(1-q)^2} \Phi_1 + \tfrac{1}{1-q} \Phi_2  \right) s_{\tableau{1}} \ ,   
\end{aligned}
\end{equation}
for three marked points
\begin{equation}
\begin{aligned}
   \sum_{k=0}^2 \Phi_k \left\langle \tfrac{\Phi^k}{1-qL};\, \Phi_1,\Phi_1 \right\rangle_{0,3,1}^{S_2} =& 
     \left( \tfrac{1-3q+3q^2+3q^3}{(1-q)^4(1+q)} \Phi_0+\tfrac{1-2q-2q^2}{(1-q)^3(1+q)} \Phi_1+\tfrac{1}{(1-q)^2} \Phi_2  \right) s_{\tableau{2}} \\
     &+\left( \tfrac{q(1-3q)}{(1-q)^4(1+q)} \Phi_0  +  \tfrac{q}{(1-q)^3(1+q)} \Phi_1 \right) s_{\tableau{1 1}} \ , \\
   \sum_{k=0}^2 \Phi_k \left\langle \tfrac{\Phi^k}{1-qL};\, \Phi_2,\Phi_2 \right\rangle_{0,3,1}^{S_2} =& 
     \left( \tfrac{1-2q+q^2+2q^3}{(1-q)^3(1+q)} \Phi_0+\tfrac{1-2q}{(1-q)^2}\Phi_1+\tfrac{1}{1-q} \Phi_2  \right) s_{\tableau{2}} \\
     &+\left( \tfrac{q(1-2q-q^2)}{(1-q)^3(1+q)} \Phi_0  +  \tfrac{q}{(1-q)^2} \Phi_1 +\tfrac{(-1)}{1-q}\Phi_2 \right) s_{\tableau{1 1}} \ ,   
\end{aligned}
\end{equation}
and for four marked points 
\begin{equation} \label{eq:eqQ14}
\begin{aligned}
   \sum_{k=0}^2 \Phi_k &\left\langle \tfrac{\Phi^k}{1-qL};\, \Phi_1,\Phi_1,\Phi_1 \right\rangle_{0,4,1}^{S_3} \\
     &\qquad=\left( \tfrac{1-2q+q^2+4q^3+4q^4+q^5}{(1-q)^4(1+q)(1+q+q^2)} \Phi_0
       +\tfrac{1-q-3q^2-3q^3-q^4}{(1-q)^3(1+q)(1+q+q^2)} \Phi_1+\tfrac{1}{(1-q)^2} \Phi_2  \right) s_{\tableau{3}} \\
     &\qquad\qquad\qquad\quad +\left( \tfrac{q(1-2q-2q^2)}{(1-q)^4(1+q+q^2)} \Phi_0 +\tfrac{q(1+q)}{(1-q)^3(1+q+q^2)}\Phi_1  \right) s_{\tableau{2 1}} \ , \\
     &\qquad\qquad\qquad\quad +\left( \tfrac{3q^3}{(1-q)^4(1+q)(1+q+q^2)} \Phi_0+\tfrac{q^2}{(1-q)^3(1+q)(1+q+q^2)} \Phi_1\right)  s_{\tableau{1 1 1}} \ , \\
    \sum_{k=0}^2 \Phi_k &\left\langle \tfrac{\Phi^k}{1-qL};\, \Phi_2,\Phi_2,\Phi_2 \right\rangle_{0,4,1}^{S_3} = 
   \left( \tfrac{1-q+2q^3+4q^4}{(1-q)^3(1+q)(1+q+q^2)} \Phi_0+\tfrac{1-2q}{(1-q)^2} \Phi_1+\tfrac{1}{1-q}\Phi_2\right) s_{\tableau{3}} \\
     &\qquad\qquad\qquad\qquad +\left( \tfrac{(-3)q}{(1-q)^3(1+q+q^2)} \Phi_0+\tfrac{q}{(1-q)^2} \Phi_1+\tfrac{(-1)}{1-q} \Phi_2  \right) s_{\tableau{2 1}} \ , \\
     &\qquad\qquad\qquad\qquad +\left( \tfrac{-1+q+4q^3+2q^4}{(1-q)^3(1+q)(1+q+q^2)} \Phi_0+\tfrac{(-1)}{(1-q)^2} \Phi_1+\tfrac{1}{1-q} \Phi_2  \right)  s_{\tableau{1 1 1}} \ .
\end{aligned}
\end{equation}
Furthermore, at degree $Q^2$ we find for two marked points 
\begin{equation}
\begin{aligned}
   \sum_{k=0}^2 \Phi_k \left\langle \tfrac{\Phi^k}{1-qL};\, \Phi_1\right\rangle_{0,2,2} &= 
      \left( \tfrac{1-3q-q^2+21q^3+21q^4}{(1-q)^7(1+q)^4} \Phi_0  +  
      \tfrac{1-3q-6q^2}{(1-q)^6(1+q)^3} \Phi_1 + \tfrac{1}{(1-q)^5(1+q)^2} \Phi_2  \right) s_{\tableau{1}} \ , \\
    \sum_{k=0}^2 \Phi_k \left\langle \tfrac{\Phi^k}{1-qL};\, \Phi_2 \right\rangle_{0,2,2} &= 
     \left( \tfrac{1-3q+18q^3+15q^4}{(1-q)^6(1+q)^3} \Phi_0  + 
             \tfrac{1-3q-5q^2}{(1-q)^5(1+q)^2} \Phi_1 + \tfrac{1}{(1-q)^4(1+q)} \Phi_2  \right) s_{\tableau{1}} \ ,   
\end{aligned}
\end{equation}
and for three marked points 
\begin{equation} \label{eq:eqQ23}
\begin{aligned}
   \sum_{k=0}^2 &\Phi_k\left\langle \tfrac{\Phi^k}{1-qL};\, \Phi_1,\Phi_1 \right\rangle_{0,3,2}^{S_2}\\ 
     &=\left( \tfrac{1-2q-3q^2+18q^3+39q^4+33q^5+15q^6}{(1-q)^7(1+q)^5} \Phi_0
     +\tfrac{1-2q-8q^2-8q^3-5q^4}{(1-q)^6(1+q)^4} \Phi_1+\tfrac{1+q+q^2}{(1-q)^5(1+q)^3} \Phi_2  \right) s_{\tableau{2}} \\
     &+\left( \tfrac{q(1-2q-3q^2+12q^3+15q^4)}{(1-q)^7(1+q)^5} \Phi_0  +  \tfrac{q(1-2q-5q^2)}{(1-q)^6(1+q)^4} \Phi_1
     +\tfrac{q}{(1-q)^5(1+q)^3}\Phi_2 \right) s_{\tableau{1 1}} \ , \\
   \sum_{k=0}^2 &\Phi_k \left\langle \tfrac{\Phi^k}{1-qL};\, \Phi_2,\Phi_2 \right\rangle_{0,3,2}^{S_2} \\
     &=\left( \tfrac{1-2q-2q^2+16q^3+31q^4+22q^5+2q^6-6q^7}{(1-q)^6(1+q)^4} \Phi_0
     +\tfrac{1-2q-7q^2-7q^3-2q^4+3q^5}{(1-q)^5(1+q)^3}\Phi_1+\tfrac{1+q+q^2-q^3}{(1-q)^4(1+q)^2} \Phi_2  \right) s_{\tableau{2}} \\
     &+\left( \tfrac{q(1-2q-8q^2-14q^3-23q^4-16q^5)}{(1-q)^6(1+q)^4} \Phi_0 +
      \tfrac{q(1+q+5q^2+7q^3)}{(1-q)^5(1+q)^3} \Phi_1 +\tfrac{(-2)q^2}{(1-q)^4(1+q)^2}\Phi_2 \right) s_{\tableau{1 1}} \ ,
\end{aligned}
\end{equation}
For reference to more invariants, we have listed the first few terms of the permutation equivariant $J$-function $J_K^\text{eq}$ with input $t = a\Phi_1 + b\Phi_2$ in Appendix~\ref{app:P2}.

Using the relationship~\eqref{eq:RelOrdKTheory} together with the dimensions~\eqref{eq:SnReps} of the representations of the symmetric groups, we can easily recover the ordinary K-theoretic invariants encoded in the $J$-function $J_K$, for instance from eq.~\eqref{eq:eqQ14} at degree $Q$ with four marked points we obtain the ordinary K-theoretic invariants
\begin{equation}
\begin{aligned}
   \sum_{k=0}^2 \Phi_k \left\langle \tfrac{\Phi^k}{1-qL};\, \Phi_1,\Phi_1,\Phi_1 \right\rangle_{0,4,1} &=
   \frac{\Phi_0+\Phi_1+\Phi_2}{(1-q)^2}   \ , \\
    \sum_{k=0}^2 \Phi_k \left\langle \tfrac{\Phi^k}{1-qL};\, \Phi_2,\Phi_2,\Phi_2 \right\rangle_{0,4,1} &= 0 \ ,
\end{aligned}
\end{equation}
or from eq.~\eqref{eq:eqQ23} at degree $Q^2$ with three marked points we get 
\begin{equation}
\begin{aligned}
   \sum_{k=0}^2 \Phi_k\left\langle \tfrac{\Phi^k}{1-qL};\, \Phi_1,\Phi_1 \right\rangle_{0,3,2} &=
     \tfrac{1-3q+18q^3+15q^4}{(1-q)^7(1+q)^3} \Phi_0 
     +\tfrac{1-3q-5q^2}{(1-q)^6(1+q)^2} \Phi_1+\tfrac{1}{(1-q)^5(1+q)} \Phi_2 \ , \\
   \sum_{k=0}^2 \Phi_k \left\langle \tfrac{\Phi^k}{1-qL};\, \Phi_2,\Phi_2 \right\rangle_{0,3,2} &=
     \tfrac{1-4q+6q^2}{(1-q)^5} \Phi_0 +\tfrac{1-3q}{(1-q)^4} \Phi_1+\tfrac{1}{(1-q)^3} \Phi_2 \ . 
\end{aligned}
\end{equation}
Upon setting $q=0$ our results confirm the invariants listed in ref.~\cite{IMT}, where they have been computed by reconstruction techniques in ordinary quantum K-theory. For instance, they are readily determined with Givental's reconstruction theorem for ordinary quantum K-theory \cite{GivER}, p.VIII of \cite{Giv15all}, i.e.,
\begin{equation} \label{eq:ReconCP2ord}
   J_K(t(\epsilon)) \,=\, e^{\frac{\sum_{\ell=0}^2 \epsilon_\ell P^{\ell r} q^{\ell r Q\partial_Q}}{(1-q)}} J_K(0) \ .
\end{equation}
Here $\epsilon=\sum_\ell \epsilon_\ell P^\ell$ is now a formal series in $Q$ of the Novikov ring $\mathbb{C}[[Q]]$ with coefficients in the polynomial ring $K(\mathbb{P}^2) \otimes \mathbb{C}[q,q^{-1}]$.\footnote{Convergence in the reconstruction fomula~\eqref{eq:ReconCP2ord} is ensured if the function $\epsilon$ lies in a proper ideal of the ring $K(\mathbb{P}^2)\otimes\mathbb{C}[q,q^{-1}]\otimes \mathbb{C}[[Q]]$.}

\subsection{The quintic Calabi--Yau 3-fold \label{subsec:EQKquintic}}
As our next example we consider the quintic Calabi--Yau 3-fold $X$ given as the degree five hypersurface in the projective space $\mathbb{P}^4$. Its classical K-theory ring $K(X)$ is generated (over $\mathbb{Q}$) by $\Phi_k = (1-P)^k$, $k=0,1,2,3$, where the line bundle $P$ is the restriction of the tautological line bundle $\mathcal{O}(-1)$ of $\mathbb{P}^4$ to the hypersurface $X$.\footnote{Note that integral generators of the K-group $K(X)$ of the quintic Calabi--Yau 3-fold $X$ are given by $(\Phi_0,\Phi_1,\frac15 \Phi_2,\frac15 \Phi_3)$.} The intersection pairing for the generators $\Phi_k$ reads
\begin{equation}
  \left( \Phi_k , \Phi_\ell \right) \,=\, \int_X \operatorname{td}(X) \operatorname{ch}(\Phi_k \otimes \Phi_l) \,=\, 
  \begin{pmatrix}
   0 & 5 &-5 & 5 \\ 5 & -5 & 5 & 0 \\ -5 & 5 & 0 & 0 \\ 5 & 0 & 0 & 0
  \end{pmatrix} \ .
\end{equation}  

The spectrum of the three-dimensional Abelian $U(1)$ gauge theory associated to the quintic 3-fold reads 
\begin{equation} \label{eq:SpecQuintic}
\hbox{
\vbox{
\offinterlineskip
\halign{\vrule width1.2pt\strut~#~\hfil\vrule width0.8pt&~#~\hfil\vrule width0.8pt&~#~\hfil~\vrule width1.2pt\cr
\noalign{\hrule height 1.2pt}
$N=2$ chiral multiplets & $U(1)$ charge & $R$-charge \cr
\noalign{\hrule height 1.2pt}
~$\varphi_i, i=1,\ldots,5$ & ~$+1$ & ~$\phantom{+}0$ \cr
~$\varphi_0= p$ & ~$-5$ & ~$+2$ \cr 
\noalign{\hrule height 1.2pt}
}}}
\end{equation}

The disk partition function on $S^1 \times_q D^2$ of this gauge theory computes the $J$-function $J_K^{\text{sym}}$ of the symmetric quantum K-theory of the quintic $X$ given in \cite{Giv15all}
\begin{equation} \label{eq:JFuncQuintic1}
   J_K^{\text{sym}}(t^\text{sym}) \,=\, (1-q) \sum_{d=0}^{+\infty} \frac{\prod_{i=1}^{5d}(1-q^iP^5)}{\prod_{i=1}^{d}{(1-q^iP)^5}} Q^d \ .
\end{equation}
Note that the $J$-function has a complicated non-vanishing input $t^\text{sym}$, which is a formal power series in the Novikov variable $Q$ with coefficient in the K-theory ring $K(X) \otimes\mathbb{C}[q,q^{-1}]$. To leading order in $Q$, it takes the form 
\begin{multline} \label{eq:inpJKQ}
   t^\text{sym} \,=\, Q  \big[ 
     (1+q)^2(1+q^2)(1+q+q^2)(1-q^5) \Phi_0 \\
     +5q^2(1+q)(1+q+2q^2+q^3+2q^4)(1+2q+3q^2+3q^3+2q^4) \Phi_1 \\
     +5(115+117q+\ldots+112q^{10}+38q^{11}) \Phi_2 \\
     +5(-230+2q+ \ldots+228q^{11}) \Phi_3  \big] 
   + \ldots \ .
\end{multline} 

We can again change the input \eqref{eq:inpJKQ} of the $J$-function~\eqref{eq:JFuncQuintic1} with Givental's reconstruction theorem according to ref.~\cite{Giv15all}(p.VIII)
\begin{equation} \label{eq:ReconQ1}
   J_K^\text{sym}(t(\epsilon)) \,=\, e^{\sum_{r=1}^{+\infty}\frac{\sum_{\ell} \Psi_r(\epsilon_\ell) P^{\ell r} q^{\ell r Q\partial_Q}}{r(1-q^r)}} J_K^\text{sym}(t^\text{sym}) \ ,
\end{equation}
where $\epsilon=\sum_\ell \epsilon_\ell P^\ell$ is a formal power series in the Novikov variable $Q$ with coefficients in the polynomial ring $K(X)\otimes \mathbb{C}[q,q^{-1}]$, and where the Adams operator $\Psi_r$ acts as
\begin{equation}
  \psi_r(Q) \,=\, Q^r \ ,  \qquad \Psi_r(q) \,=\, q^r \ .
\end{equation}  
In particular, we can use the formula~\eqref{eq:ReconQ1} to obtain the $J$-function  $J_K(0)$ with vanishing input $t(\epsilon)=0$
\begin{multline} \label{eq:JFuncQuintic2}
   J_K(0) \,=\, 
   (1-q) +\left(\tfrac{575 \Phi _2}{1-q}+\tfrac{1150 (1-2 q) \Phi _3}{(1-q)^2}\right)Q \\
   +\left(\tfrac{25 \left(9794+19496 q+9725q^2\right) \Phi _2}{(1-q) (1+q)^2}
   +\tfrac{50 \left(7380+9748 q-14760 q^2-29244 q^3-12139 q^4\right) \Phi _3}{(1-q)^2 (1+q)^3}\right) Q^2+ \ldots \ .
\end{multline}
Note that the $J$-function \eqref{eq:JFuncQuintic2} can now be used to reconstruct $J$-functions with non-vanishing inputs for both the permutation equivariant and the ordinary quantum K-theory of the quintic. In particular, for two, three and four marked points we obtain the permutation equivariant quantum K-invariants at degree one in $Q$ 
\begin{equation}
\begin{aligned}
     \sum_{k=0}^3 \Phi_k \left\langle \tfrac{\Phi^k}{1-qL};  \Phi_1 \right\rangle_{0,2,1} &= 
         \left(\tfrac{575 }{1-q}\Phi _2+\tfrac{575 (2-3 q)}{(1-q)^2} \Phi _3\right) s_{\tableau{1}} \ ,\\
     \sum_{k=0}^3 \Phi_k \left\langle \tfrac{\Phi^k}{1-qL};  \Phi_1,\Phi_1 \right\rangle_{0,3,1}^{S_2} &=
     \left(\tfrac{575}{1-q}\Phi_2+\tfrac{575 \left(2-q-2 q^2\right)}{(1-q)^2 (1+q)}\Phi_3\right) s_{\tableau{2}} 
        +\tfrac{575 q}{(1-q)^2 (1+q)} \Phi _3 \, s_{\tableau{1 1}} \ , \\
     \sum_{k=0}^3 \Phi_k \left\langle \tfrac{\Phi^k}{1-qL};  \Phi_1,\Phi_1,\Phi_1 \right\rangle_{0,4,1}^{S_3} &= 
     \left(\tfrac{575}{1-q} \Phi _2-\tfrac{575 \left(-2-q+q^2+2 q^3+q^4\right) }{(-1+q)^2 (1+q)\left(1+q+q^2\right)}\Phi _3\right) s_{\tableau{3}} \\
   &\quad+ \tfrac{575 q (1+q)}{(1-q)^2 \left(1+q+q^2\right)} \Phi _3\,s_{\tableau{2 1}} 
     - \tfrac{575 q^2}{(1-q)^2 (1+q)\left(1+q+q^2\right)} \Phi _3\,s_{\tableau{1 1 1}}
     \ ,
\end{aligned}    
\end{equation}
and degree two in $Q$
\begin{equation}
\begin{aligned}
     \sum_{k=0}^3 \Phi_k \left\langle \tfrac{\Phi^k}{1-qL};  \Phi_1 \right\rangle_{0,2,2} \,&=\, 
     \left(\tfrac{25 (19519+19496 q)}{(1-q) (1+q)} \Phi _2
     +\tfrac{25 \left(29313+19496 q-48832 q^2-38992 q^3\right)}{(1-q)^2 (1+q)^2} \Phi
   _3\right)s_{\tableau{1}} \ ,\\
     \sum_{k=0}^3 \Phi_k \left\langle \tfrac{\Phi^k}{1-qL};  \Phi_1,\Phi_1 \right\rangle_{0,3,2}^{S_2} \,&=\,
      \left(\tfrac{25 \left(29290+58511 q+29244 q^2\right)}{(1-q) (1+q)^2} \Phi _2\right. \\
      &\qquad+\left.\tfrac{25 \left(43981+78030 q-19634 q^2-97526 q^3-43866 q^4\right) }{(1-q)^2 (1+q)^3}\Phi _3\right)s_{\tableau{2}} \\
     &\quad + \left(\tfrac{25 \left(9725+19519 q+9771 q^2\right)}{(1-q) (1+q)^2} \Phi _2\right. \\
     &\qquad\left.+\tfrac{25 \left(14553+39038 q+19634q^2-19542 q^3-14668 q^4\right)}
     {(1-q)^2 (1+q)^3}\Phi _3\right) s_{\tableau{1 1}}
      \ ,\\
     \sum_{k=0}^3 \Phi_k \left\langle \tfrac{\Phi^k}{1-qL};  \Phi_1,\Phi_1,\Phi_1 \right\rangle_{0,4,2}^{S_3} \,&=\, 
      \left(\tfrac{50 (19519+19496 q) }{(1-q) (1+q)}\Phi _2\right. \\
       &\qquad\left.+\tfrac{25 \left(58603+107320 q+48717 q^2-48809 q^3-87824 q^4-38992 q^5\right)}{(1-q)^2 (1+q)^2 \left(1+q+q^2\right)} \Phi _3\right) s_{\tableau{3}} \\
       &\quad  + \left(\tfrac{25 (19496+19519 q)}{(1-q) (1+q)} \Phi _2\right. \\
       &\qquad\left.+\tfrac{25 \left(29221+78030 q+87824 q^2+39015 q^3-19496 q^4-19519 q^5\right)}
       {(1-q)^2 (1+q)^2 \left(1+q+q^2\right)} \Phi _3\right) s_{\tableau{2 1}} \\
       &\quad+\tfrac{25 q \left(9725+9725 q+9794 q^2+9771 q^3\right)}{(1-q)^2 (1+q)^2 \left(1+q+q^2\right)}
        \Phi _3\,s_{\tableau{1 1 1}}   \ ,
\end{aligned}    
\end{equation}
and the vanishing invariants
\begin{equation}
 \sum_{k=0}^3 \Phi_k \left\langle \tfrac{\Phi^k}{1-qL};  \Phi_\ell,\ldots,\Phi_\ell \right\rangle_{0,n+1,d}^{S_n} \,=\,0 
 \ \text{for $\ell=2,3$, $n\ge 1$, $d\ge 1$} \ .
\end{equation}     
These equivariant invariants furnish according to eq.~\eqref{eq:RelOrdKTheory} a refinement of the ordinary quantum K-invariants at degree one in $Q$
\begin{equation}
\begin{aligned}
  \sum_{k=0}^3 \Phi_k \left\langle \tfrac{\Phi^k}{1-qL};  \Phi_1,\Phi_1 \right\rangle_{0,3,1} \,&=\,
     \frac{575}{1-q} \Phi _2+\frac{1150}{1-q} \Phi _3
      \ ,\\
     \sum_{k=0}^3 \Phi_k \left\langle \tfrac{\Phi^k}{1-qL};  \Phi_1,\Phi_1,\Phi_1 \right\rangle_{0,4,1} \,&=\, 
      \frac{575}{1-q} \Phi _2+\frac{575 (2-q) }{(1-q)^2}\Phi _3  \ , \\
\end{aligned}    
\end{equation}
and at degree two in $Q$      
\begin{equation}
\begin{aligned}      
     \sum_{k=0}^3 \Phi_k \left\langle \tfrac{\Phi^k}{1-qL};  \Phi_1,\Phi_1 \right\rangle_{0,3,2} \,&=\,
    \frac{975375}{1-q} \Phi _2+\frac{1463350 }{1-q}\Phi _3
      \ ,\\
     \sum_{k=0}^3 \Phi_k \left\langle \tfrac{\Phi^k}{1-qL};  \Phi_1,\Phi_1,\Phi_1 \right\rangle_{0,4,2} \,&=\, 
      \frac{1950750}{1-q} \Phi _2+\frac{975375 (3-2 q)}{(1-q)^2} \Phi _3 \ .
\end{aligned}    
\end{equation}

\begin{table}[ht]
\hfil
\scriptsize{
\hbox{
\vbox{
\offinterlineskip
\halign{
  \vrule width1.2pt\strut
  \hfil~#~\hfil\vrule width0.8pt&\hfil~#~\hfil\vrule width0.8pt&
  \hfil~$#$~\vrule width0.8pt&\hfil~$#$~\vrule width0.8pt&
  \hfil~$#$~\vrule width0.8pt&\hfil~$#$~\vrule width0.8pt&
  \hfil~$#$~\vrule width1.2pt\cr
  \noalign{\hrule height 1.2pt}
  Invariant&Irrep.& d=1\hfil & d=2\hfil & d=3\hfil & d=4\hfil & d=5\hfil \cr
  \noalign{\hrule height 1.2pt}
  $\left\langle 1 \right\rangle_{0,1,d}$ & ---  & 2\,875 & 620\,750 & 317\,232\,250 & 242\,470\,013\,000 & 229\,305\,888\,959\,500\cr
  \noalign{\hrule height 1pt}
  $\left\langle \Phi_1 \right\rangle_{0,1,d}$ & --- & 
2\,875 & 1\,224\,250 & 951\,627\,750 & 969\,872\,568\,500 & 1\,146\,529\,444\,452\,500 \cr
  \noalign{\hrule height 1pt}
  $\left\langle \Phi_1, \Phi_1 \right\rangle_{0,2,d}^{S_2}$ & $\tableau{2}$ & 2\,875 & 1\,836\,375 & 1\,903\,246\,875 & 2\,424\,679\,579\,125 &    3\,439\,588\,333\,328\,750\cr
  & $\tableau{1 1}$ & 0 & 603\,500 & 951\,613\,375 & 1\,454\,803\,340\,750  &  2\,293\,058\,888\,864\,750\cr
  \noalign{\hrule height 0.6pt}
  $\left\langle \Phi_1, \Phi_1 \right\rangle_{0,2,d}$ & --- & 2\,875 & 2\,439\,875 & 2\,854\,860\,250 & 3\,879\,482\,919\,875 & 5\,732\,647\,222\,193\,500\cr
   \noalign{\hrule height 1pt}
  $\left\langle \Phi_1,\Phi_1,\Phi_1 \right\rangle_{0,3,d}^{S_3}$ & $\tableau{3}$ & 2\,875 & 2\,445\,625 & 3\,172\,078\,125 & 4\,849\,356\,706\,875 &   8\,025\,706\,111\,081\,250\cr
  & $\tableau{2 1}$ & 0 & 1\,215\,625 & 2\,537\,639\,500 & 4\,849\,350\,591\,375 &   9\,172\,235\,555\,493\,500\cr
  & $\tableau{1 1 1}$ & 0 & 0 & 317\,217\,875 & 969\,868\,907\,250 &   2\,293\,058\,888\,887\,750\cr
  \noalign{\hrule height 0.6pt}
  $\left\langle \Phi_1,\Phi_1,\Phi_1 \right\rangle_{0,3,d}$ & --- & 2\,875 & 4\,876\,875 & 8\,564\,575\,000 & 15\,517\,926\,796\,875 & 28\,663\,236\,110\,956\,000 \cr
   \noalign{\hrule height 1pt}
  $\left\langle \Phi_1 ,\ldots,\Phi_1 \right\rangle_{0,4,d}^{S_4}$ & $\tableau{4}$ & 2\,875 & 3\,054\,875 & 4\,758\,110\,000 & 8\,486\,371\,493\,250 & 16\,051\,412\,222\,151\,000\cr
  & $\tableau{3 1}$ & 0 & 1\,824\,875 & 4\,758\,089\,875 & 10\,911\,040\,660\,500 & 24\,077\,118\,333\,189\,125\cr
  & $\tableau{2 2}$ & 0 & 612\,125 & 1\,903\,235\,375 & 4\,849\,354\,258\,375 & 11\,465\,294\,444\,384\,125 \cr
  & $\tableau{2 1 1}$ & 0 & 0 & 951\,624\,875 & 3\,637\,012\,358\,000 &   10\,318\,764\,999\,954\,625\cr
  & $\tableau{1 1 1 1}$ & 0 & 0 & 0 & 242\,468\,122\,000 &   1\,146\,529\,444\,429\,500\cr
  \noalign{\hrule height 0.6pt}
  $\left\langle \Phi_1 ,\ldots,\Phi_1 \right\rangle_{0,4,d}$ & --- & 2\,875 & 9\,753\,750 & 25\,693\,725\,000 & 62\,071\,707\,187\,500 & 143\,316\,180\,554\,780\,000\cr
   \noalign{\hrule height 1pt}
  $\left\langle \Phi_1 ,\ldots,\Phi_1 \right\rangle_{0,5,d}^{S_5}$ & $\tableau{5}$ & 2\,875 & 3\,664\,125 & 6\,661\,348\,250 & 13\,578\,191\,451\,000 &  28\,892\,541\,999\,869\,500 \cr
  & $\tableau{4 1}$ & 0 & 2\,434\,125 & 7\,612\,947\,250 & 20\,367\,276\,166\,875 & 51\,364\,519\,110\,807\,875\cr
  & $\tableau{3 2}$ & 0 & 1\,221\,375 & 4\,758\,098\,500 & 14\,548\,059\,111\,000 & 40\,128\,530\,555\,337\,250\cr
  & $\tableau{3 1 1}$ & 0 & 0 & 1\,903\,244\,000 & 8\,728\,831\,088\,625 & 28\,892\,541\,999\,869\,500\cr 
  & $\tableau{2 2 1}$ & 0 & 0 & 951\,616\,250 & 4\,849\,353\,042\,750 & 17\,197\,941\,666\,574\,750\cr
  & $\tableau{2 1 1 1}$ & 0 & 0 & 0 & 969\,871\,332\,750 & 5\,503\,341\,333\,277\,125\cr
  & $\tableau{1 1 1 1 1}$ & 0 & 0 & 0 & 0 & 229\,305\,888\,913\,500\cr
  \noalign{\hrule height 0.6pt}
  $\left\langle \Phi_1,\ldots,\Phi_1 \right\rangle_{0,5,d}$ & --- & 2\,875 & 19\,507\,500 & 77\,081\,175\,000 & 248\,286\,828\,750\,000 & 716\,580\,902\,773\,900\,000\cr
  \noalign{\hrule height 1.2pt}
}}}}
\caption{\label{tab:GWquintic}Listed are the non-vanishing permutation equivariant quantum K-invariants $\left\langle \Phi_1 ,\ldots,\Phi_1 \right\rangle_{0,n,d}^{S_n}$ and the ordinary quantum K-invariants $\left\langle \Phi_1 ,\ldots,\Phi_1 \right\rangle_{0,n,d}$ up to degree $d=5$ and up to five marked points $n=5$ of the quintic Calabi--Yau 3-fold. 
}
\end{table}

Table~\ref{tab:GWquintic} below summarizes further quantum K-invariants together with their equivariant refinements, where we employ the string equation of equivariant quantum K-theory \cite{Giv15all}(p.VII), which in particular implies
\be
\left\langle 1;  \Phi_k,\ldots,\Phi_k \right\rangle_{0,n+1,d}^{S_n}
=\left\langle  \Phi_k,\ldots,\Phi_k \right\rangle_{0,n,d}^{S_n} \ .
\ee
We observe that the (ordinary) quantum K-invariants with three and more marked points at degree $d$ in $Q$, are directly related to the (rational) cohomological Gromov--Witten invariants $N_d^\text{GW}$ of the moduli space $\overline{\mathcal{M}}_{0,d}(X)$ at degree $d$ in $Q$ as
\begin{equation}
    N_d^\text{GW} \,=\, \frac{1}{d^k} \left\langle \Phi_1, \ldots, \Phi_1 \right\rangle_{0,k,d} \quad \text{for} \quad k\ge 3 \ ,
\end{equation}
with
\begin{equation}
  N_1^\text{GW} \,=\, 2\,875 \ , \quad
  N_2^\text{GW} \,=\, \frac{4\,876\,875}{8} \ , \quad
  N_3^\text{GW} \,=\, \frac{8\,564\,575\,000}{27} \ , \quad \ldots \ .
\end{equation}
A general relation of the quantum K-theory  invariants to Gopakumar--Vafa invariants will be discussed in sect.~\ref{subsec:QKvsGV}.

\section{Factorization properties and ring structures}\label{sec:tft}
The 2d A-model, which arises as the IR phase of the 2d limit of the theories considered in this note, is a topological field theory (TFT) characterized by the associative, commutative Frobenius algebra determined by the product in quantum cohomology. It has been argued above that the IR limit of the 3d gauge theory partition function computes the quantum K-theory of ref.~\cite{Giv15all}, which defines another associative, commutative Frobenius algebra representing a quantum product of vector bundles. Moreover, since the 3d sphere and disk partition functions are indices, which can be computed both in the UV and in the IR, we expect an corresponding TFT structure already for the indices of the parent gauge theory. 

\subsection{Disks with insertions and $tt^*$ overlaps functions \label{subsec:overlap}}
In the following we study the factorization properties of the gauge theory and the inner product defined on the boundary theory. For further reference we recollect the result for the disk partition function with left boundary from sect.~\ref{sec:LVlim} in the geometric form 
\be\label{ZLrep}
\begin{aligned}
Z_{L,\mu}&=\int \prod \frac{dz_a}{2\pi i z_a} e^{-S_\text{class}}\prod_{\alpha } Z^{\text{1-loop}}_\alpha \ \cdot   \ff_\mu
\\
&=
\int_X \td^\beta(X)\ \left\{e^{-J-\frac{\ch_2^\beta(X)}{\ln q}} \frac{\hGq}{\eta^{\dim X}}\right\}\ I(Q,q)\ e_\mu
=: \int_X \td^\beta(X)\,  \TT_0 \ e_\mu\ . 
\end{aligned}
\ee
The factor in the curly bracket is the perturbative contribution from tree and one-loop. The non-perturbative contributions are collected in the vortex sum 
\be
I(Q,q)=\sum_{\gamma \in H_2(X,\mathbb{Z})}  e^{-t\cdot \gamma} \ \frac{\hGq^*}{\widehat\Gamma^*_X(\gamma)}=1+\hdots \ ,
\ee
with $\gamma$ labelling the different topological sectors (see eq.~\eqref{wGclass}). For the description of the boundary we introduce the following notations. Let $E_\mu$ be a formal linear combination of left $\mathcal{N}=(0,2)$ boundary theories and $\ff_\mu$ the brane factor for it, as introduced in sect.~\ref{subsec:branef}. For each choice of $\ff_\mu$, the  partition function is a solution of the system of difference equations \eqref{DiffEqgen}. This system has $d=\dim K(X)$ linearly independent solutions at a regular point in the space $\Lambda$ parametrized by $(y,q)$. The solution depends only on the K-theory class of $\ffe_\mu$ in $\cx H=K(X)\otimes \Lambda$, represented by the cohomology class $e_\mu$ in eq.~\eqref{ZLrep}. We neglect algebraic subtleties and assume that we can  take $\Lambda=\IQ(q,y)$. Then $\cx H$ is a complex vector space of dimension $d$ over $\Lambda$.\footnote{A more careful treatment would involve the use of formal power series and freely generated modules.}

The $d$-dimensional vector of solutions represents the restriction to the unit operator $\pp_0=1$ of an operator-state correspondence. The disk partition function without insertions computes the overlap of the vacuum with a boundary state associated with $\ffe_\mu$. Overlaps with insertions of operators $\pp_{i>0}$ at the center of the disk can be generated by taking derivatives with respect to the mass parameter of a single trace operator \eqref{IIopst}:\footnote{For simplicity we often restrict to the $U(1)$ case in writing the following formulas, i.e., to a single K\"ahler parameter. For the general case one needs to simply restore indices running over a basis of $H^2(X,\IZ)$.}
\be\label{yder}
\left.(1-q)\p_{y_i} \left(\int \frac{dz}{2\pi i z} \exp\left({\frac{y_iz^{-i}}{1-q}}\right) 
e^{-S_{class}}\prod_\alpha Z_\alpha \ \cdot   \ff_\mu\right)\right|_{y_i = 0}\ .
\ee
with $i=0,...,d-1$. This expression computes the vev of a Wilson line operator wrapping $S^1$
\bea\label{Wi}
\qcor{W_i}_{\mu} = \int \frac{dz}{2\pi i z} z^{-i} 
e^{-S_\text{class}}\prod_\alpha Z_\alpha\ \cdot   \ff_\mu
=: \int_X \td^\beta(X)\,  \TT_i \ e_\mu\ . 
\eea
Upon evaluation at the poles $z^{-1}=q^{n-\eps}=Pq^n$, an insertion is represented by the operator $(Pq^\theta)^i$ acting on the integrand of eq.~\eqref{ZLrep}, as discussed in sect.~\ref{subsec:pertexp}. Alternatively, to make contact with the basis $\pp_i=(1-P)^i$, we can use shifted Wilson line operators $\Ws_i$ defined by replacing $y_iz^{-i}\to \tilde y_i(1-z^{-1})^i$ in eq.~\eqref{yder}. In the classical sector with vortex number $n=0$, successive derivatives then generate the classical K-theory ring 
\be\label{qkringclass}
\p_{\tilde y_i}\p_{\tilde y_j} \, I|_{\tilde y_k=0} = (1-Pq^\theta)^i  (1-Pq^\theta)^j \, I
\quad \buildrel n=0 \over \leadsto\quad   \Phi_i\, \cdot \,  \Phi_j \ .
\ee
The sectors with vortex number $n>0$ induce the quantum corrections to the product on the right hand side.

\def\pp{\phi}
In virtue of the difference equations, there are only $d$ independent insertions $W_i$, $i=0,\ldots,d-1$. These satisfy the related difference equations \eqref{eveq} with eigenvalues $Qq^i$. The $d\times d$ matrix of linearly independent eigenfunctions can be read as a map from bulk operators $\pp_i$ to K-theory classes of left boundary states $\ket{\ffe_\mu}$:
\bea
\TT:\ \pp_i\mapsto \ket{\ffe_i}=\TT^\mu_i\ket{\ffe_\mu}\ ,\qquad \TT^\mu_i = \TT^\mu_i(Q,y,q)\ .
\eea
Here $\{\ket{\ffe_\mu}\}$ is a basis for $\cx H$ and $\TT_i^\mu$ are the vortex sums of the partition functions with insertions in a chosen basis $\{\pp_i\}$ for the bulk operators, where we allow for a $(Q,y,q)$-dependent linear basis change compared to eq.~\eqref{Wi}. The disk diagram with a right boundary defines a related map obtained by sending $q\to\bar q$
\bea\label{That}
\TTh:\ \pp_i\mapsto \bra{\ffe_i}=\TTh^\mu_i\bra{\ffe_\mu}\ , \qquad \TTh^\mu_i = \TT^\mu_i(Q,y,\bar q)\ .
\eea
If we extend the coefficient ring to include the $Q$ parameters, we can view the maps $\TT$ and $\TTh$ as endomorphisms of $\cx H = K(X)\otimes\Lambda_Q$ with $\Lambda_Q= \IQ(q,y)[[Q]]$.\footnote{The ring~$\IQ(q,v)[[Q]]$ denotes a formal power series in $Q$ with coefficients in $\IQ(q,v)$.} More generally one can choose different coefficient rings $\Lambda_{L/R}$ resulting in distinct families $\cx H_{L/R}$ for the spaces of left/right boundaries. The maps $\TT$ and $\TTh$ can be diagrammatically represented as
\begin{center}\hskip1.2cm \includegraphics[scale=0.6]{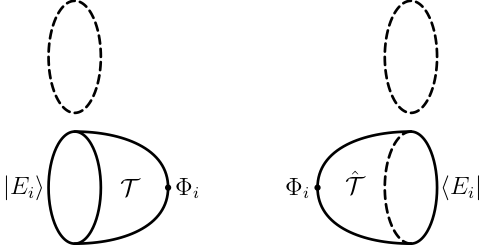}\end{center}
The extra $S^1$ direction will be often omitted in the figures below.

In addition to the disk diagrams, one has inner products defined by putting the theory on the sphere and the annulus. By the completeness of the bases $\{\pp_i\}$ and $\{\ket{\ffe_\mu}\}$, the inner products can be factorized into disk diagrams as 
\begin{equation}\label{fact}
\begin{aligned}
  \eta_{ij} = \eta(\pp_i,\pp_j) \ &=\ \TTh_i^\mu \chi_{\mu\nu}\TT^\nu_j\ , \\[2ex]
  \vcenter{\hbox{\includegraphics[scale=0.6]{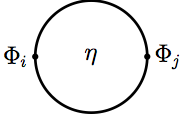}}} \ & =\ \vcenter{\hbox{\includegraphics[scale=0.6]{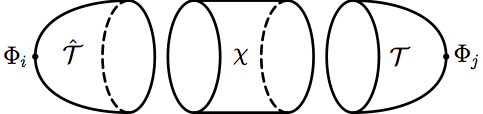}}} \ , \\[2.5ex]
   \chi_{\mu\nu} = \chi(\ffe_\mu,\ffe_\nu)&=(\TTh^{-1})_\mu^i \eta_{ij}(\TT^{-1})^j_\nu \ , \\[2ex]
   \vcenter{\hbox{\includegraphics[scale=0.6]{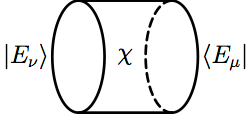}}} \ &=\ \vcenter{\hbox{\includegraphics[scale=0.6]{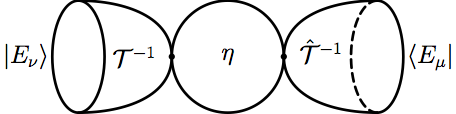}}} \ . \\[1.25ex]
\end{aligned}
\end{equation}
The factorization structure holds for the 3d theory and its 2d limit, in the UV and in the IR phase, but the explicit expressions will depend on the details. E.g. in the small radius limit we should recover the well-known structure of the 2d theory. A Wilson line insertion $\Ws_1\simeq (1-Pq^\theta)$ reduces to the differential operators $H+\hbar \theta$, or simply $\hbar\theta$ replacing $I\to I e^{H\ln Q /\hbar}$ (cpw. fn. \ref{fn14}). The operators $\pp_i\sim H^i \in H^{2*}(X,\IZ)$ represent chiral ring operators of the closed string sector, whereas the boundary states are labeled by elements $\ket{\ffe_\mu}\in K(X)$ associated with the Ramond charge of a D-brane. The respective spaces for the bulk operators and the boundary charges are isomorphic after tensoring with $\mathbb{Q}$. In 2d mirror symmetry, the matrix $\TT_i^\mu$ comprise the integrals of the holomorphic forms $\theta^k\Omega$ of the mirror manifold. 

In the IR phase of the 2d theory, one may choose bases in which the inner products \eqref{fact} are independent of the parameters, namely
\be\label{topmet}
\eta^\text{2d}_{ij}=\int_X \phi_i\wedge\phi_j\ ,\qquad
\chi^\text{2d}_{\mu\nu}=\int_X \td(X) {\ch(\ffe^*_\mu)\ch(\ffe_\nu)}\ .
\ee
Here $\chi^\text{2d}$ is the Witten index in the open string theory \cite{HIV} (cf. eq.~\eqref{HIVind}) and $\eta^\text{2d}$ is the constant metric of refs.~\cite{DVV,Cecotti:1991me}. Note that the arguments in the last reference use special properties of the super--multiplets of the 2d theory, which do not hold in 3d. Correspondingly one does not expect, that there is a basis in which the 3d sphere metric is constant over the moduli space parametrized by $Q$.

The $tt^*$ structure of refs.~\cite{Cecotti:1991me,CV13} emerges if one considers in addition the complex conjugated operators $\pp_{\bx \imath}=\pp_i^*$. To connect  the 3d sphere and disk partition functions to $tt^*$ objects, we propose the relations
\be\label{Zoverlap}
\qcor{\mu|0}\leadsto Z_{D^2\times_q S^1}(\ffe_\mu)\ ,\hskip 1.5cm
\qcor{\bx 0  | 0}\leadsto Z_{S^2\times_q S^1}\ .
\ee
Here $\qcor{\mu|i}$ denotes the $tt^*$ correlator with left boundary $\mu$ and an insertion of $\pp_i$, and $\qcor{\bx \imath  |j}$  the $tt^*$ sphere correlator with  insertions of $\pp_{\bx \imath}$ on the left and and $\pp_j$ on the right. The $tt^*$ correlators, including $\qcor{\mu|0}$, have a non-holomorphic dependence on the deformations in the general non-conformal case. To compare them to the partition functions with a holomorphic dependence, one has to take the holomorphic limit defined in ref.~\cite{CV13}, and this is meant by $\leadsto$. The equations \eqref{Zoverlap} represent 3d generalizations of similar relations for 2d sphere and disk amplitudes proposed in ref.~\cite{Jockers:2012dk} and refs.~\cite{HonOk,HoriRomo}, respectively.

\subsubsection{Cohomological inner products}
Let us now consider a general  inner product $\chi_\AA$ associated with an annulus diagram with left/right boundaries $E_{L/R}$. The geometric interpretation \eqref{RestoVolInt} of the Coulomb branch integrals  is defined on the level of cohomology. Accordingly, we represent in the following the left/right boundaries $E_{L/R}$ of the Hilbert spaces $\cx H_{L/R}$ in terms of suitable forms $e_{L/R}$ representing cohomology classes on $X$ with coefficients in $\Lambda_{L/R}$.  The general inner product is defined as 
\be
\chi_\AA(e_R,e_L) = \int_X \td^\beta(X) \, \AA \, e_L \, e_R \ ,
\ee
in terms of a cohomological integration kernel $\AA$. Note that $\chi_\AA$ takes values in the combined coefficient ring $\Lambda_{LR}=\Lambda_L \otimes \Lambda_R$. Similarly, we express the Coulomb integral for the disk diagram as
\be
  Z_{L}(e_R)=\int_X\, \td^\beta(X) \,\om_{L} \, e_R \ ,
\ee
with the cohomological integration kernel $\omega_L$. The analogous formula is defined with $L/R$ exchanged. As linear maps acting on boundary elements $e_{L/R}$, the disk diagrams $Z_{L/R}$ are elements of the dual spaces $(\cx H_{R/L})^*$, and we can compute the dual pairing\footnote{In order to define the dual paring $\chi_\AA^{-1}$, we extend the annulus diagram to a bi-linear map $\chi_\AA: ( \cx H_{R}\otimes \Lambda_L) \times (\cx H_L \otimes \Lambda_R) \to \Lambda_{LR}$ over the common coefficient ring $\Lambda_{LR}$. Then we can view the dual pairing as a bi-linear map $\chi_\AA^{-1}:  \left( (\cx H_L)^* \otimes \Lambda_R \right) \times \left((\cx H_{R})^*\otimes \Lambda_L\right) \to \Lambda_{LR}$.}
\begin{equation}\label{chidual}
\begin{aligned}
  Z_{\AA}(\om_L,\om_R) = \chi_\AA^{-1} (Z_L,Z_R) 
  =  \chi_\AA\left(\frac{\omega_L}{\AA} , \frac{\omega_R}{\AA} \right) 
   = \int_X\, \td^\beta(X) \,\frac{\omega_R \,\omega_L}\AA \ ,
\end{aligned}  
\end{equation}
which yields a spherical partition function depending on the integration kernels $\AA$ and $\omega_{L/R}$. Eq.~\eqref{chidual} is a generalized Riemann bi-linear identity for the Coulomb integrals.

Identical partition functions, arising from distinct choices of $\AA$, describe different factorizations of the spherical partition function into disk and annulus amplitudes. We will now discuss two relevant examples. Let us first determine the annulus metric $\chi_\ann$ in the basis chosen by the 3d Coulomb integral. The holomorphic limit of the $tt^*$ type factorization \eqref{ResExS} is of the form 
\be\label{ZDfac}
\begin{aligned}
Z_{S^2\times_q S^1} \ &= \ \bar Z_{D^2\times_{\bar q} S^1}(E^*_\mu)\, \chi_\ann^{\mu\nu}\, Z_{D^2\times_q S^1} (E_\nu)\ , \\[1ex]
\vcenter{\hbox{\includegraphics[scale=0.6]{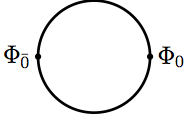}}} \ &=\ \vcenter{\hbox{\includegraphics[scale=0.6]{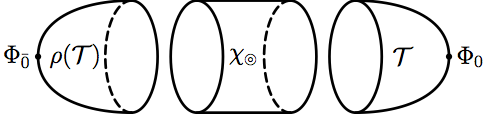}}} \ . \\[1.25ex]
\end{aligned}
\ee
For the $tt^*$ type factorization \eqref{ZDfac} we take
\be
\om_L  = \TT_0(X)\ ,\qquad \om_R  = \rho(\TT_0(X))\ ,
\ee
where
\be\label{tt0}
\TT_0(X) = \left\{e^{-J-\frac{\ch_2^\beta(X)}{\ln q}} \frac{\hGq}{\eta(q)^{\dim(X)}}\right\}\ I(Q,q) \  =\TT_0^\mu e_{L,\mu} \ ,
\ee
is the integrand in eq.~\eqref{ZLrep}, and $\{e_{L/R,\nu}\}$ furnish cohomological bases for the left/right boundaries. The map $\rho=\rho_*\circ \rho_\vee\circ \rho_q$ is the combined operation of complex conjugation of parameter $(Q,y)$, $q\to \bar q$ and duality $x\to -x$ on the Chern roots
\be \label{eq:invol}
\rho_*:\ (Q,y)\to (\bar Q,\bar y)\ ,\qquad \rho_\vee:\  x\to -x\ ,\qquad \rho_q:\ q\to \bar q\ .
\ee
This determines the $q$-dependent integration kernel $\AA_\ann$ to be
\begin{equation}
  \AA_\ann = \frac{\Gamma_{X,q}\Gamma_{X,\bar q}}{\left( \eta(q) \eta(\bar q) \right)^{\dim(X)}} \ , 
\end{equation}
and the $q$-dependent annulus metric $\chi_{\ann}$ becomes in the bases $\{ e_{L/R,\mu} \}$ 
\be\label{annm}
\chi_{\ann,\mu\nu}=\int_X \td^\beta(X) \frac{\Gamma_{X,q}}{\eta(q)^{\dim(X)}}\frac{\Gamma_{X,\bar q}}{\eta(\bar q)^{\dim(X)}}\, e_{R,\mu}\, e_{L,\nu}\ .
\ee
On the other hand,  the holomorphic sphere metric $\eta$ corresponds to taking
\be
\om_L = \TT_0(X)\ ,\qquad \om_R = e^{i\pi c^\beta_1/\ln q} \rho_q(\TT_0(X))\ ,
\ee
where the factor $e^{i\pi c_1/\ln q}$ is the 3d lift of an analogous factor in the definition of the 2d holomorphic sphere metric discussed in refs.~\cite{GGIritani,Iritani2}. Using eqs.~\eqref{fact} and \eqref{annm}, we obtain 
\be\label{eta00}
\eta_{00}(Q,q)=\int_X \td^\beta(X) \, I(Q,q) \, \hat I(Q,q)\, 
\ee
where the hat on a function $f$ is again short for $\rho_q(f)$, i.e., $\hat I(Q,q):= I(Q,\bar q)$, cpw. eq.~\eqref{That}.

To extend the inner product to insertions, we first note that the following $Q$-independent change of basis simplifies the annulus metric $\chi_\ann$ to the constant one
\be
e^{-J-\frac{\ch_2^\beta(X)}{\ln q}}\frac{\Gamma_{X,q}\, e^R_\mu }{\eta(q)^{\dim(X)}}=: M_\mu^r\Phi_r\, , \qquad 
e^{-\rho_q(J)+i\pi c_1^\beta/\ln q+\frac{\ch_2^\beta(X)}{\ln q}}\frac{\Gamma_{X,\bar q}\, e^L_\mu }{\eta(\bar q)^{\dim(X)}}=: \hat M_\mu^r\Phi_r\, , \qquad 
\ee
where $\Phi_r=(1-P)^r$ as before. The basis change transforms  $\chi_{\ann,\mu\nu} = (M\chi\hat M^T)_{\mu\nu}$ to the standard inner product  with integration kernel $\AA\equiv1$, i.e., 
\be\label{chistd}
\chi_{rs} = \int_X \td^\beta(X) \Phi_r\Phi_s\ .
\ee
Moreover, expanding $I(Q,q)=I^\alpha\Phi_\alpha$ with $I^\alpha = (\hat M^T \cdot \TT_0)^\alpha$, we obtain
\be
Z_{L,\mu}=M_\mu^r \int_X \td^\beta(X) I(Q) \Phi_r = (M\cdot \chi \cdot I)_\mu=(\chi_{\ann}\cdot \TT_0)_\mu\ ,
\ee
and a similar relation holds for $Z_{R,\mu}$. Combining this basis change with non-trivial insertions in the 3d partition function, we arrive at the generalization of the holomorphic sphere metric
\be\label{etaij}
\eta_{ij}(Q,q) = \TTh_i^\alpha\chi_{\ann \alpha\beta}\TT_j^\beta \ = \  \hat I_i^\alpha\chi_{\alpha\beta}I_j^\beta \ = \  (\hat I_i, I_j)_X\ ,
\ee
where $(A,B)_X$ is the standard inner product \eqref{chistd} and $I_i^\alpha = (\hat M^T \cdot \TT_i)^\alpha$.
The two expressions on the r.h.s. give two different representations of the sphere factorization \eqref{fact} into disk correlators represented by either the Coulomb branch expressions $\TT_i^\alpha$, containing perturbative terms,  or the vortex sums $I_i^\alpha$ with perturbative terms stripped off.

The above argument started from determining the kernel $\AA$ for the annulus metric $\chi_\ann$ on the Coulomb branch imposing the 3d  factorization condition \eqref{ZDfac}. As will be discussed below, eq.~\eqref{etaij} matches the inner product for the WDVV relation on quantum K-theory determined in ref.~\cite{GivWDVV} for a particular choice of basis, which confirms the proposed 3d/quantum K-theory correspondence.

\subsubsection{Towards a Mukai pairing on loop space }
The cohomological computation above should be related to a loop space generalization of the Mukai pairing of ref.~\cite{MR2141853}, which involves (on the level of K-theory)\footnote{In ref.~\cite{MR2141853} the Chern homomorphism is formulated for the bounded derived category of coherent sheaves $D^b(X)$.} a modified Chern homomorphism $\mu:K(X)\to H^*(X,\IQ)$ that assigns to a K-theory element $E\in K(X)$ a cohomology class $\mu(E)$, called the Mukai vector. The compatibility of $\mu$ with the Grothendieck--Riemann--Roch formula for a proper morphism $\pi: X\to Y$ requires $\pi_*(\mu(E) \td(X))= \mu(\pi_!E)\td(Y)$, where $\pi_*$ and $\pi_!$ are the cohomological and K-theoretic push-forwards of $\pi$, respectively. More specifically, $\mu$ can be chosen such that the K-theoretic inner product $\chi^K(E,F)$ equals the cohomological  inner product 
\be
\chi^K(E,F)=\int_X\mu_R(E)\mu_L(F)\ .
\ee
Here $\mu_R$ is the dual of $\mu_L$ up to a correction factor \cite{MR2141853}, $\mu_R(E)=\tau(\mu_L(E))e^{c_1(X)/2}$, with $\tau(\omega_k)=(-)^k\omega_k$ for a $2k$-form $\omega_k$.\footnote{We restrict the discussion to even cohomology $H^{2*}(X,\IQ)$.}  The 2d Gamma class is a particular solution to this problem, see eq.~\eqref{eq:2dprod}. 

The cohomological expressions obtained above suggest the following generalization to the 3d case. Let $\cx H=K(X)\otimes \Lambda_q$ be the basic space of boundary states over a suitable coefficient ring $\Lambda_q$. That is to say we view the boundary K-theory elements as $(q,y)$-dependent classes on the $S^1$ fixed point set $X$, such that we tentatively set $\Lambda_q=\IQ(q,y)$ or a suitable extension thereof. We now define two maps $\mu_{L/R}: \cx H\to H^{2*}(X,\Lambda_q)$ as
\be
\begin{aligned}
\mu_L:&\  \ E_L\mapsto \Ch(E_L)\hGq\Gamma_{X,\beta}\rho_L\ , \\
\mu_R:&\ \ E_R\mapsto \Ch(E^*_R)\hGq^*\Gamma_{X,\beta}^* e^{-\ch^\beta_2(X)/\ln q}\rho_R \ ,
\end{aligned}
\ee
where $\Gamma_{X,\beta}$ is the ordinary Gamma class \eqref{hG}, in 3d normalization. The factor $\Ch(E)$ contains all the dependence on the argument and agrees with $\ch_{S^1}(E)$ for $\ch_2(E)=0$. The factors $\rho_{L/R}$ parametrize a universal  ambiguity in lifting the cohomological expressions to $\cx H$. A bilinear inner product $\chi^K: \ \cx H \times \cx H \to \Lambda_q$ is now given by 
\be\label{chiKLX}
\chi^K(E_R,E_L)=\int_X \mu_R(E_R)\mu_L(E_L)\ .
\ee
In the special case $\Ch(E_{L/R})=1$, the integrand reduces to 
\be
\mu_L(1)\mu_R(1)=\td^\beta(X)\hGq \Gamma_{X,\bar q}\, (\rho_L\rho_R)\ .
\ee
For the choice $\rho_L\rho_R=(\eta(q)\eta(\bar q))^{-{\dim(X)}}$ one obtains the kernel $\AA_\ann$.

To include left/right disk partition functions, the above structure needs to be generalized, by allowing for distinct spaces $\cx H_{L/R}$ for the left/right boundaries.\footnote{This amounts to consider more general Fourier--Mukai transforms.} Here $\cx H_{L/R}=K(X)\otimes \Lambda_{L/R}$ differ only by the coefficient ring. This generalization is needed, as the disk partition functions define maps 
\be
T:\cx H_L^b \to \cx H_L\ ,\qquad \hat {T}:\cx H_R^b \to \cx H_R\  .
\ee
Here $\cx H_{L/R}^b= K(X)\otimes \Lambda_{L/R}^b$ is the space of bulk operators and $\cx H_{L/R}=K(X) \otimes \Lambda_{L/R}$, where $\Lambda_{L/R}$ is an extension of $\Lambda_q$ by $\Lambda_{L/R}^b$. E.g., for the holomorphic sphere metric we take $\Lambda_L^b=\Lambda_R^b=\Lambda_q[[Q]]$, whereas for the $tt^*$ metric we consider $\Lambda_L^b=\Lambda_q[[Q]]$ and $\Lambda_R^b=\Lambda_q[[Q^*]]$.

Extending by linearity, one obtains maps $\mu_{L/R}:\cx H_{L/R}\to H^{2*}(X,\Lambda_{L/R})$ and the inner product $\chi:\cx H_L\times \cx H_R \to \Lambda_{LR}$ with $\Lambda_{LR}=\Lambda_L\otimes \Lambda_R$, which can be used to glue disk partition functions. In this way, the left disk partition function with insertions can be written as the inner product
\be
Z_{L,i}(E_R)=\chi(E_R,T_i)\ ,
\ee
together with the assignment
\be
\Ch(T_i)=e^{-J}\frac{I_i(Q,q)}{\Gamma_q^*}e^{c_1^\beta/2} \ ,
\ee
for a K-theory class $T_i\in \cx H$ and $e_R=\Ch(E_R)$. Similarly, one can write the holomorphic sphere and $tt^*$ sphere metric as 
\be
\eta_{ij}=\chi(\hat T_j,T_i)\ ,\qquad \eta_{i\bar\jmath} = \chi(\bar T_j,T_i)\ ,
\ee
by assigning
\be
\Ch(\hat T_i)=\rho_q(\Ch(T_i))\, e^{+i\pi c_1^\beta/\ln(q)}\ , \qquad
\Ch(\bar T_i)=\rho(\Ch(T_i))\ .
\ee
with the maps defined in eq.~\eqref{eq:invol}.

The cohomological account provided by the 3d path integral is unsatisfactory in two respects. First, the argument does not fix the ambiguity in the factors $\rho_{L/R}$ and the precise form of the modified Chern character. More importantly one should show that the left hand side of eq.~\eqref{chiKLX} is indeed equal to the K-theoretic inner product on $K_{S^1}(LX)$. To this end one needs to generalize the argument of \cite{MR2141853} to a derived category of sheaves on the loop space of $X$.

\subsection{Flatness equations}
Eq.~\eqref{Zoverlap} relates the 3d gauge theory partition function to the overlap functions of ref.~\cite{CV13} in the holomorphic limit. Including insertions \eqref{yder} one obtains a more general relation between 3d disk correlators with insertions and the holomorphic limit of the 3d $tt^*$ overlap functions $\qcor{\mu | i}$. In a special flat basis of operators and deformations, these represent the flat sections of a holomorphic Gauss--Manin connection on the bundle with fiber $K(X)$ varying over the parameter space $(Q,y)$ \cite{CV13}. The flat holomorphic sections $\Pi^\mu_i$ are thus related to the vortex sums $\TT_i^\mu$, or equivalently, the vortex sums $I_i^\mu$ with the perturbative part stripped off, by an equation of the form 
\be\label{basischange}
\Pi^\mu_i(Q,t,q)=U_i^k(Q,y,q)\, I^\mu_k(Q,y,q)\, .
\ee 
Here the matrix $U(Q,y,q)$ represents a linear change of basis for the operators and $t_i=t_i(Q,y,q)$ is a reparametrization of the deformations $y_i$, such that the $\Pi_i^\mu$ fulfill the flatness equations
\be\label{Ider}
\big((1-q)\delta^j_k\p_{t_i} - C_{ij}^{\ k} \big)\ \Pi_k^\mu  =  0\ ,
\ee
with $C_{ij}^{\ k}$ the structure constants for the chiral ring
\be\label{Ccr}
\Phi_i*\Phi_j = C_{ij}^{\ k} \Phi_k = \Phi_i \otimes \Phi_j + \cx O(Q)\ .
\ee
The variable change from $y_i$ to $t_i$ represents a reparametrization of the UV quantities in terms of IR variables and is the 3d equivalent of the mirror map. The matrix $\Pi_k^\mu$ is the 3d equivalent of the period matrix of 2d mirror symmetry in a flat basis. 

By the 3d/quantum K-theory relation proposed in sect.~\ref{sec:qdiffsys}, the vortex sums compute K-theory correlators, now expressed in the IR variables. The problem of finding the flat coordinates $t(Q,y,q)$ starting from the K-theory correlators has been solved in refs.~\cite{GivER,Giv15all,IMT}, and can be applied to the gauge theory side after making the appropriate identifications. The basis transformation \eqref{basischange} is obtained by a Birkhoff factorization of $I^\mu_k$, and the flat coordinates are determined by the expansion \eqref{Jexp}, which has been used already to obtain the flat coordinates for the examples in sect.~\ref{sec:EQK}. An explicit example for the computation of the $q$-period matrix $\Pi_i^\mu$ and the flat coordinates will be given for Calabi--Yau targets in sect.~\ref{sec:app}. In the following we review the results of of refs.~\cite{GivER,Giv15all,IMT} and connect them to the gauge theory side.

We restrict to a description of the simplest perturbations by single trace operators \eqref{IIopst}; for the generalization to multi-traces one has to consider the general theory treated in part~VII of ref.~\cite{Giv15all}. The basic objects in quantum K-theory are the correlators (cpw. eq.~\eqref{Jexp}) 
\be
T_i^\mu= \delta_i^\mu+\qcord{\frac{\Phi^\mu}{1-qL},\Phi_i}_{0,2}\  ,
\ee
where 
\be
\qcord{\ldots}_{0,m}=\sum_{\beta\geq 0,n\geq 0}\frac  1 {n!}\qcor{\ldots,t^n}_{0,m+n,\beta}Q^\beta\ ,\qquad t=t_k(q)\Phi_k\ ,
\ee
denotes a perturbed correlator. The correlators with $i=0$ enter the $J$-function computed in sect.~\ref{sec:EQK} as 
\be\label{JTrel}
J=(1-q)T^\mu_0\, \Phi_\mu\ .
\ee
Indices on the basis elements $\Phi$ are raised and lowered with the constant metric \eqref{chistd}. Similarly to the $\TT_i^\mu$, the matrix $T_i^\mu$ of K-theory correlators defines a map $T(Q,t,q)\in \textrm{End}(\cx H)$. It is shown in refs.~\cite{GivWDVV, Lee:2001mb} that $T$ is a fundamental solution to the equations
\be\label{fundsol}
(1-q)\p_\ell\,  T = T \, \Phi_\ell*,\qquad
(1-P_aq^{\theta_a}) \, T = T \, \Ws_a \cdot \ .
\ee
Here $*$ denotes the K-theoretic quantum product in the $t$-directions and similarly $W_a $ stands for a multiplication induced by the difference operator, which we already identified with the action of a (shifted) Wilson line operator in the gauge theory.

The two types of deformations combine into a system of a differential connection in the $t$-directions and a difference connection in the $Q$-directions acting on sections of the bundle with fibre $K(X)$ 
\be\label{mixedconn}
\begin{aligned}
\nabt_\ell &= (1-q)\p_\ell - C_\ell\ , &\qquad&\ell=0,\ldots,\dim(K(X))-1\ , \\
\nabQ_a &=  1-P_aq^{\theta_a} - D_aq^\theta\ , &\qquad&a=1,\ldots,\dim (H^2(X))\ ,
\end{aligned}
\ee
with $\p_\ell = \frac{\p}{\p t_\ell}$ and $\theta_a = Q_a\frac{\p}{\p Q_a}$. Eq.~\eqref{fundsol} implies the flatness of the connection  \cite{GivTon,IMT}
\be
[\nabt_\ell,\nabt_k] = [\nabt_\ell,\nabQ_a] = 
[\nabQ_a,\nabQ_b] =0\ .
\ee
The K-theoretic product $*$ is identified with the product of field operators in the IR limit of the 3d gauge theory. The matrix $T_i^\mu$ is the transpose of the period matrix $\Pi_i^\mu$ of the vortex sums in the IR basis. This is the IR equivalent of the UV correspondence between the 3d disk partition functions and the K-theoretic $I$-function found in sect.~\ref{sec:qdiffsys}.\footnote{On the level of maps, UV and IR in the gauge theory corresponds to quasi-maps and stable maps in quantum K-theory, respectively. See refs.~\cite{WitPhases, MP} for the discussion of the 2d case.} Thus the full $(Q,y)$ dependence of the 3d partition functions deformed by massive and Wilson line insertions is determined by the {\it combined} system of differential/difference equations \eqref{mixedconn}. 

Note that the special insertion $\Phi_\mu/(1-qL)$ in the first slot of the K-theory correlator is the operator that creates the hole in the disk and determines the class of the boundary in the gauge theory:
\begin{equation}
   \scriptstyle{| E_\mu \rangle}  \ 
   \vcenter{\hbox{\includegraphics[scale=0.6]{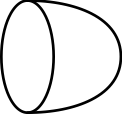}}} \
   \leftrightarrow \ \tfrac{\phi^\mu}{1-qL}
   \qquad\qquad
   \vcenter{\hbox{\includegraphics[scale=0.6]{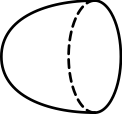}}}\,
   \vcenter{\hbox{\includegraphics[scale=0.6]{FigDiskLeft.png}}} \
   \leftrightarrow \ \tfrac{\sum_\mu\phi_\mu  \phi^\mu }{1-N_+N_-}
\end{equation}
Moreover, factorization onto boundary states in the 3d gauge theory translates on the K-theory side to a factorization locus on the moduli space of maps, representing a domain curve that splits at a node \cite{GivLee, GivTon}. The factorization at the node involves the insertion $\Phi^\mu\times \Phi_\mu/(1-N_+N_-)$ in the correlator, where the denominator is the contribution from the deformation smoothing the node. $N_\pm$ are the classes of the duals of the tangent lines on the two components connected by the node. Upon smoothing the node, the insertions create left and right boundaries of the disks connected by a cylinder metric \eqref{chistd}.

With the above identifications, we obtain similar diagrams as in sect.~\ref{subsec:overlap} with $\TT$ replaced by $T$. Gluing a half-sphere with right boundary to the $t$-derivative of a disk correlator with left boundary gives the inner product for the Froebenius algebra
\begin{equation}
  (1-q)^{-1}\, (C_\ell)_i^m\eta_{mk}\ =\ (\hat T_k,\p_\ell T_i)_X\ =\
  \phi_k \vcenter{\hbox{\includegraphics[scale=0.55]{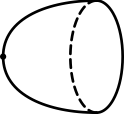}}}\,
  \p_\ell \,  \vcenter{\hbox{\includegraphics[scale=0.55]{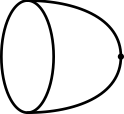}}}\,
  \phi_i \ = \ \phi_k \, \vcenter{\hbox{\includegraphics[scale=0.55]{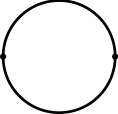}}}\, {\phi_\ell\phi_i} \ .
\end{equation}
The WDVV equations \cite{GivWDVV,Giv15all} ensure the existence of a K-theoretic potential $F(Q,t)=\qcord{1}_{0,1}$ such that
\be
\eta_{ij}=(\hat T_i,T_j)_X=\p_i\p_j F(Q,t),\qquad (C_\ell)_i^m=\eta^{mk}\p_i\p_\ell\p_k F(Q,t).
\ee
Note that the $q$-dependence drops out of the sphere correlators in the flat basis, but the metric is still $Q$-dependent. The power series expansions of $\eta$ and $C_\ell$ in $Q$ have integral coefficients which represent the degeneracies of 3d BPS states in the IR frame.

\subsection{Defect entropies and vortex counting \label{subsec:defent}}
The connection matrices $D_a$ for the finite shifts in the $Q$-directions can be similarly associated with a sphere diagram 
\begin{equation}
  (1-D_a)_i^m\eta_{mk} \ =\
  \phi_k \vcenter{\hbox{\includegraphics[scale=0.55]{FigDiskRightDot.png}}}\,
  P q^{\theta_a} \,  \vcenter{\hbox{\includegraphics[scale=0.55]{FigDiskLeftDot.png}}}\, \phi_i 
  \ = \  \vcenter{\hbox{\includegraphics[scale=0.7]{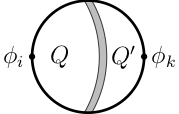}}} \ .
\end{equation}
This correlator represents a defect that separates two regions with different FI parameters $Q$ and $Q'$, where $Q'_a= Q_bq^{\delta_{ab}}$. More generally, we can define defect entropies connecting two regions with FI parameters $Q'/Q=q^\ell$ as (restricting again to the one modulus case to avoid cluttering of notation)
\be\label{edef}
\cx E_{i,j}(\ell)=(\hat T_j,(Pq^\theta)^{\ell} T_i)_X\ .
\ee
The connection matrices $D$ correspond to $\ell=1$.
\\[2mm]

\subsubsec{Example: $X=\IP^{N-1}$}
Let us first discuss the projective space as an example. In this case the mirror map is trivial, i.e. $T_k=I_k=(Pq^\theta)^kI(Q,q)$. The sphere correlator with trivial insertions $i=j=0$  for this case has been studied in ref.~\cite{GivLee}\footnote{See also ref.~\cite{IMT} for an interpretation of the K-theory correlators with non-trivial insertions.} and has been related to a holomorphic Euler characteristic
\be\label{egiv}
\cx E_{0,0}(\ell) = \sum_{d=0}^\infty \chi_H(\qm_d,\cx O(-\ell))\ ,
\ee
Let us explain and rederive this formula from a simple vortex counting \cite{DGH}, using a 3d version of the  arguments of ref.~\cite{NekABCD}, where the instanton partition function of a 5d gauge theory was studied. The relevant compactification for the moduli space of the non-perturbative BPS configurations in the gauge theory has been described in refs.~\cite{MP,WitPhases} for the 2d theory. The instantons of the 2d theory, which become the vortices in the 3d theory, are degree $d$ maps described by $N$ holomorphic sections $(f_1(z),\hdots, f_{N}(z))$ of $\cx O_{\IP^1}(d)$ without common factor, modulo overall rescaling by $\IC^*$. The compactification of the moduli space allows for point-like instantons represented by a $N$-tuple $f_i=Q(z)\tilde f_i(z)$ with common factor $Q(z)$ of degree $d'\leq d$. This gives the moduli space $\qm_d$ of quasi-maps of degree $d$. The real mass parameters $y_i$ and the rotations in the $z$-plane with weight $q$ define a $H=T^N\times S^1$ action on the sections as
\be
f_i(z)=\sum_{n=0}^d z^n a_{in}\ \to \ y_i\, f_i(zq)\ .
\ee
The $N(d+1)$ coefficients $a_{in}$ serve as homogeneous coordinates of weight $y_iq^{-n}$ on the moduli space. Applying the reasoning of ref.~\cite{NekABCD}, the number of holomorphic sections of degree $d$ is given by the $H$ equivariant character 
\be\label{chpn}
\Ch_{H}(\qm_d(\IP^{N-1}))=\frac{1}{\prod_{i=1}^N\prod_{n=0}^d (1-Py_iq^{-n})}\ .
\ee
Here we introduced the weight $P$ for the diagonal $\IC^*$ action on $\IP^{N-1}$ by redefining $y_i\to Py_i$ and imposing $\prod_i y_i = 1$. For $N=1$, there is only a single function $f_1(z)=Q(z)$ representing point-like vortices, and one recovers the result of ref.~\cite{DGH} for $X=$pt.
For $N>1$ there are instantons of finite size. The generating function will be denoted by\footnote{The character $\hat{\cx I}$ is the counting function discussed in sect.~\ref{subsec:mon} and should not be confused with Givental's $I$-function for $\IP^{N-1}$, which is formally obtained from $\hat{\cx I}$ by replacing the weight $P$ by the Chern character $P=\ch(\cx O(-1))$, which fulfills the relation $(1-P)^N=0$. } 
\be
\hat {\cx I}(X)=\sum_{d=0}^\infty Q^d \Ch_{H}(\qm_d(X))\ .
\ee
In the GLSM, the diagonal $U(1)$ corresponding to the homogeneous action on $\IP^{N-1}$ is gauged and one needs to project onto gauge invariants. More generally, an observable associated with the insertion of a Wilson line of charge $\ell$ is obtained by a projection onto the term $\sim P^\ell$ of $\hat{\cx I}(X)$ compensating for the background charge.\footnote{For general boundary conditions, the perturbative term in the 3d partition function carries also a non-trivial representation, and gives another contribution to the background charge.} The projection can be implemented by the contour integral
\be\label{Chint}
\frac{1}{2\pi i} \oint \frac {dP}P P^{-\ell} \ \hat{\cx I}(X) \ ,
\ee
along the circle $|P|=1$. 
The coefficient of $Q^d$ yields the equivariant character $\tr_{H^0(\qm_d,\cx O(\ell))} h$, where $h$ takes into account the action of the group $H$. For $\ell\ge 0$ this counting agrees with the equivariant holomorphic Euler characteristic
\be \label{eq:holEu}
  \chi_H(h)=\sum_p (-)^p \tr_{H^p(\qm_d,\cx O(\ell))}h \ ,
\ee
because in this case all summands with $p>0$ are zero. For $\ell < 0$ the integral~\eqref{Chint} represents the $T^N\times S^1$ fixed point localization formula for the equivariant holomorphic Euler characteristic~\eqref{eq:holEu} if the contour of integration encircles the poles of the integrand from the zeroes of the denominator in $\hat{\cx I}(X)$ \cite{GivLee}. Deforming the contour to enclose instead the poles at $P=0$ and $P=\infty$, we obtain for all $\ell$ a relation between the defect entropies and the monopole expansion of the vortex sum discussed in sect.~\ref{subsec:mon}, namely
\be
\chi_H(\qm_d(X),\cx O(\ell))= \hat{\cx I} |_{Q^dP^\ell}-\frac{q^{Nd(d+1)}}{\prod_i(-\Lambda_i)^{d+1}} \ \tilde{\cx I}|_{Q^dP^{-\ell-N(d+1)}}\ ,
\ee
with $\tilde{\cx I}=\hat{\cx I}|_{\Lambda_i\to \Lambda_i^{-1},q\to q^{-1}}$. For the generating function of the holomorphic Euler characteristics~\eqref{eq:holEu} we find for all $\ell$ the relation
\be\label{Ifact}
\sum_{d=0}^\infty Q^d\chi_H(\qm_d(X),\cx O(\ell))=\big(\hat I_0\ , (Pq^{\theta})^{-\ell} I_0)_{X} = \cx E_{0,0}(-\ell) \ .
\ee
For $\ell \ge 0$ it simplifies to 
\be\label{Ient}
 \hat{\cx I} = \sum_{\ell=0}^\infty P^\ell {\cx E}_{0,0}(-\ell) \  ,
\ee
due to a vanishing residue at infinity.

The above reasonings generalize to defect entropies for the sphere correlator with non-trivial insertions. The result can be written as
\be \label{egen}
\cx E_{i,j}(\ell) = \cx E_{0,0}(i+j+\ell)|_{Q\to Qq^{-j}}\ .
\ee
Under a reflection in $q$, $\cx E_{i,j}(\ell)(q^{-1}) = \cx E_{i,j}(\ell)(q)|_{Q\to Qq^{j-i-\ell}}$. The quantities $\cx E_{i,j}(\ell)$ depend only on the K-theory charges of the insertions  and  on the total background charge up to shifts of $Q$.

In the massless limit $y_i=1$, the defect entropies are closely related to the index \eqref{HIVind}, counting massless open strings between D-branes with RR charge in $K(X)$. Since $\qm_0=X$, this is the same as the leading term of the $Q$-expansion of the entropy \eqref{egen} for the difference bundle $E_a^*\otimes  E_b=P^{-(i+j+\ell)}$. Ref.~\cite{HIV} considers special bases of D-branes, so-called strong exceptional collections of sheaves on $X$, for which most of the Ext groups in eq.~\eqref{HIVind} vanish. Two such collections for $\IP^{N-1}$ are given by $\cx R= \{R_1,\hdots R_{N}\}$ with $R_a=P^{1-a}$ and $\cx S= \{S^1,\hdots S^{N}\}$ with $S^a = (-)^{N-a}\Lambda^{a-1}TX\otimes \cx O(-N+1-a)$. These can be related to bosonic/fermionic maps of the 2d GLSM for $X$ \cite{Mayr:2000as,Govindarajan:2001kr} and are dual in the sense that $(S^{a},R_b)_X = \delta^a_b$. Moreover, one has
\be
(\chi_R)_{ab} := (R_a^*,R_b)_X=\frac{1}{(1-h)^N},\quad 
(\chi_S)^{ab} := (S^{a*},S^b)_X={(1-h)^N} = (\chi_R^{-1})^{ab},
\ee 
where $h$ is an $N\times N$ matrix with unit entries above the diagonal and zeroes otherwise. E.g., for $\IP^2$, one has 
\be
\chi_R=\begin{pmatrix} 1&3&6\\0&1&3\\0&0&1\end{pmatrix}\, ,\qquad 
\chi_S=\begin{pmatrix} 1&-3&3\\0&1&-3\\0&0&1\end{pmatrix}\, .
\ee
The entries count the number of bosonic/fermionic maps between the basis elements of $\cx R$ and $\cx S$ with sign $(-1)^F$. The subleading terms of the $Q$-expansions of the defect entropies for the elements of $\cx R$ can be written in a similar form. Defining $(\chi^{3d}_R)_{ab}=\cx E_{0,-(b-1)}(a-1)=(\chi_R)_{ab}+\cx O(Q)$ one finds 
\bea
(\chi^{3d}_R)_{ab}= &(\chi_R \cdot \frac{1}{1-Q\, \Delta\cdot \chi_R})_{ab}
&\buildrel \IP^2  \over \longrightarrow \left(
\begin{array}{ccc}
 \frac{1}{1-Q} & \frac{3}{(1-Q) (1-q Q)} & \frac{6 +3q Q}{(1-Q) (1-q Q) \left(1-q^2 Q\right)} \\
 0 & \frac{1}{1-q Q} & \frac{3}{(1-q Q) \left(1-q^2 Q\right)} \\
 0 & 0 & \frac{1}{1-q^2 Q} \\
\end{array}
\right)
\nonumber \\
((\chi^{3d}_R)^{-1})^{ab}= &(\chi_S \cdot (1-Q\,  \chi_R\cdot \Delta))^{ab}&\buildrel \IP^2  \over \longrightarrow \begin{pmatrix} 1-Q&-3&3\\0&1-Qq&-3\\0&0&1-Qq^2\end{pmatrix}\ .\nonumber
\eea
where $\Delta =\textrm{diag}(1,q,...,q^{N-1})$. The extra factors compared to the 2d result seem to be related to the modes of the monopole operator in the 3d theory. Indeed the entries in the first line of $\chi^{3d}_R$ coincide with the coefficients of the sum $\cx I=\hat{\cx I}|_{q\to q^{-1}}$ by \eqref{Ient}.
It would be interesting to derive these results from lifting the discussion of ref.~\cite{HIV} from sheaves on $X$ to sheaves on the loop space $LX$. 
\ \\

\subsubsec{Example: Quintic hypersurface in $\IP^{4}$}
The vortex counting can be generalized to hypersurfaces by introducing constraints as in ref.~\cite{NekABCD}. A degree $\ell$ hypersurface $X_\ell$ in $\IP^{N-1}$ is defined by the zero of a section of $\cx O(\ell)_{\IP^{N-1}}$, which pulls back to a section of $\cx O(\ell d)_{\IP^1}$  for a degree $d$ map. Requiring that the section vanishes at $d\ell+1$ point as in ref.~\cite{MP} gives $d\ell +1$ constraints of weight $q^n$, $n=0,...,\ell d$, which contribute a numerator
\be
\Ch_{H}(\qm_d(X_\ell))=\frac{\prod_{n=0}^{\ell d} (1-P^\ell q^{-n})}{\prod_{i=1}^N\prod_{n=0}^d (1-Py_iq^{-n})}\ .
\ee
The generating function $\hat{\cx I}(X_\ell)$ function satisfies again an equation of the form \eqref{Ifact}. For $\ell^2\geq N$ there are poles at $P=\infty$ and a non-trivial UV/IR map. Flowing to the IR, the gauged vortices associated with quasi-maps are replaced by the vortices of the non-linear sigma model \cite{WitPhases,MP}, which correspond to the stable maps of ref.~\cite{Kont}. Correspondingly, there are now two versions for the entropy, counting IR vortices as in eq.~\eqref{edef}, or UV vortices if one replaces  $T_k$ in this formula by $I_k$.

In sect.~\ref{sec:app} we study in detail the case of Calabi--Yau 3-folds and write a closed formula for the connection matrix $D$ in terms of Gopakumar--Vafa and K-theoretic invariants. The entropies for $\ell\neq1$ can be computed from the $J$-function. For the quintic 3-fold, the leading series for some entropies $\cx E_{a}:=\cx E_{0,-a}(0)$ at $t=0$ are
\begin{small}
\be
\begin{aligned}
\cx E_{-3}&=-35-609\,250 \left(3 q^2+4 q+3\right) Q^2q^{-4}-750 (2\,537\,651 q^5+4\,229\,426 q^4\\
   &\qquad +5\,075\,302 q^3+5\,075\,302 q^2+4\,229\,426 q+2\,537\,651) Q^3{q^{-7}} + \ldots \ , \\
\cx E_{-2}&=-15-{612125 Q^2}{q^{-2}}-125 \left(7612953 q^2+10150696 q+7612953\right) Q^3q^{-4}+\ldots \ ,\\
\cx E_{-1}&=-5-{242468139250 Q^4}{q^{-2}}-{458611777775250 (q+1) Q^5}{q^{-3}}+\ldots \ ,\\
\cx E_0&=\cx F(Q,0)=2\,875 Q + 620\,750 Q^2 + 317\,232\,250 Q^3 + 242\,470\,013\,000 Q^4 \\
    &\qquad+229\,305\,888\,959\,500 Q^5+\ldots \ , \\
\cx E_{1}&=5+5\,750 (q+1) Q+1\,000 \left(1\,845 q^2+2\,437 q+1\,845\right) Q^2\\
    &\qquad+250 \left(5\,075\,440 q^3+7\,612\,953 q^2+7\,612\,953 q+5\,075\,440\right) Q^3+\ldots \ , \\
\cx E_{2}&=15+2\,875 \left(3 q^2+4 q+3\right) Q \\
   &\qquad+125 \left(24\,554 q^4+38\,992 q^3+44\,073 q^2+38\,992 q+24\,554\right) Q^2\\
   &\qquad\qquad+125 \left(17\,763\,902 q^6+30\,451\,812 q^5+38\,064\,765 q^4 +40\,602\,784 q^3\right.\\
   &\qquad\qquad\qquad\left.+38\,064\,765 q^2+30\,451\,812 q+17\,763\,902\right) Q^3 + \ldots \ .
\end{aligned}
\ee
\end{small}
The holomorphic sphere metric $\eta_{ij}$  is, in this language, the matrix for the identity defect with non-trivial insertions. Differently to the 2d case, it depends on the FI parameters $Q$ and is given, up to order $\cx O(Q^4)$, by
\begin{footnotesize}
\bea
\phantom{1}\hskip-15pt
\begin{pmatrix}
 2\,875 Q+620\,750 Q^2+317\,232\,250 Q^3& 5+2\,875 Q+1\,224\,250 Q^2+951\,627\,750 Q^3&  -5 & 5 \\
 5+2\,875 Q+1\,224\,250 Q^2+951\,627\,750 Q^3& -5+2\,875 Q+2\,439\,875 Q^2+2\,854\,860\,250
   Q^3& 5 & 0 \\
 -5 & 5 & 0 & 0 \\
 5 & 0 & 0 & 0 \\
\end{pmatrix}
\nonumber\\[3pt]
\eea
\end{footnotesize}

\section{Applications to Calabi--Yau manifolds} \label{sec:app}
From the target point view of string theory and M-theory, the case where $X$ is a Calabi--Yau manifold is distinguished.  In the following we study some details of this situation, mainly for dimension $\dim(X)=3$, which is the first case with an interesting IR theory. For $\dim(X)<3$, the IR theory is the classical K-theory. The case of Calabi--Yau $n$-folds with $n>3$ is also interesting and can be treated similarly.

\subsection{Quantum K-theory invariants and Gopakumar--Vafa invariants\label{subsec:QKvsGV}}\def\ngv{\mathfrak{n}}
The integral quantum K-theory invariants for a Calabi--Yau 3-fold $X$ count degeneracies of BPS operators in the world-sheet 3d theory. On the other hand the Gopakumar--Vafa invariants \cite{GV98} count degeneracies of BPS states in the 5d target space theory obtained by an M-theory compactification on $X$. In this and the next section we observe universal relations between world-sheet and target-space invariants for a set of one moduli Calabi--Yau manifolds.

That such a relation exists in principle, follows from the more general results of refs.~\cite{GivTon,Giv15all,Ton16}, where a relation between quantum K-theory invariants and cohomological Gromov--Witten  invariants has been described for general target $X$  in terms of a quantum Hirzebruch--Riemann--Roch formula. The latter equates a single quantum K-theory correlation function to a sum of correlation functions of the so-called fake quantum K-theory computed on the orbifold strata of the moduli stack. This gives also a relation between Gopakumar--Vafa and quantum K-theory invariants, which is however quite implicit and technical in practice. 

For the special case of Calabi--Yau manifolds, we instead find a relatively simply and explicit relation between integral world-volume and target space invariants below. It would be interesting to understand this relation from the point of world-sheet/target space duality. 

As a simple class of examples, we consider the Calabi--Yau 3-fold hypersurfaces $W\IP^{4}_{(k_1,k_2,k_3,k_4,k_5)}[d]$ of degree $d=\sum_{i=1}^5 k_i$ in the weighted projective spaces $W\IP^{4}_{(k_1,k_2,k_3,k_4,k_5)}$ and Calabi--Yau 3-fold complete intersections $\IP^{k+3}[d_1,\ldots,d_k]$ of codimension $k$ in projective spaces $\IP^{k+3}$, namely
\be\label{3f1m}
\begin{aligned}
  &\IP^4[5]\ , &&W\IP^4_{(2,1^4)}[6]\ ,  &&W\IP^4_{(4,1^4)}[8]\ ,  &&W\IP^4_{(5,2,1^3)}[10] \ , \\
  &\IP^5[2,4] \ , &&\IP^5[3,3] \ , &&\IP^6[2,2,3] \ , &&\IP^7[2,2,2,2] \ .
\end{aligned}
\ee
These are one-moduli cases, with the first case being the quintic considered in sect.~\ref{sec:EQK}. The Gromov--Witten potential and the Gopakumar--Vafa invariants  at genus zero for these manifolds have been computed in refs.~\cite{KT,Libgober:1993hq}.

In the next section we will give a closed expression for the $n$$>$2-point functions of ordinary quantum K-theory in terms of the Gromov--Witten potential $\cx F$ for $X$, by studying the chiral ring equations in the $t$-directions. In this section we consider the permutation equivariant case. We concentrate on the dependence on a perturbation $t_1\Phi_1$, since the dependence on the parameters $t_{a\neq 1}$ is the classical one for $a>2$ (see eq.~\eqref{cy3pv}), and for $t_0$ it is fixed  by the K-theoretic string equation \cite{Lee:2001mb},\cite{Giv15all}(p.VII). 

To display the general structure of the quantum K-theory correlators, we write them as 
\be\label{lown}
\left\langle\frac{\Phi_\alpha}{1-qL};\, \Phi_1^r\right\rangle_{0,r+1} = \begin{cases}
\frac{1}{1-q}\sum_kQ^k f^{(r)}_{\alpha,k}&\alpha=0,1\ , \\
0&\alpha=2,3 \ . \end{cases}
\ee
By explicit computation, we observe that the functions $f^{(r)}_{\alpha,k}$ at degree $k$ can be expressed for all 3-folds $X$ in eq.~\eqref{3f1m} in terms of the Gopakumar--Vafa invariants  $\ngv_{n\leq k}$ of $X$. For $r=0$ we find
\bea
&&f_{0,1}^{(0)}=\frac{\ngv_1 (3 q-1)}{q-1}\ ,\quad 
   f_{0,2}^{(0)}=\frac{\ngv_1 \left(-3 q^4+9q^2-4\right)}{(q-1) (q+1)^3}+\frac{\ngv_2 (3 q-1)}{q-1}\ ,\nonumber\\
&&f_{0,3}^{(0)}=\frac{\ngv_1 \left(-8q^6+19 q^3-9\right)}{(q-1) \left(q^2+q+1\right)^3}+\frac{\ngv_3 (3q-1)}{q-1}\ ,\nonumber\\
&&f_{0,4}^{(0)}=\frac{\ngv_2 \left(-3 q^4+9 q^2-4\right)}{(q-1)
   (q+1)^3}+\frac{\ngv_1 \left(-15 q^8+33 q^4-16\right)}{(q-1) (q+1)^3\left(q^2+1\right)^3}+\frac{\ngv_4 (3 q-1)}{q-1}\ ,\\
&&f_{1,1}^{(0)}=\ngv_1\ ,\quad f_{1,2}^{(0)}=-\frac{\ngv_1 \left(q^2-2\right)}{(q+1)^2}+2\ngv_2\ ,
   \quad f_{1,3}^{(0)}=\frac{\ngv_1 \left(3-2 q^3\right)}{\left(q^2+q+1\right)^2}+3\ngv_3\ ,\nonumber\\
&&f_{1,4}^{(0)}=-\frac{2 \ngv_2 \left(q^2-2\right)}{(q+1)^2}+\frac{\ngv_1
   \left(4-3 q^4\right)}{(q+1)^2 \left(q^2+1\right)^2}+4 \ngv_4\ .\nonumber
\eea
Similar  expressions for $n$-point functions with $n>1$ are given in app.~\ref{app:1mcy}. 

The functions $f_{1,k}^{(r)}$ for an insertion of $\Phi_1$ in the first slot are finite in the small radius limit $q\to1$. For $r=0$, they reproduce the multi-cover formula of quantum cohomology \cite{Aspinwall:1991ce,Candelas:1990rm}. On the other hand we observe that for $r>0$, the information about the $S_n$ representations of the permutation equivariant quantum K-theory partially survives in the 2d limit
\bea
\sum_{k>0} Q^kf^{(r)}_{1,k}|_{q=1}  =
\sum_{k>0}  Q^k N^{GW}_k \ k \sum_{\mu(R)=r} \dim_{R,k}\cdot  R\ .
\eea
Here $N^{GW}_k$ are the Gromov--Witten invariants, $\dim_{R,k}$ is the dimension of the representation $R$ in $SU(k)$ and $R$ runs over the Young tableaux with $r$ boxes. 

In the $q\to 0$ limit,  the 1-point function takes the simple form
\be
\langle{\Phi_1}\rangle_{0,1} = Q\p_Q \cx F_{q=0},\qquad 
\cx F_{q=0}=\sum_{k>0} \frac{\ngv_kQ^k}{1-Q^k}\ ,
\ee
which is the expected form for the 5d theory.
\\

\subsubsec{Calabi--Yau $r$-folds}
The computation of the 3d world-volume theory invariants for dimension $r>3$ is similar. We checked that the $n\leq 4$-point functions for the degree $N$ hypersurface in $\IP^{N-1}$ for $N=6,7,8$ can be expressed up to degree 3 by the {\it same} formulas, i.e., eq.~\eqref{lown} and app.~\ref{app:1mcy}, if we replace the Gopakumar--Vafa invariants as
\be
\ngv_k\to k^{-3}\, n_k\ .
\ee
Here $n_k$ are the numbers of rational curves defined and computed  in ref.~\cite{GMP}.

\subsection{Flatness equations and chiral rings for Calabi--Yau 3-folds}
In this section we study the flatness equations and ring structures discussed in sect.~\ref{sec:tft} for Calabi--Yau targets, with a focus on 3-folds.

\subsubsection{$t$-directions}
The $J$-function describes the action of $T$ on the unit $\Phi_0=1$ in $K(X)$:
\be
J=(1-q)T\Phi_0 = J^\alpha \Phi_\alpha\  ,
\ee
where the r.h.s. is the expansion in a given basis $\{\Phi_\alpha\}$ of $K(X)$.
The action of $T$ on the other elements $\Phi_\ell\in K(X)$ can be expressed through the $t$-derivatives of $J$ as
\be
(1-q)\p_\ell J = T\Phi_\ell*\Phi_0 = T\Phi_\ell \ .
\ee
In the basis $\{\Phi_\alpha\}$, $T$ can be viewed as a matrix, whose transpose $\Pi = T^T$ is the 3d analogue of the ``period matrix'' well-known from 2d mirror symmetry
\be
\Pi = \big(\Pi_\ell^\alpha\big)\ ,\qquad \Pi_\ell^\alpha = (T\Phi_\ell)^\alpha\ .
\ee
The chiral ring relations then take the familiar form 
\be\label{Pider}
(1-q)\p_\ell \Pi = C_\ell \cdot \Pi\ ,
\ee
In reverse, starting from the $J$-function of the quantum K-theory, the ring structure constants can be obtained from the $q$-period matrix as 
\be\label{Cs}
C_\ell=(1-q)\p_\ell \Pi \cdot \Pi^{-1}\ .
\ee

The above equations hold in general. We now specialize to the Calabi--Yau 3-fold hypersurfaces \eqref{3f1m} for concreteness; the higher dimensional case works out similarly. The ring structure constants obtained from the $J$-function for $\ell=0,2,3$ are the classical ones. In the basis $\Phi_\ell = (1-P)^\ell$, $\ell=0,1,2,3$,
\be \label{eq:Cclas}
C_0 = \begin{pmatrix} 1&0&0&0\\0&1&0&0\\0&0&1&0\\0&0&0&1 \end{pmatrix} ,\ \ 
C_2 = \begin{pmatrix} 0&0&1&0\\0&0&0&1\\0&0&0&0\\0&0&0&0 \end{pmatrix} ,\ \ 
C_3 = \begin{pmatrix} 0&0&0&1\\0&0&0&0\\0&0&0&0\\0&0&0&0 \end{pmatrix} .
\ee
The only multiplication that is modified in the quantum theory is
\be \label{eq:Cqu}
C_1(Q,t) 
= \begin{pmatrix} 0&1&0&0\\0&0&C_{ttt}&\cc\\0&0&0&1\\0&0&0&0 \end{pmatrix} 
= \begin{pmatrix} 0&1&0&0\\0&0&1&0\\0&0&0&1\\0&0&0&0 \end{pmatrix} +\cx O(Q)\ ,
\ee
where $t\equiv t_1$. The $t$-ring structure constants are independent of the twisting parameter $q$, i.e. the $t$-ring does not depend on the $S^1$ radius $\beta$ of the compactification. 

Geometrically, the ring structure constants encode a quantum deformation of the tensor product of vector bundles. For the line bundle $P$, the deformation of the classical tensor product `$\otimes$'  to the quantum tensor product `$*$' is
\be
 \Phi_1*\Phi_1- \Phi_1 \otimes \Phi_1 = P*P- P\otimes P = 
\left[C_{ttt}(Q) - 1\right] (1 - P)^2 + \cc (1-P)^3 \ .
\ee
Since the quantum corrections are order $H^2$ and higher, they do not modify the lower degree terms in the tensor products. More generally, the upper triangular form of the structure constants in the chosen K-theory basis $\{\Phi_\ell\}$ is a peculiarity of the Calabi--Yau case and goes back to the ghost charge conservation in the cohomological theory. As any vector bundle $\mathcal{V}$ can be expanded as $\mathcal{V}=\operatorname{rk}(\mathcal{V}) \Phi_0 + \cx O(\Phi_1)$, the quantum corrections preserve, e.g., the rank, but modify the higher Chern characters of the classical tensor product. On the other hand the quantum tensor product does not preserve the rank of the classical tensor product of vector bundles for targets without ghost number conservation in the cohomological theory, as e.g. the Fano varieties considered in ref.~\cite{IMT}.
\\

\subsubsec{$t$-Differential equations and $q$-period vector}
Iterating the system of first order  equations \eqref{Pider}, one obtains differential operators  in the $t$-parameters that annihilate the  period vector $\Pi_0=J(Q,t,q)$:
\bea\label{tdeq}
&&\cx D_a \Pi_0 = 0\ ,\nonumber\\[2mm]
&&\cx D_1 = 
\p_1\,  [1+(1-q)\mu]^{-1}\, \p_1 \, [C_{ttt}]^{-1}\, \p_1^2\ , \qquad \mu = \p_1(\cc \, C_{ttt}^{-1})\ ,\\[2mm]
&&\cx D_{a>1} = \{\p_2\p_1^2,\p_2\p_2,\p_3\p_1,\p_3\p_2,\p_3\p_3\}\ . \nonumber 
\eea
Note that the $q$-dependence of the differential operators is only in the prefactor of the term $(1-q)\mu$ in $\cx D_1$, and it would vanish without the special entry $\cc$. In the 2d limit $q\to 1$, this term is subleading, and one obtains back the ordinary Picard--Fuchs equation of the 3-fold $X$ in flat coordinates. In particular it follows from this limit that 
\be\label{Cttt}
\kappa C_{ttt}(Q,t)= \p_t^3 \cx F(Qe^t) =\kappa+\cx O(Q)\ ,
\ee
where $\cx F$ is the prepotential of the  Gromov--Witten theory for $X$ and $\kappa=\int_X H^3$.\footnote{The 2d prepotential for the above examples has been computed by mirror symmetry in ref.~\cite{KT,Libgober:1993hq}. }

In sect.~\ref{sec:qdiffsys} we argued, that the {\it difference} equation annihilating the 3d partition function reduces to the Picard--Fuchs equation of the 2d theory in the small radius limit. In the above we obtained the same Picard--Fuchs equation in the small radius limit of the $t$-flatness equation.~This is another illustration of the fact, that $Q$ and $t$ deformations become equivalent in the 2d limit, at least at the level of the holomorphic quantities considered in this paper.

Differently to the 2d case, the $t$-differential equations alone do not determine $\Pi_0$. Any linear polynomial $f= a_0+a_1 t_1$ with arbitrary integration constants $a_{0,1}(Q,q)$ solves $\cx D_a\Pi_0=0$. These terms correspond to the 1- and 2-point functions and they are not fixed by the $t$-ring structure constants; however, they appear in the multiplication rule of the Wilson line operators, as discussed below. On the other hand a term $f=a_2 t^2$ generates a solution only if $a_2=C_{ttt}(Q)$, times some $q$ dependent function, which can be fixed from the classical terms. The vector of independent solutions for \eqref{tdeq} is
\be\label{cy3pv}
(\Pi_0)^\alpha=\begin{pmatrix}
\qq\\
t_1\\
t_2 +\frac1{\qq}(\frac 12 {t_1^2}+\cx F^q_t(z))+p_1(Q,t,q)\\
t_3 + \frac 1 {\qq}t_1 t_2 +\frac{1}{\qq^2}(\frac{1}{3!}t_1^3+(1-3q)\cx F^q(z)+(qt_1+\qq)\cx F^q_t(z))+p_2(Q,t,q)
\end{pmatrix}\ ,
\ee
where we have used the large volume limit to fix the classical terms, $\qq = 1-q$, $z = Qe^t$, $\kappa \cx F^q$ is the quantum part of $\cx F$, $f_t = \p_t f$, and $p_k$ are degree $k$ polynomials in $t=t_1$ determined by the $n$-point functions with $n<3$.\footnote{The term of $p_2$ quadratic in $t$ is also fixed by the lower order terms.} Using the explicit result for the $J$-function, the source term $\mu$ for the $q$-dependence of $\cx D$ can be written, to the computed order in $Q$,  as 
\be
\cc=(1-t)\cx F^q_{ttt}+\cx F^q_{tt} +N_1(\ln (1-Q)+Q/(1-Q))+\delta \cc(Q)\ ,
\ee
where $N_1$ is the number of rational curves of degree one, and $\delta \cc(Q)$ is a $Q$-series of $\cx O(Q^4)$ determined by the low $n$-point functions \eqref{lown}.
In the small radius  limit, one obtains from \eqref{cy3pv} the period vector of the 2d theory, which reads, after a rescaling of the basis (cpw. \eqref{pbtmp})
\be
\Pi_0^{GW} \sim (1,t/\hbar,\cx F_t/\hbar^2,-\cx F_0/\hbar^3)^T\ ,
\ee
with $\cx F_0=2\cx F-t\cx F_t$.
The special form of $\Pi_0^{GW}$ for the Calabi--Yau case was imposed by $\cx N=2$ special geometry of the $t$-deformation space. It would be interesting to find a similar interpretation for the $q$-period $\Pi_0$ in ref.~\eqref{cy3pv}.

Eq.\eqref{cy3pv} gives a simple and explicit expression for the K-theoretic $n$-point functions for $n>2$ in terms of the Gromov--Witten prepotential $\cx F$. The polynomials $p_{1,2}$ are determined by the $n$-point functions for low $n$ given in app.~\ref{app:1mcy}. It was also proven in \cite{IMT}, that the correlation functions of ordinary quantum K-theory are polynomials in $t$ and $e^t$. The above formulas suggest that substantial simplifications occur for Calabi--Yau targets. 

\subsubsection{$Q$-directions}
The action of the difference connection gives another first order system for the $q$-period matrix:\footnote{To simplify the expressions, we use here the convention that $\Pi$ contains the factor $P^{\frac{\ln Q}{\ln q}}$, leading to the replacement $1-Pq^\theta\to 1-q^\theta$; see also fn.~\ref{fn14}.}
\be
\delta_a \Pi = D_a \Pi\ ,\qquad \delta_a = 1-q^{\theta_a}\ .
\ee
The matrices $D_a=\delta_a \Pi \cdot \Pi^{-1}$ capture the multiplication of Wilson line operators related to the defects discussed in sect.~\ref{subsec:defent}. For the class of one modulus 3-folds, $a=1$ and the matrix $D=D_1$ computed from the $J$-function has the form
\be
D(Q,t,q)= \begin{pmatrix} 0&1&a&b\\0&0&x&c\\0&0&0&1\\0&0&0&0 \end{pmatrix} = \begin{pmatrix} 0&1&0&0\\0&0&1&0\\0&0&0&1\\0&0&0&0 \end{pmatrix} +\cx O(Q)\ .
\ee
The non-zero entries $x,a,b,c$ are functions of all parameters $(Q,t,q)$.

One may iterate the first order system to obtain a difference equation for the $q$-period vector
\be
\DL\Pi_0=0,\qquad \DL = \delta [(1+\delta \nu)]^{-1}\delta[(x+\delta a)]^{-1}\delta^2\ ,
\ee
with  
\be
\nu = \frac{c+\tilde a +\delta b}{x+\delta a}, \qquad \tilde f(Q) := f(Qq)\ .
\ee From \eqref{DifftoDiff}, the leading behavior of the difference operator $\delta$ in the 2d limit is 
\be
\delta \to (1-q) \theta\ ,\qquad \theta = Q\frac d {dQ}\ .
\ee
The entries $x,a,b,c$ have a finite $q\to 1$ limit. In the 2d limit one can drop the terms with a factor of $\delta$ in the square brackets and obtains the differential equation
\be
\DL\Pi_0 = 0 \quad \longrightarrow \quad \theta^2\, x^{-1}_{q=1}\, \theta^2\ \Pi_0^{GW}=0\ .
\ee
This is just another time the Picard--Fuchs equation for the 3-fold in flat coordinates, this time written in the $Q$-variable.\footnote{In the formalism with doubled number of parameters $(Q,t)$ for $H^2(X)$, the mirror map acts only on the $t$-parameters, but not on $Q$, with $Qe^{t(Q,\tau)}$ parametrizing the 2d theory; see sect.~\ref{subsec:mm}. } In particular it follows that $x_{q=1} =C_{ttt}(Qe^t)$. 

More generally we can use the solution vector \eqref{cy3pv} to compute the structure constants $D$ in terms of $\cx F$ and the polynomials $p_{1,2}$. We find 
\bea
x &=&  1+d_Q \cx F^q_{tt}+\delta p_{1t}\ , \nonumber \\
a &=& \qq^{-1}(d_Q \cx F^q_t-\cx F^q_{tt} +\delta p_1-\qq p_{1t})\ , \nonumber \\
c &=& \qq^{-1}((1-2q)d_Q \cx F^q_t+\qq(1-t)d_Q\cx F^q_{tt}+\tilde{\cx F^q}_{tt}+
\qq\delta p_{2t}+\qq\tilde p_{1t}-t\delta p_{1t})\ ,\\
b&=& \qq^{-1}(\qq^{-1}(1-3q)d_Q \cx F^q+(1-t)d_Q\cx F^q_{t}+
\qq^{-1}(\tilde{\cx F_t^q}-(1-2q)\cx F_t^q-\qq(1-t)\cx F^q_{tt})\nonumber\\
&&+\delta p_2+\tilde p_1-\qq p_{2t}-\qq^{-1}t(\delta p_1-\qq p_{1t})\ .\nonumber \eea
Here a subscript $t$ denotes a $t_1$-derivative and 
\be
d_Q=\frac{1-q^\theta}{\!\! 1-q}\ .
\ee
To summarize, the $Q$- and $t$-multiplications in ordinary quantum K-theory can be written in closed form in terms of the Gromov--Witten prepotential $\cx F$  and the 1- and 2-point functions for the class of Calabi--Yau targets considered above.

\subsection{K-theoretic mirror map and integrality \label{subsec:mm}}
The change from the UV coordinates $(\tau,Q,q)$ to the flat coordinates $(t,Q,q)$ of the IR theory is often called the mirror  map. Compared to the 2d theory, the  interesting novelty of the 3d case is, that this map connects two integral expansions, the UV expansion related to the  index \eqref{ind} and the IR expansion related to quantum K-theory. It should be emphasized that the underlying indices of the 3d gauge theory are independent of the RG flow --  this is the reason, why mirror symmetry, regarded as a map from a UV to the IR theory, works at all. The mirror map reformulates the UV index in terms of variables and boundary conditions adapted to the IR regime. Generally, there can be a mixing of global $U(1)$ currents and the $U(1)_R$ current along the flow.

In the following we study the integrality of the mirror map and its relation to 3d BPS invariants. In 2d, the mirror map from the algebraic coordinate $z$ to the flat K\"ahler coordinates $t$ near a large volume point has the simple form \cite{Candelas:1990rm,MB}\footnote{We restrict again to the one modulus case and use the example of the quintic in explicit formulas below.}
\be\label{2dmm}
t(z)=\frac{\omega_1(z)}{\omega_0(z)}=\ln (z)+770 z+717\,825 z^2+\frac{3\,225\,308\,000 z^3}{3}+\ldots\ ,
\ee
where $\omega_1\sim \ln(z) +\cx O(z)$ is the period with single logarithmic behavior and $\omega_0\sim 1+\cx O(z)$ is the fundamental period. The mirror map \eqref{2dmm}  is written in the formalism with a single set of parameters for $H^2(X)$, making use of the dependence of the 2d theory on $Qe^t$. In the 3d theory, one needs to keep both types of deformations $(Q,\tau)$; the algebraic coordinate $z$ in \eqref{2dmm} then corresponds to $e^\tau$. 

We now turn to the K-theoretic mirror map in the 3d theory. The flat coordinates~$t$ are defined in symmetric quantum K-theory by the expansion \eqref{Jexp} as
\cite{Giv15all}
\be\label{3dmm}
t(\tau,Q,q) = J(\tau,Q,q)|_{\cx K_+}-(1-q)\ .
\ee
Despite of this simple definition, the actual transformation from $(Q,\tau)$ to the flat parameters $t$ is $q$-dependent and complicated. In sect.~\ref{sec:EQK} we divided the computation of the K-theoretic $J$-functions into the steps:
$$
Z^X_{3d}(Q,q) \xrightarrow{\ \text{vortex sum}\ } J_K^{\text{sym}}(\delta t, Q, q)\xrightarrow{\ \text{mirror map}\ } J_K(0,Q,q) 
\longrightarrow J_K^{\text{ord,sym,eq}}(t,Q,q) 
$$
The  vortex sum $\s(Q,q)$ obtained from the 3d partition function gives a $J$-function of the symmetric quantum K-theory at non-zero perturbation $\delta t(Q, q)$. Note that the starting point  is the {\it  unperturbed} 3d UV partition function without extra massive modes,  which nevertheless acquires a non-zero input $\delta t(Q,q)$ in the IR, see eq.~\eqref{eq:inpJKQ}. The non-zero input arises from point-like solitons in the UV theory, which can be absorbed into a field redefinition after flowing to the IR theory \cite{MP,WitPhases,Losev:1999nt}, both in 3d and 2d. This field redefinition is described by the mirror map and corresponds to the next step from $J_K^\text{sym}(\delta t)$ to $J_K(0)$. The last step indicates that starting from the $J$-function with zero input $J_K(0)$ one may finally obtain the $J$-functions of the ordinary, symmetric, or equivariant theory by perturbing with the appropriate single- or multi-trace operators. 

The 2d mirror map \eqref{2dmm} involves two operations: the correct normalization of the basis elements, such that they are constant over the deformation space, and the choice of the flat coordiante $t$. The first is achieved by divison by $\omega_0$ and the flat coordinate is encoded in the period $\omega_1$. It is helpful to distinguish the two steps also in the K-theoretic mirror map
\be\label{3dmmsteps}
J_K^{\text{sym}}(\delta t, Q, q) \xrightarrow{\text{\ normalization\ }}  J_K^{\text{sym,norm}}(\delta t, Q, q) \xrightarrow{\text{\ flat coordinates\ }} J_K(0,Q,q) \ .
\ee
 The necessary basis change from the non-constant basis element $\tilde \Phi_0=\omega_0(Q,q)\Phi_0+\ldots$ to the constant unit $\Phi_0=1$ is easy to read off: it is the first $q$-period in \eqref{seps}
\bea\label{om0q}
\omega_0 &=& \s(Q,q,\eps=0)=\sum_{k=0}^\infty Q^k\frac{\prod_{n=1}^{5k}(1-q^{n})}{\prod_{n=1}^{k}(1-q^{n})^5} \nonumber\\
&=& \sum_{(k,r)\geq 0} N^{UV}_{n,r}Q^kq^r = 1+\sum_{k>0}Q^k p_k(q) \ .
\eea
The integral coefficients $N^{\text{UV}}_{k,r}$ of a term $Q^kq^r$ in an expansion of $\omega_0$ in a series in $Q$ and $q$ are the degeneracies of 3d BPS operators with vortex charge $k$ and spin $r$, in the sector with full Dirichlet boundary conditions. Note that each term of $\omega_0$ for fixed vortex number $k$ is a {\it polynomial} $p_k(q)$ in $q$, as indicated in \eqref{om0q}. From the point of the 3d field theory this means that the input arises from a finite number of unpaired fermionic modes. Dividing by $\omega_0$ to normalize the unit $\Phi_0$ is multiplication by a power series with integral coefficients determined by the BPS degeneracies.

In the 2d limit $q\to 1$,  $\omega_0$ reduces to the fundamental period (c.f., eq.~\eqref{tmpper}). Its integral coefficients are given by the 3d BPS degeneracies, summed over the spin quantum number $r$:
\be
\omega_0^{2d}= \sum_{k\geq 0}N^{\text{UV}}_kQ^k,\qquad N^{\text{UV}}_k=\sum_{r\geq 0}N^{\text{UV}}_{k,r}\ .
\ee
E.g., the first terms for the $q$-period of the quintic are
\be
\begin{aligned}
\omega_0 &= 1+Q\prod_{\ell=0}^4\sum_{r=0}^\ell q^r + Q^2\prod_{\ell=0}^4\left(\sum_{r=0}^\ell q^{2r}\right)\left(\sum_{r=0}^{2\ell} q^{r}\right) +\ldots \\
&\xrightarrow{\ q\to 1\ }  \omega_0^{2d}= 1 + 120 Q + 113\,400 Q^2+\ldots \ .
\end{aligned}
\ee
The full basis change $\tilde \Phi_n = U_{\alpha n}(Q,q)  \Phi_\alpha$ for the other elements can be found with the help of a Birkhoff factorization. This step has been described in detail in ref.~\cite{IMT}, where it has been used to normalize the operators in ordinary quantum K-theory. The same technique can be applied to the symmetric quantum K-theory to obtain the $J$-function $J^{\text{sym,norm}}_K(\delta t)$ in the constant basis $\{\Phi_\alpha\}$. As in the case of the coefficient of $\Phi_0$ spelled out above, the normalization process involves multiplication by power series with integral coefficients determined by the numbers of 3d BPS operators.

The second step in \eqref{3dmmsteps} describes the choice of flat coordinates. The map between $J^\text{sym,norm}_K(\delta t)$ and $J_K(0)$ is given by the transformation \eqref{eq:ReconCP2}, restricted to the totally symmetric representations. To compare with \eqref{2dmm}, we concentrate again on the dependence on the parameter $t=t_1$ associated to the deformation $\Phi_1=(1-P) = H+\cx O(H^2)$, with $H$ the hyperplane class. Since the correlator terms of the $J$-function at zero input start at order $H^2$, $J(0)=(1-q)+\cx O(H^2)$, the mirror map for $t_1$ is determined by the action of the operator on the classical term 
\be\label{t1tf}
e^{\sum_{k>0} \frac {\sum_{a=0}^3 \psi^k(\tilde \eps_a)(Pq^\theta)^{ka}}{k(1-q^k)}}\ \big( (1-q)+\cx O(H^2) \big)  = (1-q)\ e^{\sum_{k>0}\frac {\psi^k(\eps_0)}{k(1-q^k)}}\ 
(1- \Delta t \cdot H )+\cx O(H^2)\ . \nonumber
\ee
The exponential factor with argument $\eps_0 = \sum_a\tilde \eps_a$ is fixed by the normalization of $\Phi_0$, setting $\tilde \eps_0=0$. The coefficient of $H$  is 
\be\label{Deltat}
\Delta t = \sum_{k>0} \frac{\psi^k(\eps_1)}{1-q^k}\ ,\qquad \eps_1 = \sum_{a} a \tilde \eps_a\ .
\ee
In quantum K-theory, the terms with $k>1$ represent correlators, see eqs.~\eqref{Jexp},\eqref{Kdef}. The shift $\delta t_K$ of $t_1$ representing the 3d mirror map is therefore
\be
\delta t_K= (1-q)\Delta t|_{k=1}=\eps_1\ .
\ee
The shift $\delta t_K$ is fixed by requiring the $H^1$ term on the r.h.s. of \eqref{t1tf} to match the one in $J^\text{sym,norm}_K(\delta t)$ obtained from the 3d vortex sum, going backwards at the second arrow in \eqref{3dmmsteps}. Since the normalization procedure preserves integrality, this is an integral series 
\be \label{ktmmq}
\eps_1=-\sum_{r>0} n^{\text{UV}}_r Q^r\ .
\ee
The explicit coefficients for the quintic are 
$$
\begin{aligned}
  &n^{\text{UV}}_1 = 770 \ , \quad
  n^{\text{UV}}_2 = 717\,440\ , \quad
  n^{\text{UV}}_3 = 1\,075\,102\,410\ , \quad
  n^{\text{UV}}_4 = 1\,973\,656\,926\,400\ , \\
  &n^{\text{UV}}_5 = 4\,062\,154\,117\,561\,250\ , \quad
  n^{\text{UV}}_6 = 8\,998\,533\,447\,740\,749\,920\ , \quad \ldots \ .
\end{aligned}
$$
Now consider instead the small radius limit $q\to 1$ of  $\Delta t$, which should reproduce the 2d mirror map:
\be\label{Deltatlim}
\delta t_{GW}=\lim_{q\to 1} (1-q)\sum_k \frac{\psi^k(\eps_1)}{1-q^k} = \sum_{r>0} n^{UV}_r \ln(1-Q^r)\ .
\ee
The r.h.s. is the series part of the r.h.s. of eq.~\eqref{2dmm}. Adding the log term and exponentiating, the 2d mirror map is expressed in terms of the integers  $n^{\text{UV}}_r$ as
\be\label{mmexp}
e^{t}(Q)=Qe^{-\delta t_{GW}} = Q  \prod_{r>0}(1-Q^r)^{-n^{\text{UV}}_r}\ ,
\ee
The fact that the exponentiated mirror map \eqref{mmexp} has integral coefficients was observed long time ago and proven for the quintic in ref.~\cite{Lian} using $p$-adic methods. The new aspect of the 3d derivation of this fact is the connection \eqref{mmexp} of these  integral coefficients to 3d BPS degeneracies. The assumptions entering the above argument, and therefore BPS formula \eqref{mmexp}, hold also for Calabi--Yau $n$-folds with $n>3$.

\section{Outlook and open questions}\label{sec:outlook}
The correspondence between 3d gauge theory and permutation equivariant quantum K-theory proposed in this note raises a number of interesting questions, some of which have already been mentioned, for instance, a comparison of the gluing prescription for point vertices of ref.~\cite{Giv15all} with the gluing of the topological vertex, the description of a derived category of E-branes and its relation to elliptic cohomology, or the $tt^*$ geometry related to the $q$-period vector. 

Another important direction is the generalization of the correspondence to higher genus. On the side of quantum K-theory, a higher genus definition exist \cite{Giv15all}(p.IX). From the relation between the K-theoretic vertex and the topological vertex we have already noted that the 3d theory resums the genus expansion. For an illustration consider the case of a Calabi--Yau 3-fold $X$. The partition function in the large radius limit gives 
\be\label{ZCY3}
 Z_{S^1\times_q S^2}\sim\kappa \frac{(t+\bar t)^3}{3!\hbar^3}-\chi \, \ln(q)^3 q \frac d {dq} \ln M(q)+\beta^2\frac{c_2}{12\hbar}(t+\bar t)\ ,
\ee
where
$\kappa = \int_X H^3$, $\chi=\int_X c_3(X)$,  and $c_2=\int_X c_2(X) H$. The first term is the classical volume of the manifold. The second term reproduces in the 2d limit the perturbative correction to the K\"ahler potential \eqref{zeta3}, i.e., we obtain the large radius limit of the genus zero K\"ahler potential of topological string theory on $X$. 

Interestingly, the 3d corrections to the 2d limit are related to known higher genus quantities of the topological string on $X$. The MacMahon function $M(q)$ is known to compute the all genus contribution of the constant maps to the topological string \cite{GV98}
\be
\sum_{g=0}^\infty  \left.F_g\gstring^{2g-2}\right|_\textrm{const maps}=\frac{\chi}2 \ln M(q)=-\frac{\chi}2  \frac{\zeta(3)}{\gstring^{2}}+O(\gstring^0)\ ,
\ee
where $\gstring$ is the string coupling and $q=e^{i\gstring}$. Matching the $q$ parameters of the 3d theory and the topological string gives the identification
\be\label{qlambda}
\gstring =i\beta \hbar \quad  \xrightarrow{\ \hbar = -2\pi i\ } \quad \gstring= \frac{\beta}{2\pi}\ .
\ee
The special choice for $\hbar$ made in the second step is the natural value in the $A$-model.\footnote{See sect.~10 of ref.~\cite{CK}.} With this identification the linear term in eq.~\eqref{ZCY3} corresponds to the string 1-loop term, which is indeed the only $t$-dependent term present at large volume.

The particular combination of higher genus terms in eq.~\eqref{ZCY3} can be obtained in ${\cal N}=2$ 4d supergravity from the standard relation 
\be \label{Ksugra}
-i e^{-{\cal K}}=X^A\bar \cF_A-\bar X^A \cF_A\ .
\ee
Here $X^A=(X^0,X^0t)$ are the homogeneous variables for the one modulus case, $\cF_A=\p \cF/\p X^A$. If one uses  the all genus prepotential
\be
\cF_\text{top}=(X^0)^2\left(\kappa \frac{t^3}{3!} -\frac{c_2}{24}t\right)+\frac{\chi}2\ln M(q)+\frac{c_2}{24}t\ ,
\ee
for constant maps in \eqref{Ksugra}, and identifies $X^0=1/\gstring$, one obtains
\be\label{DefK3d}
Z_{S^2\times S^1}=e^{-{\cal K}(\beta)}\ .
\ee
The K\"ahler potential ${\cal K}(\beta)$ defined in this way depends on the radius, or the string coupling. The identification \eqref{qlambda} between the $S^1$ radius and the string coupling constant $\gstring$ on the world-sheet is reminiscent of a similar relation in M-theory in  target space \cite{WittenM}. Prepotentials including higher genus and non-perturbative corrections play an important role in the study of black holes, see e.g. refs.~\cite{Ooguri:2004zv,LopesCardoso:1999fsj}.

\bigskip
\bigskip

\noindent{\bf Acknowledgments:} We would like to thank 
Ilka Brunner,
Tudor Dimofte,
Sergey Galkin,
Andreas Gerhardus,
Ken Intriligator,
Urmi Ninad,
Alexander Tabler
and Martin Vogrin
for discussions and correspondences. 
The work of P.M. is supported by the Deutsche Forschungsgemeinschaft and the German Excellence Initiative. 
\newpage\appendix 
\section{Appendix}

\subsection{3d partition functions for 3d GLSM \label{app:3dpf}} \label{app:3dpar}
\def\X{\Upsilon}\def\tD{\tilde D}\def\tN{\tilde N}\def\kh{{\hat k}}\def\nh{{\hat n}}\def\x{u}\def\gq{{f}}
In this section we give some details on the computation of the 3d partition functions used in sect.~\ref{sec:qdiffsys}.\\

\subsubsecb{$U(1)$ partition function on $S^1\times_q S^2$}
First consider the $S^1 \times_q S^2$ partition function for a hypersurface in a weighted projective space $W\IP^{N}$. This is a $U(1)$ theory with $N+1$ fields $\varphi_\alpha$ of $U(1)$ charges $q_\alpha$ and $R$-charges $\Delta_\alpha$. Here $\alpha=0$ refers to the field of negative charge equal to the degree of the hypersurface constraint and with $R$-charge $\Delta_0=2-\sum_{\alpha>0}\Delta_\alpha$. The fields $\alpha>0$ represent the homogeneous coordinates on $W\IP^{N}$ of weights $q_\alpha$. A canonical choice  for the $R$-charges for the compact case is 
\be\label{canRch}
\Delta_0=2\ , \qquad \Delta_{\alpha>0}=0\, .
\ee
 
\noindent 
The 1-loop determinant for a chiral field $\varphi_\alpha$ with charges $(q_\alpha,\Delta_\alpha,f_{\alpha r})$ under gauge, $R$- and global symmetries is \cite{Imamura:2011su,Dimofte:2011py,Kapustin:2011jm,Benini:2013yva} 
\be
Z_\phi=(q^\frac{1-\Delta_\alpha}{2}z^{-q_\alpha}y_r^{-f_{\alpha r}})^{-mq_\alpha/2}\frac{(z^{-q_\alpha}q^{-mq_\alpha/2+1-\Delta_\alpha/2}y_r^{-f_{\alpha r}},q)_\infty}{(z^{q_\alpha}q^{-mq_\alpha/2+\Delta_\alpha/2}y_r^{f_{\alpha r}},q)_\infty}\ ,
\ee
where $z=e^{ih}$ a $U(1)$ Wilson line on $S^1$, $m$ the magnetic flux on $S^2$ and $(x,q)_\infty$ the $q$-Pochhammer symbol. $q$ is the chemical potential for combined $U(1)_R$ and $S^2$ rotations and $y_r$ are chemical potentials for the global symmetries.\footnote{Some of these correspond to the toric ${T}^N$ action of equivariant quantum K-theory/quantum cohomology theory.} We mostly set $y_r=1$ in the following and restore the dependence on the $y_r$ by an appropriate shift of $h$ when needed.
Then the  poles are at
$$
z^{q_\alpha}=q^{mq_\alpha/2-\Delta_\alpha/2-k+\epsilon}\ ,\qquad k\geq mq_\alpha\ .
$$
There is always a field $\varphi_*$ of minimal $U(1)$ charge one. The factor $Z_{\varphi_\alpha}$ at the poles from the field $\varphi_*$ with charges $(q_*=1,\Delta_*)$ can be witten in terms of $q$-Gamma functions as 
\be
Z_{\varphi_\alpha}=q^{r_\alpha}(1-q)^{s_\alpha}\frac{\Gamma_q(\x_\alpha(k))}{\Gamma_q(1-\x_\alpha(n))}\ ,
\ee
where  $n=k-m$ and the arguments and exponents are
\begin{align}\label{exponents}
\x_\alpha(k)&=\mu_\alpha-kq_\alpha\ , 
&\mu_\alpha&=q_\alpha\left(\epsilon-\frac{\Delta_*}{2}\right)+\frac{\Delta_\alpha}{2}\ ,\\
s_\alpha&=\x_\alpha(k)+\x_\alpha(n)-1\ ,
&r_\alpha&=\frac 1 4 s_\alpha \left(\x_\alpha(n)-\x_\alpha(k)\right)\ .
\end{align}
Using the identity 
$$
\Gamma_q(x-k)\Gamma_q(1-x+k)=(-)^kq^{\frac{k(k+1)} 2-kx}\Gamma_q(x)\Gamma_q(1-x)
$$
the product of the contributions of all fields can be recast in the form
$$
\prod_\alpha Z_{\varphi_\alpha}=\tilde \Upsilon\cdot  \Omega_{k,q}\cdot \Omega_{n,\bar q}
$$
with\\[-8mm]
\begin{small}
\begin{eqnarray*}
\Omega_{k,q}&=&(-)^{k(c_1+q_0)}(1-q)^{-c_1\kh}q^{\frac 1 4 a(k)}\frac{\Gamma_q(\x_0(k))}{\prod_{\alpha>0}\Gamma_q(1-\x_\alpha(k))}\, \\
a(k)&=&\sum_{\alpha>0} (\x_\alpha(k)^2-\x_\alpha(k))-(\x_0(k)^2-\x_0(k))\ ,\\
\tilde \Upsilon&=&\frac{\prod_{\alpha>0}\Gamma_q(\mu_\alpha)\Gamma_q(1-\mu_\alpha)}{\Gamma_q(\mu_0)\Gamma_q(1-\mu_0)}\cdot (1-q)^{-N+1}q^{-\frac 1 2 a(0)}(-)^{c_1\hat\eps}  \ .
\end{eqnarray*}
\end{small}
\hskip-4pt Here $\bar q = q^{-1}$, $c_1=\sum_\alpha q_\alpha$, $\kh = k-\hat \eps$, $\hat\eps=\eps-\Delta_*/2$.\\

\noindent
The classical action gets a contribution from FI-terms and CS terms 
\begin{eqnarray*}
e^{-S_{class}(z,m)}&=&e^{4\pi\xi ih/\ln q + im\theta} \  e^{\kappa_g ihm}\ e^{\kappa_Rm\beta\hbar/2 }\\
&=&Q^{\kh}\bar Q^{\nh}q^{-\frac{\kappa_g}{2}(\kh^2-\hat n^2)}
q^{-\frac{\kappa_R}{2}(\kh-\hat n)}\ ,\nonumber
\end{eqnarray*}
where $\kappa_g$ and $\kappa_R$ are the CS coefficients for the gauged $U(1)$  and mixed gauge/$U(1)_R$ CS terms and 
$$
Q=e^{-2\pi \xi + i\theta}\ .
$$
Collecting the $k$-dependent terms and summing over $k$ one gets
\be
\s(Q,q,\eps)=\sum_{k=0}^\infty \big(\frac Q {(1-q)^{c_1}})^{\kh}(-)^{k(c_1+q_0)}q^{d(k,\eps)}
\frac{\Gamma_q(\x_0(k))}{\Gamma_q(\mu_0)}\prod_{\alpha>0}\frac{\Gamma_q(1-\mu_\alpha)}{\Gamma_q(1-\x_\alpha(k))} \ ,
\ee
where we have included a constant normalization factor such that $\s(Q,q,0)=1+\cx O(Q)$. The exponent of the $q$ factor is 
$$
d(k,\eps)=\kh^2(\frac{t_2}{2}-\frac{\kappa_g}{2})+\kh(\frac{t_1}4-\frac{\kappa_R}2)\ ,
$$
where 
\be\label{tdef}
t_2=\frac 12 \sum_{\alpha}\sigma_\alpha q^2_\alpha\, ,\qquad 
t_1=\sum_{\alpha}\sigma_\alpha q_\alpha(1-\Delta_\alpha)\ ,
\ee
and $\sigma_\alpha=+1$ (-1) for $\alpha>0$ ($\alpha=0$). The coefficient $t_2$ is the numerical coefficient of the second Chern character of $X$ and moreover $t_1$ agrees with that of the  first Chern class for the canonical choice of $R$ charges \eqref{canRch}. The exponent vanishes for the special choice of CS terms
$$
\kappa_g=t_2\ ,\qquad \kappa_R = t_1/2 \quad \Rightarrow \quad d(k,\eps)=0\ .
$$
Similarly the $n$-dependent terms give the series $\s(\bar Q,\bar q,\eps)$. The remaining $k,n$-independent terms can be collected into the folding factor
\be
\FFs = \frac{1-\mu_0}{\prod_{\alpha>0}\mu_\alpha}\ \td_\beta (X) \hGq/{\hGqb}\ ,
\ee
where 
\be
\hGq = \frac{\prod_{\alpha>0}\Gamma_q(1+\mu_\alpha)}{\Gamma_q(1+(1-\mu_0))}\ ,\qquad 
{\hGqb}= \frac{\prod_{\alpha>0}\Gamma_q(1-\mu_\alpha)}{\Gamma_q(1-(1-\mu_0))}\ ,
\ee
and
\be
\td_\beta(X) = \frac{1-q^{-(1-\mu_0)}}{1-\mu_0}\ \prod_{\alpha>0}\frac{\mu_\alpha}{1-q^{-{\mu_\alpha}}} \ .
\ee
There is an involution symmetry $\hGq \leftrightarrow {\hGqb}$ generated by a sign flip of $\eps$ and $\Delta_{\alpha>0}$. 

Collecting all the terms above, the partition functions is.
\be
Z_{S^1\times_q S^2}=\ln(q)\ \oint_0 \frac{d\eps}{2\pi i } \s(\bar Q,\bar q,\eps)\, \FFs\, \s(Q,q,\eps)\ .
\ee 
 For the canonical choice of $R$ charges, 
\be
\mu_{\alpha>0}= \eps q_\alpha\ ,\qquad 
1-\mu_0=-q_0\eps\ ,
\ee 
and we obtain the expressions \eqref{DefOmq},\eqref{FFEx}.\\

\subsubsecb{Partition function on $S^1\times_q D^2$} 
To describe a complete intersection hypersurface $X$ in a toric variety, we consider a $U(1)^n$ gauge theory with $\tN$ chiral fields with Neumann boundary conditions and $\tD$ chiral fields with Dirichlet boundary condition, as defined in \cite{YS}. The first and second Chern characters of $X$ are determined by the charges $q_\alpha^i$ of the fields as 
\bea
c_1(J)\!\!&=c_iJ_i\ ,\qquad \ \ \ c_i &= \sum_{\alpha\in N,D} q_\alpha^i\ ,\quad  i=1,...,n \ ,\\
ch_2(J)&=c_{ij}J_iJ_j\ , \qquad c_{ij} &= \frac12 \left(\sum_{\alpha\in N}-\sum_{\alpha\in D}\right) q_\alpha^iq^j_{\alpha}\ ,
\eea
where ${J_i}$ denotes a basis for $H^2(X,\mathbb{Z})$ in the K\"ahler cone. A field $\varphi_\alpha$ with Dirichlet boundary conditions  and negative $U(1)^n$ charges $q_\alpha^i$, $i=1,...,n$ implements a hypersurface constraint of degree $| q_\alpha^i|$. 

The 1-loop determinants for a field  of charges $(q_\alpha,\Delta_\alpha,\gq_{\alpha r})$ with N(eumann) boundary conditions are \cite{YS}
\be\label{zn}
Z^N_{\varphi_\alpha} = \frac{q^{-\frac 1 {24} -\frac14((\x'_\alpha)^2-\x'_\alpha)}}{(z^{q_\alpha}q^{\Delta_\alpha/2}y_r^{\gq_{\alpha r}},q)_\infty},
\ee
where $z_\alpha^{q_\alpha}=\prod_{i=1}^nz_{i\alpha}^{q_\alpha^i}$ and $\x'_\alpha$ is the $q$-exponent of the argument of the $q$-Pochhammer symbol 
$$
q^{\x'_\alpha}=z^{q_\alpha}q^{\Delta_\alpha/2}y_r^{\gq_{\alpha r}}\, .
$$
For fields with D(irichlet) boundary\footnote{In the notation of \cite{YS}, we have set $2\beta=2\beta_2=(\beta\hbar)_{\textrm here}$. In the anomalous term in their eq.~(4.13) there appears to be a typo: the terms $\sim a$ and $\sim M_l$ should have the reversed sign and this was used here.}
\bea\label{zd}
Z^D_{\varphi_\alpha} &=& q^{\frac 1 {24} +\frac14((\x'_\alpha)^2-\x'_\alpha)} \, (z^{-q_\alpha}q^{1-\Delta_\alpha/2}y_r^{-\gq_{\alpha r}},q)_\infty\ ,\nonumber \\
q^{\x'_\alpha}&=&z^{-q_\alpha}q^{1-\Delta_\alpha/2}y_r^{-\gq_{\alpha r}}.
\eea
The following computation is similar to the previous one up to small modifications. To simplify the exposition, we set $y_r=1$ and assume a canonical choice for the $R$-charges 
\be\label{Rchcan}
\Delta_{\alpha}=0(2) \textrm{ for } N(D)\ .
\ee
In a large volume phase we pick an integration contour that sums up the poles at 
\be
z^{q_{\alpha}}=q^{-\kh_\alpha},\qquad \kh_\alpha = k_\alpha-\eps_\alpha = \sum_iq_\alpha^i(k_i-\eps_i)\, , \quad 0\leq k_i \in \mathbb{Z}\ .
\ee
The partition function then takes the form 
\be\label{Zgen}
Z_{S^1\times_q D^2} = (\ln q)^n \, \int \prod_{i=1}^n \frac{d\eps_i}{2\pi i}\,  \FFd \cdot \s(Q,q,\eps)\ .
\ee
The holomorphic series is
\be\label{OmD2}
\s(Q,q,\eps)=\sum_{k_i=0}^\infty (\prod_i \big(\frac{Q_i}{(1-q)^{c_i}}\big)^{\kh_i})(-)^{c_1(k)}q^{d(k,\eps)}\ 
\prod_{\alpha \in D}\frac{\Gamma_q(1-\kh_\alpha)}{\Gamma_q(1+\eps_\alpha)}\
\prod_{\alpha\in N}\frac{\Gamma_q(1-\eps_\alpha)}{\Gamma_q(1+\kh_\alpha)}\ .
\ee
Here 
\be\label{dgen}
d(k,\eps)=\frac12(\kh_i\kh_j(c_{ij}-\kappa_{ij})+\kh_i(\frac12 c_i-\kappa_i))\ ,
\ee
where $\kappa_{ij}$ and $\kappa_i$ are the CS couplings for $U(1)_i\times U(1)_j$ and $U(1)_i\times U(1)_R$. The folding factor is 
\be \label{ResD2}
\FFd=(-\eta(q))^{\tD-\tN}\, q^{-ch_2(\eps)-\frac12c_1(\eps)}\  \hGq \  \td_\beta X \, \frac{\prod_D (-\beta\eps_\alpha)}{\prod_N \beta\eps_\alpha}\ ,
\ee
where $\eta(q)=q^{1/24}\prod_{r=1}^\infty(1-q^r)$, the $q$-Gamma class of $X$ is
\be
\hGq = \frac{\prod_{N}\Gamma_q(1+\eps_\alpha)}{\prod_D\Gamma_q(1-\eps_\alpha))},
\ee 
and the $\beta$-dependent Todd class is
\be
\td_\beta(X) = \prod_{N}\frac{\beta\eps_\alpha}{1-q^{-{\eps_\alpha}}}\ \prod_D\frac{1-q^{\eps_\alpha}}{-\eps_\alpha\beta}\ .
\ee

\subsection{Some $q$-functions \label{app:qfunc}}
For the convenience of the reader we collect in this appendix some definitions and a few formulas on $q$-functions used in the computations. A good reference for background material is ref.~\cite{Rahman}.\\

\subsubsec{$q$-Pochhammer}
Assuming $|q|<1$, we have the identities among $q$-Pochhammer symbols
\bea
(x,q)_k &=& \prod_{n=0}^{k-1}(1-xq^n) \ , \qquad (q)_k = (q,q)_k = \prod_{n=1}^k (1-q^n) \ , \\
\label{qpsum}
(x,q)_\infty &=&\sum_{k=0}^\infty \frac{(-x)^kq^{k(k-1)/2}}{(q)_k}=\prod_{l=0}^\infty(1-xq^l) \ , \\
\label{qpoch}\frac{1}{(x,q)_\infty}&=&\exp\left(\sum_{k=1}^{+\infty} \frac{x^k}{k(1-q^k)}\right)
= \sum_{k=0}^\infty \frac{x^k}{(q)_k}=\frac{1}{\prod_{l=0}^\infty(1-xq^l)} \ . 
\eea
In terms of the logarithmic derivative $\theta=x\partial_x$, we find 
\be\label{phdeq}
\begin{aligned}
(1-yq^{\theta-a})\, \left[\frac{y^{-\frac{\ln x}{\ln q}} x^a}{(x,q)_\infty}\right]&=\frac{y^{-\frac{\ln x}{\ln q}} x^{a+1}}{(x,q)_\infty}\ ,\\
(1-yq^{-\theta+a}) \left[y^{+\frac{\ln x}{\ln q}}x^a (qx,q)_\infty\right]&=y^{+\frac{\ln x}{\ln q}}x^{a+1} (qx,q)_\infty\ . 
\end{aligned}
\ee\\

\subsubsec{$q$-Gamma}
For $q<1$, 
\be\label{defgammaq}
\Gamma_q(x)=\frac{(q,q)_\infty}{(q^x,q)_\infty}(1-q)^{1-x}\ .
\ee
and for $q>1$
\be
\Gamma_q(x)=\Gamma_{\bx q}(x)q^{(x-2)(x-1)/2}\ ,
\ee
where $\bx q = q^{-1}$. One has
\bea\label{qGammaid}
\Gamma_q(x+1)&=&\frac{1-q^x}{1-q}\Gamma_q(x) \ ,\\
\Gamma_q(1+x)\Gamma_q(1-x)&=&
\prod_{n=1}^\infty \frac{(1-q^n)^2}{(1-q^ne^y)(1-q^ne^{-y})}=e^{2\sum_{k=1}^\infty \frac{y^{2k}}{2k!}G_{2k}^0(q)},
\eea
with $y=x \ln q$.\\

\subsubsec{Theta-functions }
\bea
\label{theta}\theta(x,q)&=&(x,q)_\infty(q/x,q)_\infty\, 
\eea

\subsubsec{$q$-Polygamma function}
The $q$-Polygamma function
\be
\psi_q(k,x)=\frac{d^{k+1}}{dx^{k+1}}\ln \Gamma_q(x)\ ,
\ee
has for $x=1$ the small $q$ expansion 
\be
\psi_q(k,1)=\ln(q)^{k+1}d_k(q)- \delta_{k,0}\ln(1-q) \ ,
\ee
where 
\be
d_k(q)=\sum_{n=1}^\infty \sigma_{k}(n)q^n,\qquad \sigma_k(n)= \sum_{d|n}d^k\ .
\ee
For all $k$ one notices  the infinite product formula 
\be
d_k(q)=\sum_{n=1}^\infty \frac{n^k q^n}{1-q^n}=q\frac{d}{dq}\ln\left(\prod_{n=1}^{+\infty}(1-q^n)^{-n^{k-1}}\right) \ .
\ee
For $k=1,2$, the argument of the logarithm on the r.h.s. is the counting functions of 2d- and 3d-partitions 
\be\label{MacM}
k=1: \  \prod_{n=1}^{+\infty}(1-q^n)^{-1} = \frac{q^{1/24}}{\eta(q)}\ ,\qquad
k=2: \  \prod_{n=1}^{+\infty}(1-q^n)^{-n}= M(q) \ ,
\ee 
where $M(q)$ denotes the MacMahon function.\\

\subsubsec{Eisenstein series}
\be \label{DefGk}
G_{2k}(\tau)=\gamma_{2k}+\sum_{n=1}^\infty \sigma_{2k-1}(n)q^n,\qquad \gamma_{2k}=-\frac{B_{2k}}{4k}.
\ee
Here $q=e^{2\pi i \tau}$ and $B_{2k}$ are the Bernoulli numbers. The modular transformation of $G_{2k}$ is
\be
G_{2k}(\frac{a\tau+b}{c\tau+d})=(c\tau+d)^{2k} G_{2k}(\tau)-\delta_{k,1}\frac{c(c\tau+d)}{4\pi i}\  .
\ee
The $q$-Polygamma functions for odd $k$ are almost modular:
\be\label{evenpg}
\psi_q(2k-1,1) =\ln(q)^{2k}G_{2k}^0(\tau)\ .
\ee
where $G_{2k}^0(\tau)=G_{2k}(\tau)-\gamma_{2k}$ are the Eisenstein series with constant term $\gamma_{2k}$ removed. The constant terms are related to the characteristic function of the A-roof genus as 
\be
\hat A(x)=\frac{(x/2)}{\sinh(x/2)}=\exp\left[{2\sum_{k=1}^\infty \frac{x^{2k}}{2k!}\gamma_{2k}}\right]\ .
\ee

\subsubsec{$q$-Gamma genus}
The expansion of the logarithm of $q$-Gamma is, with $y = x/\ln q$:
\bea\label{lngammaq}
\ln \Gamma_q(1+y) &=& \sum_{k=1}^\infty \frac{y^{k}}{k!}\psi_q(k-1,1) =L_e + L_o\ ,\nonumber\\
L_e&=&\sum_{k=1}^\infty \frac{G_{2k}^0}{2k!}x^{2k}\ ,\\
L_o&=&-x\frac{\ln(1-q)}{\ln q}+
\sum_{k=0}^\infty \frac{d_{2k}}{(2k+1)!}x^{2k+1}\ .\nonumber
\eea

\subsection{Solutions to $q$-difference system \label{app:sol}}
The difference equation \eqref{DiffEq1} for the degree $N$ hypersurface $X$ in $\mathbb{P}^{N-1}$ can be factorized as $\DL =(1-q^\theta)\DL_X$
with 
\be\label{diffeqfac}
\DL_X = (1-q^\theta)^{N-1}-Q\sum_{j=0}^{N-1} q^{(\theta+1)j}\prod_{j=1}^{N-1}(1-q^{N\theta + j})\ .
\ee
Around the large volume point $Q=0$, a basis of $N-1$ solutions is given by the first $N-1$ coeffients $\omega_i$ of the $\eps$-expansion of the vortex sum 
\bea
\s(Q,q,\eps)&=&c(\eps)\sum_{k=0}^\infty Q^{k-\eps} \frac{\Gamma_q(1+N(k-\eps))}
{\Gamma_q(1+k-\eps)^N}=\sum_{i=0}^{N-2}\omega_i(Q,q)(-\eps)^i\ .\nonumber
\eea
The leading terms in $Q$ are
\bea
 (\omega_i^{LV})_{c(\eps)=\hGq^*} &=& \frac{1}{i!}(\ln Q)^i\\
(\omega_i^{LV})_{c(\eps)=1} &=&\begin{pmatrix}
1 \\ \ln Q  \\ \frac 1 2 (\ln Q)^2 +c_2\psi_1\\ 
\frac 1 {3!} (\ln Q)^3 +\ln Q c_2\psi_1-\frac 12 c_3 \psi_2\\ 
\frac 1 {4!} (\ln Q)^4+\frac 12 (\ln Q)^2 c_2\psi_1-\frac 12 \ln Q c_3 \psi_2+\frac 1 {12}(6c_2^2\psi_1^2+(2c_4-c_2^2))\psi_3\\... 
\end{pmatrix} ,\nonumber
\eea
where  it is understood that  $\omega^{LV}_i$ is set to zero if $i>N-2=\textrm{dim}X$. The subleading terms in the last expansion arise from the series expansion of the $q$-Gamma class $1/\hGq^*(X)$.  Here $c_k$ denotes the numerical coefficient of the $k$-th Chern class of $X$ and $\psi_k=\psi_q(k,1)$. The series expansion in $z$ has the general form
\bea\label{LVPV}
(\omega^{LV}) &=& s_0\ \begin{pmatrix}
1 \\
\ell +s_1\\
\frac 12 \ell^2+s_2\\
\frac 1{3!}\ell^3+s_2\ell+s_3\\
\frac 1{4!}{\ell^4}+\frac 12 s_2\ell^2+s_3\ell +s_4 \\
\vdots
\end{pmatrix}\, \qquad \ell = \ln Q +s_1\ ,
\eea
where 
\be
s_i(Q,q) =(\ln q)^i \tilde s_i(Q,q)\ ,
\ee
with $\tilde s_i(Q,q)$ a power series in both $Q$ and $q$, starting at $\cx O(Q^1)$ for $i>0$. 

Near the Landau-Ginzburg point $1/Q=0$, the natural variable is $\psi=Q^{-1/N}$. 
A series solution of eq.~\eqref{diffeqfac}  is given by 
\be\label{solLG}
\omega^{LG}_0(\psi)= \sum_{k=1}^\infty \psi^{k}\frac{(-)^kq^{k(k-1)/2}}{\Gamma_q(k)\Gamma_q(1-\frac k N)^N}\ .
\ee
A basis of $N-1$ linearly dependent solutions is provided by the $q$-periods
\be\label{solLGg}
\omega^{LG}_k(\psi)  = \omega^{LG}_0(\psi\eta^k) \  , \qquad k=0,...,N-2\ ,\qquad \eta = e^{-2\pi i k/N}\ .
\ee
The series $\omega^{LG}_{N-1}(\psi)$ is also a solution, but linearly dependent:
\be
\sum_{k=0}^{N-1} \omega^{LG}_{k}(\psi)=0\ .
\ee

\section{More invariants} \label{app:Inv}
\subsection{One modulus Calabi--Yau 3-folds \label{app:1mcy}}
Below we give the results for the $n$-point functions at low $n$
for the one moduli Calabi--Yau 3-folds $X$ in eq.~\eqref{3f1m}. The computation has been described  in sect.~\ref{sec:EQK} for the quintic. To display the general structure of the quantum K-theory invariants, and to save some space, we express these invariants in terms of the integral Gopakumar--Vafa  invariants $\ngv_k$ of $X$.\footnote{The explicit numbers $\ngv_k$ can be found in the tables of ref.~\cite{KT}.} For the $r+1$ point functions we write
\be
\left\langle\frac{\Phi_\alpha}{1-qL}; \Phi_1^r\right\rangle_{0,r+1} = \begin{cases}
0&\alpha=2,3 \ , \\
\frac{1}{1-q}\sum_kQ^k f^{(r)}_{\alpha,k}&\alpha=0,1\ ,\end{cases}
\ee
where the functions $f^{(r)}_{\alpha,k}$ at degree $k$ depend on $\ngv_{n\leq k}$.\\

\noindent For the 1-pt functions we find, supressing the superscript on $f$
\bea
&&f_{0,1}=\frac{\ngv_1 (3 q-1)}{q-1}\ ,\quad f_{0,2}=\frac{\ngv_1 \left(-3 q^4+9
   q^2-4\right)}{(q-1) (q+1)^3}+\frac{\ngv_2 (3 q-1)}{q-1}\ ,\nonumber\\
&&f_{0,3}=\frac{\ngv_1 \left(-8
   q^6+19 q^3-9\right)}{(q-1) \left(q^2+q+1\right)^3}+\frac{\ngv_3 (3
   q-1)}{q-1}\ ,\nonumber\\
&&f_{0,4}=\frac{\ngv_2 \left(-3 q^4+9 q^2-4\right)}{(q-1)
   (q+1)^3}+\frac{\ngv_1 \left(-15 q^8+33 q^4-16\right)}{(q-1) (q+1)^3
   \left(q^2+1\right)^3}+\frac{\ngv_4 (3 q-1)}{q-1}\ ,\nonumber\\
&&f_{1,1}=\ngv_1\ ,\quad f_{1,2}=\frac{\ngv_1 \left(2-q^2\right)}{(q+1)^2}+2
   \ngv_2\ ,\quad f_{1,3}=\frac{\ngv_1 \left(3-2 q^3\right)}{\left(q^2+q+1\right)^2}+3
   \ngv_3\ ,\nonumber\\
&&f_{1,4}=-\frac{2 \ngv_2 \left(q^2-2\right)}{(q+1)^2}+\frac{\ngv_1
   \left(4-3 q^4\right)}{(q+1)^2 \left(q^2+1\right)^2}+4 \ngv_4\ .\nonumber
\eea

\noindent For the 2-pt functions 
\bea
&&f_{0,1}=\frac{\ngv_{1} (2 q-1) \tableau{1}}{q-1},\quad f_{0,2}=\frac{\ngv_{1} \left(3
   q^2-2\right) \tableau{1}}{(q-1) (q+1)^2}+\frac{2 \ngv_2 (2 q-1)
   \tableau{1}}{q-1},\nonumber \\
&&f_{0,3}=\frac{\ngv_{1} (q-1) \left(4 q^3-3\right)
   \tableau{1}}{\left(q^3-1\right)^2}+\frac{3 \ngv_3 (2 q-1)
   \tableau{1}}{q-1},\nonumber \\
&&f_{0,4}=\frac{\ngv_{1} (q-1) \left(5 q^4-4\right)
   \tableau{1}}{\left(q^4-1\right)^2}+\frac{2 \ngv_2 \left(3 q^2-2\right)
   \tableau{1}}{(q-1) (q+1)^2}+\frac{4 \ngv_4 (2 q-1)
   \tableau{1}}{q-1}\nonumber\\
&&f_{1,1}=\ngv_{1} \tableau{1},\quad f_{1,2}=\frac{\ngv_{1} \tableau{1}}{q+1}+4
   \ngv_2 \tableau{1},\quad f_{1,3}=\frac{\ngv_{1} \tableau{1}}{q^2+q+1}+9
   \ngv_3 \tableau{1},\nonumber\\
&&f_{1,4}=\frac{\ngv_{1} \tableau{1}}{q^3+q^2+q+1}+\frac{4
   \ngv_2 \tableau{1}}{q+1}+16 \ngv_4 \tableau{1}\ .\nonumber
\eea

\noindent The functions for $r=2$ are
\bea
&&f_{0,1}=\frac{\ngv_{1} \left(\left(q^2+q-1\right) \tableau{2}-q
   \tableau{1 1}\right)}{q^2-1} 
\nonumber\\
&&f_{0,2}=\frac{\ngv_{2} \left(\left(q^2-2 q-1\right)
   \tableau{1 1}+\left(3 q^2+2 q-3\right)
   \tableau{2}\right)}{q^2-1}\nonumber\\&&\hskip1.3cm +\frac{\ngv_{1} \left(\left(q^4+q^3-4 q^2-q+2\right)
   \tableau{1 1}+\left(q^3+4 q^2-q-3\right) \tableau{2}\right)}{(q-1)
   (q+1)^3}\nonumber\\
&&f_{0,3}=\frac{3 \ngv_{3} \left(\left(q^2-q-1\right) \tableau{1 1}+\left(2
   q^2+q-2\right) \tableau{2}\right)}{q^2-1}\nonumber\\&&\hskip1.3cm +\frac{\ngv_{1} \left(\left(q^5+q^4-3
   q^3-q^2-q+2\right) \tableau{1 1}+\left(q^6+q^5+q^4+3 q^3-q^2-q-3\right)
   \tableau{2}\right)}{\left(q^2-1\right) \left(q^2+q+1\right)^2}\nonumber\\
&&f_{0,4}=\frac{2 \ngv_{4}
   \left(\left(3 q^2-2 q-3\right) \tableau{1 1}+\left(5 q^2+2 q-5\right)
   \tableau{2}\right)}{q^2-1}+\nonumber\\&&\hskip1.3cm \frac{\ngv_{2} \left(\left(3 q^4+4 q^3-8 q^2-4
   q+3\right) \tableau{1 1}+\left(q^4+4 q^3+8 q^2-4 q-7\right)
   \tableau{2}\right)}{(q-1) (q+1)^3}\nonumber\\&&\hskip0.6cm +\frac{\ngv_{1} \left(\left(q^8+q^7+q^6+q^5-6
   q^4-q^3-q^2-q+4\right) \tableau{1 1}+\left(q^7+q^6+q^5+6 q^4-q^3-q^2-q-5\right)
   \tableau{2}\right)}{(q-1) (q+1)^3 \left(q^2+1\right)^2}\ .\nonumber
\eea
\bea
&&f_{1,1}=\ngv_{1} \tableau{2},\quad f_{1,2}=\frac{\ngv_{1} \left(\left(q^2+q-1\right)
   \tableau{1 1}+(q+2) \tableau{2}\right)}{(q+1)^2}+2 \ngv_{2}
   (\tableau{1 1}+3 \tableau{2}),\nonumber\\
&&f_{1,3}=\frac{\ngv_{1} \left(q
   \tableau{1 1}+\left(q^2+1\right) \tableau{2}\right)}{q^2+q+1}+9 \ngv_{3}
   (\tableau{1 1}+2 \tableau{2}),\nonumber\\
&&f_{1,4}=\frac{2 \ngv_{2} \left(\left(3 q^2+4
   q-1\right) \tableau{1 1}+\left(q^2+4 q+5\right)
   \tableau{2}\right)}{(q+1)^2}\nonumber\\
&&\hskip1.3cm
+\frac{\ngv_{1} \left(\left(q^4+q^3+q^2+q-1\right)
   \tableau{1 1}+\left(q^3+q^2+q+2\right) \tableau{2}\right)}{(q+1)^2
   \left(q^2+1\right)}+8 \ngv_{4} (3 \tableau{1 1}+5 \tableau{2})
\nonumber
\eea

\noindent The functions for $r=3$ are, restricting to the simpler case with insertion $\Phi_1$ in the first slot
\begin{small}
\bea
&&f_{1,1}=\ngv_{1} \tableau{3},\quad
f_{1,2}=\ngv_{1} \left(\frac{q
   \tableau{2 1}}{q+1}+\frac{2
   \tableau{3}}{q+1}\right)+\ngv_{2} \left(4
   \tableau{2 1}+8 \tableau{3}\right),\nonumber\\
&&f_{1,3}=\ngv_{1}
   \left(\frac{\left(2-q^3\right)
   \tableau{1 1 1}}{\left(q^2+q+1\right)^2}+\frac{\left(q^4+3 q^3+3
   q^2+2 q-1\right)
   \tableau{2 1}}{\left(q^2+q+1\right)^2}+\frac{\left(q^4+q^3+3 q^2+2
   q+3\right) \tableau{3}}{\left(q^2+q+1\right)^2}\right)\nonumber\\
&&\hskip1.3cm+\ngv_{3}
   \left(24 \tableau{2 1}+3 \tableau{1 1 1}+30
   \tableau{3}\right)\nonumber\\
&&f_{1,4}=\ngv_{1} \left(\frac{q^3
   \tableau{1 1 1}}{q^3+q^2+q+1}+\frac{\left(q^3+q^2+2 q+1\right)
   \tableau{2 1}}{q^3+q^2+q+1}+\frac{\left(q^3+2 q^2+2\right)
   \tableau{3}}{q^3+q^2+q+1}\right)\nonumber\\
&&\hskip1.3cm+\ngv_{2} \left(\frac{(12 q+8)
   \tableau{2 1}}{q+1}+\frac{4 q
   \tableau{1 1 1}}{q+1}+\frac{4 (q+4)
   \tableau{3}}{q+1}\right)+\ngv_{4} \left(80
   \tableau{2 1}+16 \tableau{1 1 1}+80
   \tableau{3}\right)\nonumber
\eea
\end{small}

\noindent The functions for $r=4$ are
\begin{small}
\bea
&&f_{1,1}=\ngv_{1} \tableau{4},\nonumber \\
&&f_{1,2}=\ngv_{1}
   \left(\frac{\tableau{2 2}}{(q+1)^2}+\frac{q (q+2)
   \tableau{3 1}}{(q+1)^2}+\frac{\left(q^2+2 q+2\right)
   \tableau{4}}{(q+1)^2}\right)+\ngv_{2} \left(2
   \tableau{2 2}+6 \tableau{3 1}+10
   \tableau{4}\right),\nonumber \\
&&f_{1,3}=\ngv_{1} \left(\frac{q (q+1)
   \tableau{2 2}}{q^2+q+1}+\frac{\left(2 q^2+2 q+1\right)
   \tableau{3 1}}{q^2+q+1}+\frac{\tableau{2 1 1}}{q^2+q+1}+\frac{\left(q^2+q+3\right) \tableau{4}}{q^2+q+1}\right)\nonumber\\
&&\hskip1.3cm +\ngv_{3}
   \left(18 \tableau{2 2}+45 \tableau{3 1}+9
   \tableau{2 1 1}+45 \tableau{4}\right),\nonumber\\
&&f_{1,4}=\ngv_{2}
   \Big(\frac{8 \left(q^2+2 q+2\right) \tableau{2 2}}{(q+1)^2}+\frac{6
   \left(4 q^2+8 q+3\right) \tableau{3 1}}{(q+1)^2}+\frac{2 \left(4
   q^2+8 q+3\right) \tableau{2 1 1}}{(q+1)^2}+\frac{2
   \tableau{1 1 1 1}}{(q+1)^2}\nonumber\\
&&\hskip1.3cm+\frac{2 \left(8 q^2+16 q+11\right)
   \tableau{4}}{(q+1)^2}\Big)+\ngv_{1} \Big(\frac{\left(q^2+2
   q+2\right) \tableau{2 2}}{(q+1)^2}+\frac{\left(2 q^4+q^2-2\right)
   \tableau{1 1 1 1}}{(q+1)^2 \left(q^2+1\right)^2}\nonumber\\
&&\hskip1.3cm+\frac{q \left(3
   q^5+6 q^4+10 q^3+12 q^2+8 q+6\right) \tableau{3 1}}{(q+1)^2
   \left(q^2+1\right)^2}+\frac{\left(q^6+2 q^5+q^4+4 q^3+2 q^2+2 q+3\right)
   \tableau{2 1 1}}{(q+1)^2 \left(q^2+1\right)^2}\nonumber\\
&&\hskip1.3cm+\frac{\left(2 q^6+4
   q^5+5 q^4+8 q^3+7 q^2+4 q+5\right) \tableau{4}}{(q+1)^2
   \left(q^2+1\right)^2}\Big)\nonumber\\
&&\hskip1.3cm+\ngv_{4} \left(80 \tableau{2 2}+180
   \tableau{3 1}+60 \tableau{2 1 1}+4
   \tableau{1 1 1 1}+140 \tableau{4}\right)\nonumber
\eea
\end{small}
\ \\[-1cm]

\subsection{The projective line}
Below we collect some permutation equivariant invariants for $X=\IP^1$ and the ordinary invariants to which they sum up by \eqref{eq:RelOrdKTheory}. The invariants for the ordinary quantum K-theory have been computed before in \cite{IMT}. Using the standard basis $\Phi_\alpha=(1-P)^\alpha$, $\Phi^\alpha= \chi^{\alpha\beta}\Phi_\beta$, with the pairing $\chi_{\alpha\beta}=(\Phi_\alpha,\Phi_\beta)=\begin{pmatrix}1&1\\0&1\end{pmatrix}$,  we abbreviate the correlators of degree $k$ with $r+1$ marked points and $r$ permutation symmetric insertions as
\be
\left\langle\frac{\Phi^\alpha}{1-qL}; \Phi_1^r\right\rangle_{0,r+1,k}^{S_r} = f^{(r)}_{\alpha,k}\ ,\quad \alpha=0,1\, ,\quad k=1,2,3,\ldots \ .
\ee
\\[4mm]
\noindent $\bullet$ For  $r=0$:
\begin{small}
\bea
f_{0,1}&=&\frac{1}{1-q},\qquad f_{0,2}=-\frac{1}{(q-1)^3 (q+1)^2},
\qquad f_{0,3}=-\frac{1}{(q-1)^5 (q+1)^2 \left(q^2+q+1\right)^2},
\nonumber\\
f_{0,4}&=&-\frac{1}{(q-1)^7 (q+1)^4
   \left(q^2+1\right)^2 \left(q^2+q+1\right)^2},
\nonumber\\
f_{0,5}&=&-\frac{1}{(q-1)^9 (q+1)^4 \left(q^2+1\right)^2 \left(q^2+q+1\right)^2
   \left(q^4+q^3+q^2+q+1\right)^2}
\nonumber\\
f_{1,1}&=&-\frac{2 q}{(q-1)^2},\quad f_{1,2}=-\frac{2 q (2 q+1)}{(q-1)^4 (q+1)^3},\quad f_{1,3}=-\frac{2 q \left(3 q^3+4 q^2+3 q+1\right)}{(q-1)^6 (q+1)^3
   \left(q^2+q+1\right)^3},\nonumber\\
f_{1,4}&=&-\frac{2 q \left(4 q^5+5 q^4+7 q^3+5 q^2+3 q+1\right)}{(q-1)^8 (q+1)^5 \left(q^2+1\right)^3
   \left(q^2+q+1\right)^3},\nonumber\\
f_{1,5}&=&-\frac{2 q \left(5 q^9+11 q^8+19 q^7+24 q^6+26 q^5+22 q^4+16 q^3+9 q^2+4 q+1\right)}{(q-1)^{10}
   (q+1)^5 \left(q^2+1\right)^3 \left(q^2+q+1\right)^3 \left(q^4+q^3+q^2+q+1\right)^3}
\nonumber\eea
\end{small}
\\[2mm]
\noindent $\bullet$ For  $r=1$:
\begin{small}
\bea
f_{0,1}&=&-\frac{\tableau{1}}{q-1},\quad f_{0,2}=-\frac{\tableau{1}}{(q-1)^3
   (q+1)},\quad f_{0,3}=-\frac{\tableau{1}}{(q-1)^5 (q+1)^2 \left(q^2+q+1\right)},\nonumber\\
f_{0,4}&=&-\frac{\tableau{1}}{(q-1)^7 (q+1)^3
   \left(q^2+1\right) \left(q^2+q+1\right)^2},\nonumber\\
f_{0,5}&=&-\frac{\tableau{1}}{(q-1)^9 (q+1)^4 \left(q^4+q^3+q^2+q+1\right)
   \left(q^4+q^3+2 q^2+q+1\right)^2}\nonumber\\
f_{1,0}&=&\tableau{1},\quad f_{1,1}=-\frac{q \tableau{1}}{(q-1)^2},\quad f_{1,2}=-\frac{q (3 q+2) \tableau{1}}{(q-1)^4
   (q+1)^2},\nonumber\\
f_{1,3}&=&-\frac{q \left(5 q^3+7 q^2+6 q+2\right) \tableau{1}}{(q-1)^6 (q+1)^3 \left(q^2+q+1\right)^2},\nonumber\\
f_{1,4}&=&-\frac{q
   \left(7 q^5+9 q^4+13 q^3+10 q^2+6 q+2\right) \tableau{1}}{(q-1)^8 (q+1)^4 \left(q^2+1\right)^2
   \left(q^2+q+1\right)^3},\nonumber\\
f_{1,5}&=&-\frac{q \left(9 q^9+20 q^8+35 q^7+45 q^6+50 q^5+43 q^4+32 q^3+18 q^2+8 q+2\right)
   \tableau{1}}{(q-1)^{10} (q+1)^5 \left(q^4+q^3+q^2+q+1\right)^2 \left(q^4+q^3+2 q^2+q+1\right)^3}
\nonumber\eea
\end{small}
\\[2mm]
\noindent $\bullet$ For  $r=2$:
\begin{small}
\bea
f_{0,1}&=&\frac{\tableau{2}}{1-q},\quad f_{0,2}=-\frac{q \tableau{11}+\left(q^2+q+1\right)
   \tableau{2}}{(q-1)^3 (q+1)^2},\quad f_{0,3}=-\frac{q \tableau{11}+\left(q^2+1\right)
   \tableau{2}}{(q-1)^5 (q+1)^2 \left(q^2+q+1\right)},\nonumber\\
f_{0,4}&=&-\frac{q \left(q^2+q+1\right)
   \tableau{11}+\left(q^4+q^3+q^2+q+1\right) \tableau{2}}{(q-1)^7 (q+1)^4 \left(q^2+1\right)
   \left(q^2+q+1\right)^2},\nonumber\\
f_{0,5}&=&-\frac{q \left(q^2+1\right) \tableau{11}+\left(q^4+q^2+1\right)
   \tableau{2}}{(q-1)^9 (q+1)^4 \left(q^4+q^3+q^2+q+1\right) \left(q^4+q^3+2 q^2+q+1\right)^2}
\nonumber\\
f_{1,0}&=&
\frac{\tableau{2}-\tableau{11}}{q+1},\quad f_{1,1}=\frac{q
   \left(\tableau{11}-\tableau{2}\right)}{(q-1)^2 (q+1)},\quad f_{1,2}=-\frac{q \left(q (2 q+1)
   \tableau{11}+\left(2 q^3+4 q^2+5 q+2\right) \tableau{2}\right)}{(q-1)^4 (q+1)^3},\nonumber\\
f_{1,3}&=&-\frac{q \left(q
   \left(4 q^3+6 q^2+5 q+2\right) \tableau{11}+\left(4 q^5+6 q^4+10 q^3+9 q^2+6 q+2\right)
   \tableau{2}\right)}{(q-1)^6 (q+1)^3 \left(q^2+q+1\right)^2},
\nonumber\eea
\end{small}
\\[2mm]
\noindent $\bullet$ For  $r=3$:
\begin{small}
\bea
f_{0,1}&=&\frac{\tableau{3}}{1-q},\quad f_{0,2}=-\frac{q \tableau{21}+\left(q^2+1\right)
   \tableau{3}}{(q-1)^3 (q+1)},\nonumber\\ 
f_{0,3}&=&-\frac{q^3 \tableau{111}+(q+1)^2 \left(q^2+1\right) q
   \tableau{21}+\left(q^6+q^5+2 q^4+2 q^3+2 q^2+q+1\right) \tableau{3}}{(q-1)^5 (q+1)^2
   \left(q^2+q+1\right)^2},\nonumber\\
f_{0,4}&=&-\frac{q^3 \tableau{111}+\left(q^4+q^3+q^2+q+1\right) q
   \tableau{21}+\left(q^6+q^4+q^3+q^2+1\right) \tableau{3}}{(q-1)^7 (q+1)^3 \left(q^2+1\right)
   \left(q^2+q+1\right)^2},\nonumber\\
f_{1,0}&=&\frac{-\tableau{21}+\tableau{111}+\tableau{3}}{q^2+q+1}
\nonumber\eea
\end{small}
\\[2mm]
\noindent $\bullet$ For  $r=4$:
\begin{small}
\bea
f_{0,1}&=&\frac{\tableau{4}}{1-q},
\quad f_{0,2}=-\frac{q \left(\left(q^2+q+1\right) \tableau{31}+q \tableau{22}\right)+\left(q^4+q^3+q^2+q+1\right) \tableau{4}}{(q-1)^3 (q+1)^2},\nonumber\\
f_{0,3}&=&
-\frac{q^3   \tableau{211}+\left(q^3+q\right) q \tableau{22}+\left(q^4+q^3+q^2+q+1\right) q
   \tableau{31}+\left(q^6+q^4+q^3+q^2+1\right) \tableau{4}}{(q-1)^5 (q+1)^2
   \left(q^2+q+1\right)},\nonumber\\
f_{1,0}&=&\frac{\tableau{4}-\tableau{31}+\tableau{211}-\tableau{1111}}{q^3+q^2+q+1}
\nonumber\eea
\end{small}
\\[2mm]

\subsection{The projective surface}\label{app:P2}
In this appendix we expand on sect.~\ref{sec:P2} and collect for reference further permutation equivariant quantum K-invariants, which arise from the $J$-function $J_K^\text{eq}$ of the projective surface $\IP^2$ with the parameter dependent input
\be
t = a\Phi_1 + b\Phi_2 \ ,
\ee
which arises from the equivariant $J$-function \eqref{eq:ReconCP2} together with eq.~\eqref{eq:epsP2}. The equivariant correlators in the expansion of the $J$-function with input $t$ can also be interpreted as linear combinations of $S_n \times S_m$ equivariant correlators. Namely, the permutation equivariant correlators are not multi-linear but instead obey for the given input the equivariant correlator identity \cite{Giv15all}(p.I)
\be
\left\langle\frac{\Phi^\alpha}{1-qL}; (a\Phi_1 + b\Phi_2)^r\right\rangle_{0,r+1,k}^{S_r}
= \sum_{n=0}^r \left\langle\frac{\Phi^\alpha}{1-qL}; (a\Phi_1)^n; (b\Phi_2)^{r-n}\right\rangle_{0,r+1,k}^{S_n\times S_{r-n}} \ ,
\ee
in terms of the insertions of the elements $a\Phi_1 \equiv \Phi_1^{\oplus a}$ and $b\Phi_1 \equiv \Phi_1^{\oplus b}$ of the K-group $K(\IP^2)\otimes\mathbb{C}$. Introducing the abbreviation for the equivariant correlators
\be
\left\langle\frac{\Phi^\alpha}{1-qL}; (a\Phi_1 + b\Phi_2)^r\right\rangle_{0,r+1,k}^{S_r}
=f_{\alpha,k}^{(r)} \ ,
\ee
we list the first few correlators in the following:
\paragraph{For $r=0$:}
$$
\begin{aligned}
 f_{0,1}^{(0)} &= \frac{10 q^2-5 q+1}{(q-1)^4} \ , \quad
 f_{0,2}^{(0)} = -\frac{28 q^4+24 q^3-2 q^2-3 q+1}{(q-1)^7 (q+1)^5} \ , \\
 f_{0,3}^{(0)} &=  \frac{55 q^8+143 q^7+193 q^6+154 q^5+68 q^4+10 q^3-5 q^2-q+1}{(q-1)^{10} (q+1)^5 \left(q^2+q+1\right)^5} \ , \\
 f_{1,1}^{(0)} &= \frac{4 q-1}{(q-1)^3} \ , \quad
 f_{1,2}^{(0)} =-\frac{7 q^2+3 q-1}{(q-1)^6 (q+1)^4} \ , \quad
 f_{1,3}^{(0)} = \frac{10 q^4+13 q^3+9 q^2+2 q-1}{(q-1)^9 (q+1)^4 \left(q^2+q+1\right)^4} \ , \\
 f_{2,1}^{(0)} &= \frac{1}{(q-1)^2} \ , \quad
 f_{2,2}^{(0)} =-\frac{1}{(q-1)^5 (q+1)^3} \ , \quad
 f_{2,3}^{(0)} = \frac{1}{(q-1)^8 (q+1)^3 \left(q^2+q+1\right)^3} \ .
\end{aligned}
$$
\paragraph{For $r=1$:}
\begin{footnotesize}
$$
\begin{aligned}
 f_{0,1}^{(1)} &= \frac{ \left(6 a q^2-4 a q+a-3 b q^3+6 b q^2-4 b q+b\right)\tableau{1}}{(q-1)^4} \ , \\
 f_{0,2}^{(1)} &= \frac{\left(-21 a q^4-21 a q^3+a q^2+3 a q-a+15 b q^6+18 b q^5-15 b q^4-21 b q^3+b q^2+3 b q-b\right)\tableau{1}}{(q-1)^7 (q+1)^4} \ , \\
 f_{0,3}^{(1)} &=  -\frac{\tableau{1}}{(q-1)^{10} (q+1)^5   \left(q^2+q+1\right)^4} (-45 a q^8-120 a q^7-171 a q^6-143 a q^5-67 a q^4-11 a q^3 \\
   &\qquad+5 a q^2+a q-a+36 b q^{11}+99 b q^{10}+150 b q^9+96 b q^8-33 b q^7-138 b q^6-137 b q^5\\
   &\qquad-67 b q^4-11 b q^3+5 b q^2+b q-b)\ , \\
 f_{1,1}^{(1)} &= \frac{\left(3 a q-a-2 b q^2+3 b q-b\right)\tableau{1}}{(q-1)^3} \ , \quad
 f_{1,2}^{(1)} = \frac{\left(-6 a q^2-3 a q+a+5 b q^4+3 b q^3-6 b q^2-3 b q+b\right)\tableau{1}}{(q-1)^6 (q+1)^3} \ , \\
 f_{1,3}^{(1)} &=\frac{\left(9 a q^4+12 a q^3+9 a q^2+2 a q-a-8 b q^7-11 b q^6-9 b q^5+6 b q^4+12 b q^3+9 b q^2+2 bq-b\right)\tableau{1}}
     {(q-1)^9 (q+1)^4 \left(q^2+q+1\right)^3} \ , \\
 f_{2,1}^{(1)} &= -\frac{(-a+b q-b)\tableau{1}}{(q-1)^2} \ , \quad
 f_{2,2}^{(1)} = \frac{\left(-a+b q^2-b\right)\tableau{1}}{(q-1)^5 (q+1)^2} \ , \quad
 f_{2,3}^{(1)} = -\frac{\left(-a+b q^3-b\right)\tableau{1}}{(q-1)^8 (q+1)^3 \left(q^2+q+1\right)^2} \ .
\end{aligned}
$$
\end{footnotesize}
\paragraph{For $r=2$:}
\begin{footnotesize}
$$
\begin{aligned}
 f_{0,1}^{(2)} &= \frac{ \tableau{1 1}}{2 (q-1)^4 (q+1)}
   (3 a^2 q^3-2 a^2 q+a^2-2 a b q^4+4 a b q^3-4 a b q+2 a b-3 a q^3-6 a q^2+4 a q-a\\
   &\qquad-b^2 q^4+2 b^2 q^3-2 b^2 q+b^2+3b q^4-6 b q^2+4 b q-b)
   +\frac{\tableau{2}}{2 (q-1)^4 (q+1)}(3 a^2 q^3-2 a^2q \\
   &\qquad+a^2-2 a b q^4+4 a b q^3-4 a b q+2 a b+3 a q^3+6 a q^2-4 a q+a-b^2 q^4+2 b^2 q^3-2 b^2 q+b^2\\
   &\qquad-3 b q^4+6 b q^2-4 bq+b)\ ,\\
 f_{0,2}^{(2)} &= -\frac{ \tableau{1 1}}{2 (q-1)^7 (q+1)^5}
   (15 a^2 q^6+48 a^2 q^5+51 a^2 q^4+15 a^2 q^3-5 a^2 q^2-a^2 q+a^2-20 a b q^8\\
   &\qquad-70 a b q^7-62 a b q^6+42 a b q^5+90a b q^4+30 a b q^3-10 a b q^2-2 a b q+2 a b-15 a q^6-18 a q^5\\
   &\qquad-27 a q^4-21 a q^3+a q^2+3 a q-a+6 b^2 q^9+14 b^2 q^8-5 b^2q^7-31 b^2 q^6-9 b^2 q^5+21 b^2 q^4\\
   &\qquad+9 b^2 q^3-5 b^2 q^2-b^2 q+b^2-6 b q^9+18 b q^8+51 b q^7+27 b q^6-21 b q^5-45 b q^4-27b q^3\\
   &\qquad+b q^2+3 b q-b)-\frac{\tableau{2}}{2 (q-1)^7 (q+1)^5}
   (15 a^2 q^6+48 a^2q^5+51 a^2 q^4+15 a^2 q^3-5 a^2 q^2-a^2 q\\
   &\qquad+a^2-20 a b q^8-70 a b q^7-62 a b q^6+42 a b q^5+90 a b q^4+30 a b q^3-10 a b q^2-2 a b q+2 a b\\
   &\qquad+15 a q^6+18 a q^5+27 a q^4+21 a q^3-a q^2-3 a q+a+6 b^2 q^9+14 b^2 q^8-5 b^2 q^7-31 b^2 q^6\\
   &\qquad-9 b^2q^5+21 b^2 q^4+9 b^2 q^3-5 b^2 q^2-b^2 q+b^2+6 b q^9-18 b q^8-51 b q^7-27 b q^6+21 b q^5\\
   &\qquad+45 b q^4+27 b q^3-b q^2-3 bq+b) \ , \\
 f_{1,1}^{(2)} &= \frac{\tableau{1 1}}{2 (q-1)^3 (q+1)}(2 a^2 q^2+a^2 q-a^2-2 a b q^3+2 a b q^2+2 a b q-2 a b-2 a q^2-3 a q+a-b^2 q^3\\
   &\qquad+b^2 q^2+b^2 q-b^2+3 b q^3-b q^2-3b q+b)+\frac{\tableau{2}}{2 (q-1)^3 (q+1)} 
   (2 a^2 q^2+a^2 q-a^2-2 a b q^3\\
   &\qquad+2 ab q^2+2 a b q-2 a b+2 a q^2+3 a q-a-b^2 q^3+b^2 q^2+b^2 q-b^2-3 b q^3+b q^2+3 b q-b)\ , \\
 f_{1,2}^{(2)} &=-\frac{ \tableau{1 1}}{2 (q-1)^6 (q+1)^4}
   (5 a^2 q^4+13 a^2 q^3+10 a^2 q^2+a^2 q-a^2-8 a b q^6-22 a b q^5-10 a b q^4\\
   &\qquad+20 a b q^3+20 a b q^2+2 a b q-2 a b-5a q^4-3 a q^3-6 a q^2-3 a q+a+3 b^2 q^7+5 b^2 q^6-5 b^2 q^5\\
   &\qquad-11 b^2 q^4+b^2 q^3+7 b^2 q^2+b^2 q-b^2-3 b q^7+9 b q^6+15 b
   q^5-b q^4-9 b q^3-9 b q^2-3 b q+b)\\
   &\qquad-\frac{\tableau{2}}{2 (q-1)^6 (q+1)^4}
   (5a^2 q^4+13 a^2 q^3+10 a^2 q^2+a^2 q-a^2-8 a b q^6-22 a b q^5-10 a b q^4\\
   &\qquad+20 a b q^3+20 a b q^2+2 a b q-2 a b+5 a q^4+3 aq^3+6 a q^2+3 a q-a+3 b^2 q^7+5 b^2 q^6-5 b^2 q^5\\
   &\qquad-11 b^2 q^4+b^2 q^3+7 b^2 q^2+b^2 q-b^2+3 b q^7-9 b q^6-15 b q^5+b q^4+9
   b q^3+9 b q^2+3 b q-b) \ , \\
 f_{2,1}^{(2)} &=-\frac{(a-1) (-a+2 b q-2 b) \tableau{1 1}}{2 (q-1)^2}-\frac{(a+1) (-a+2 b q-2 b)\tableau{2}}{2 (q-1)^2} \ , \\
 f_{2,2}^{(2)} &=-\frac{\tableau{1 1}}{2 (q-1)^5 (q+1)^3}(a^2 q^2+2 a^2 q+a^2-2 a b q^4-4 a b q^3+4 a b q+2 a b-a q^2-a\\
 &\qquad+b^2 q^5+b^2 q^4-2 b^2 q^3-2 b^2 q^2+b^2 q+b^2-bq^5+3 b q^4+2 b q^3-2 b q^2-b q-b)\\
 &\qquad-\frac{\tableau{2} }{2 (q-1)^5 (q+1)^3} (a^2
   q^2+2 a^2 q+a^2-2 a b q^4-4 a b q^3+4 a b q+2 a b+a q^2+a+b^2 q^5\\
  &\qquad+b^2 q^4-2 b^2 q^3-2 b^2 q^2+b^2 q+b^2+b q^5-3 b q^4-2 bq^3+2 b q^2+b q+b)\ .
\end{aligned}
$$
\end{footnotesize}
\paragraph{For $r=3$:}
\begin{footnotesize}
$$
\begin{aligned}
 f_{0,1}^{(3)} &= \frac{\tableau{2 1}}{3 (q-1)^4
   \left(q^2+q+1\right)}(a^3 q^4-a^3 q^3-a^3 q+a^3+3 a^2 b q^4-3 a^2 b q^3-3 a^2 b q+3 a^2 b\\
   &\qquad+3 a b^2 q^4-3 a b^2 q^3-3 a b^2 q+3 a b^2-aq^4-5 a q^3-6 a q^2+4 a q-a+9 b q^4-9 b q^3)\\
   &\qquad+\frac{\tableau{1 1 1}}{6 (q-1)^4 (q+1) \left(q^2+q+1\right)}
   (a^3 q^5-a^3 q^3-a^3 q^2+a^3+3 a^2 b q^5-3 a^2 b q^3-3 a^2 b q^2\\
   &\qquad+3 a^2 b-3 a^2 q^5-12a^2 q^4-3 a^2 q^3-3 a^2 q^2+6 a^2 q-3 a^2+3 a b^2 q^5-3 a b^2 q^3-3 a b^2 q^2\\
   &\qquad+3 a b^2+6 a b q^5-12 a b q^4+6 a b q^3-6 ab q^2+12 a b q-6 a b+2 a q^5+12 a q^4+22 a q^3\\
   &\qquad+4 a q^2-6 a q+2 a+6 b^2 q^5-12 b^2 q^4+6 b^2 q^3-6 b^2 q^2+12 b^2 q-6b^2-18 b q^5+18 b q^3)\\
   &\qquad+\frac{\tableau{3}}{6 (q-1)^4 (q+1) \left(q^2+q+1\right)}(a^3 q^5-a^3 q^3-a^3 q^2+a^3+3 a^2 b q^5-3 a^2 b q^3-3 a^2 b q^2\\
   &\qquad+3 a^2 b+3 a^2 q^5+12 a^2 q^4+3 a^2 q^3+3 a^2 q^2-6a^2 q+3 a^2+3 a b^2 q^5-3 a b^2 q^3-3 a b^2 q^2\\
   &\qquad+3 a b^2-6 a b q^5+12 a b q^4-6 a b q^3+6 a b q^2-12 a b q+6 a b+2 aq^5+12 a q^4+22 a q^3\\
   &\qquad+4 a q^2-6 a q+2 a-6 b^2 q^5+12 b^2 q^4-6 b^2 q^3+6 b^2 q^2-12 b^2 q+6 b^2-18 b q^5+18 bq^3) \ ,\\
 f_{1,1}^{(3)} &= \frac{ \tableau{2 1}}{3 (q-1)^3 \left(q^2+q+1\right)}(a^3 q^3-a^3+3 a^2 b q^3-3 a^2 b-3 a b^2 q^4+3 a b^2 q^3+3 a b^2 q-3 a b^2\\
   &\qquad-a q^3-3 a q^2-3 a q+a+3 b q^4-3 bq)
   +\frac{\tableau{1 1 1}}{6 (q-1)^3 (q+1)
   \left(q^2+q+1\right)}(a^3 q^4+a^3 q^3\\
   &\qquad-a^3 q-a^3+3 a^2 b q^4+3 a^2b q^3-3 a^2 b q-3 a^2 b-3 a^2 q^4-9 a^2 q^3-6 a^2 q^2-3 a^2 q+3 a^2 \\
   &\qquad-3 a b^2 q^5+3 a b^2 q^3+3 a b^2 q^2-3 a b^2+3 a bq^5+3 a b q^4-6 a b q^3-3 a b q^2-3 a b q+6 a b \\
   &\qquad+2 a q^4+8 a q^3+12 a q^2+4 a q-2 a+6 b^2 q^5-6 b^2 q^3-6 b^2 q^2+6 b^2-6b q^5-6 b q^4\\
   &\qquad+6 b q^2+6 b q)
   +\frac{\tableau{3}}{6 (q-1)^3 (q+1) \left(q^2+q+1\right)}(a^3 q^4+a^3 q^3-a^3 q-a^3+3 a^2 b q^4\\
   &\qquad+3 a^2 b q^3-3 a^2 b q-3 a^2b+3 a^2 q^4+9 a^2 q^3+6 a^2 q^2+3 a^2 q-3 a^2-3 a b^2 q^5+3 a b^2 q^3\\
   &\qquad+3 a b^2 q^2-3 a b^2-3 a b q^5-3 a b q^4+6 a b q^3+3a b q^2+3 a b q-6 a b+2 a q^4+8 a q^3\\
   &\qquad+12 a q^2+4 a q-2 a-6 b^2 q^5+6 b^2 q^3+6 b^2 q^2-6 b^2-6 b q^5-6 b q^4+6 b q^2+6 bq) \ , \\
 f_{2,1}^{(3)} &= -\frac{(a-1) (a+1) (-a+3 b q-3 b) \tableau{2 1}}{3 (q-1)^2}-\frac{(a-2)(a-1)(-a+3 b q-3 b) \tableau{1 1 1}}{6 (q-1)^2} \\
   &\qquad-\frac{(a+1)(a+2)(-a+3 b q-3 b)\tableau{3}}{6 (q-1)^2} \ .
\end{aligned}
$$
\end{footnotesize}

\newpage
\bibliographystyle{JHEP}
\bibliography{JM_QK}
\end{document}